    \documentclass[twocolumn,trackchanges]{aastex701}

    \newcommand{\COO}{$^{12}$CO(1-0)}

    \newcommand{\kms}{km~s$^{-1}$}
    \newcommand{\noicm}{{\tt noICM}}
    
    \newcommand{\icmpw}{{\tt ICM-P3}}
    \newcommand{\icmps}{{\tt ICM-P7}}

    \newcommand{\kpc}{\rm kpc}

    \defcitealias{2004AJ....127.3375V}{\scshape V04}
    
    \usepackage{graphicx}	% Including figure files
    \usepackage{amsmath}	% Advanced maths commands
    \usepackage{newtxtext,newtxmath}
    \usepackage{makecell,tabularx}
    \usepackage{xspace}

    \usepackage{newtxtext,newtxmath}
    \usepackage{makecell,tabularx}
    \usepackage[T1]{fontenc}
    
    \DeclareRobustCommand{\VAN}[3]{#2}
    \let\VANthebibliography\thebibliography
    \def\thebibliography{\DeclareRobustCommand{\VAN}[3]{##3}\VANthebibliography}

\begin{document}
    
    \title[Radio Continuum Properties of NGC~4522]{The impact of ram pressure on the radio spectral index and magnetic field of NGC~4522:\\ A high-resolution VLA continuum study}

    \author[orcid=0000-0001-5033-7208,sname='Choi']{Woorak Choi}
    \affiliation{Department of Physics and Astronomy, McMaster University, Hamilton, ON L8S 4M1, Canada}
    \affiliation{Department of Astronomy, Yonsei University, 50 Yonsei-ro, Seodaemun-gu, Seoul 03722, Republic of Korea}
    \email[show]{woorak.c@gmail.com}  
    
    \author[orcid=0000-0003-1440-8552,sname='Chung']{Aeree Chung}
    \affiliation{Department of Astronomy, Yonsei University, 50 Yonsei-ro, Seodaemun-gu, Seoul 03722, Republic of Korea}
    \email[show]{achung@yonsei.ac.kr}  
    
    \author[orcid=0000-0003-2896-3725,sname='Kim']{Chang-Goo Kim}
    \affiliation{Department of Astrophysical Sciences, Princeton University, 4 Ivy Lane, Princeton, NJ 08544, USA}
    \email{}  
    
    \author[orcid=0000-0002-3810-1806,sname='Lee']{Bumhyun Lee}
    \affiliation{Department of Astronomy, Yonsei University, 50 Yonsei-ro, Seodaemun-gu, Seoul 03722, Republic of Korea}
    \email{}  
    
    \author[orcid=0000-0002-7422-9823,sname='Cortese']{Luca Cortese}
    \affiliation{International Centre for Radio Astronomy Research, University of Western Australia, M468, 35 Stirling Highway, Crawley, WA 6009, Australia}
    \email{}  
    
    \author[orcid=0000-0003-1845-0934,sname='Brown']{Toby Brown}
    \affiliation{Herzberg Astronomy and Astrophysics Research Centre, National Research Council of Canada, 5071 West Saanich Rd, Victoria, BC, V9E 2E7, Canada}
    \email{}  
    
    \author[orcid=0000-0002-7625-562X,sname='Catinella']{Barbara Catinella}
    \affiliation{International Centre for Radio Astronomy Research, University of Western Australia, M468, 35 Stirling Highway, Crawley, WA 6009, Australia}
    \email{}  
    
    \author[orcid=0000-0002-6155-7166,sname='Emsellem']{Eric Emsellem}
    \affiliation{European Southern Observatory, Karl-Schwarzschild-Stra{\ss}e 2, Garching, 85748, Germany}
    \affiliation{Univ Lyon, Univ Lyon1, ENS de Lyon, CNRS, Centre de Recherche Astrophysique de Lyon UMR5574, 69230 Saint-GenisLaval, France}
    \email{}  
    
    \author[orcid=0000-0001-9557-5648,sname='Fraser-McKelvie']{A. Fraser-McKelvie}
    \affiliation{European Southern Observatory, Karl-Schwarzschild-Stra{\ss}e 2, Garching, 85748, Germany}
    \email{}  
    
    \author[orcid=0000-0003-0378-4667,sname='Sun']{Jiayi Sun}
    \affiliation{Department of Physics and Astronomy, University of Kentucky, 506 Library Drive, Lexington, KY 40506, USA}
    \email{}  
    
    \author[orcid=0000-0002-9405-0687,sname='Watts']{Adam Watts}
    \affiliation{International Centre for Radio Astronomy Research, University of Western Australia, M468, 35 Stirling Highway, Crawley, WA 6009, Australia}
    \email{}  
    
    %\collaboration{all}{MAUVE team}
    
    \begin{abstract}
    
    We present high-resolution Very Large Array (VLA) continuum observations at S-band ($3$~GHz, $560$~pc scale) and X-band ($10$~GHz, $200$~pc scale) of the ram-pressure-stripped Virgo galaxy NGC~4522, to investigate the characteristics of its radio continuum, spectral index, and magnetic field under the influence of the intracluster medium (ICM). The total radio continuum shows an asymmetry that extends northwest, mirroring the \ion{H}{1} gas distribution, but showing distinct features in the extraplanar regions. The spectral index steepens systematically from $\alpha\sim-0.6$ in the main disk to $\alpha\sim-1.1$ in the outer disk. We find that the spectral index behavior of the outer disk is mainly due to an ICM shock that can re-accelerate electrons and a significant reduction of thermal emission. Intriguingly, extraplanar clouds exhibit exceptionally flat spectral indices ($\alpha\sim-0.2$ to $0$), resulting from a combination of significantly enhanced thermal emission and pronounced spectral aging of the non-thermal component. Although some of these regions correlate with H$\alpha$, others do not. We propose that the mixing between the ICM and interstellar medium (ISM) is an alternative mechanism that enhances thermal emission independently of star formation. Polarized continuum emissions are highly asymmetric, preferentially distributed along the ICM wind side, and the polarization fraction increases radially outward from the galactic midplane, indicating that the polarized emission is strongly influenced by the ICM wind. Our results show how and where the ICM substantially affects the ISM, and also demonstrate that high-frequency observations are crucial for analyzing the radio continuum of ram pressure stripping galaxies.
    
    \end{abstract}
    
    \keywords{\uat{Galaxies}{573} --- \uat{Galaxy clusters}{1772} --- \uat{Interstellar medium}{847} --- \uat{Radio interferometry}{1346} --- \uat{Radio continuum emission}{1340} --- \uat{Spectral index}{1553} --- \uat{Extragalactic magnetic fields}{507}}
    
    
    \section{Introduction}
    
    Galaxy clusters represent the densest large-scale environments in the universe, and galaxies near cluster cores tend to be redder and more elliptical than those in cluster outskirts \citep{dressler1980}. Among various environmental mechanisms, ram pressure stripping (RPS), caused by the intracluster medium (ICM), can efficiently remove the interstellar medium (ISM) of galaxies on short timescales, leading to passive galaxy evolution \citep{crowl2008}. Observations of neutral hydrogen (\ion{H}{1}) provide compelling evidence for RPS, revealing truncated and disturbed gas disks as well as gaseous tails in cluster galaxies \citep[e.g.,][]{1990AJ....100..604C,2009AJ....138.1741C,2019MNRAS.487.4580R,Wang_2021_wallaby}. Star formation (SF) in these galaxies is likewise affected, showing both SF suppression due to gas loss \citep[e.g.,][]{2004ApJ...613..851K,vulcani2020_sfr_resolved} and localized SF enhancement from gas compression \citep[e.g.,][]{abramson2011,2014ApJ...780..119K,vulcani_2020_sfr}.

    For galaxies undergoing RPS, radio continuum observation, which is a powerful tracer of the magnetized ISM, provides an especially insightful diagnostic tool. The dominant component of this emission (especially at frequencies below $\sim10$~GHz) is typically nonthermal synchrotron radiation \citep[e.g.,][]{gioia1982_rc_spec,condon1992_rc_review,2001SSRv...99..243B}, which arises from cosmic-ray electrons (CRe) spiraling in galactic magnetic fields. Consequently, radio continuum observations provide a direct probe of magnetic field strength and structure. The CRe themselves are primarily accelerated in the supernova remnants, linking the radio emission to the star formation cycle over timescales of $\sim 100~{\rm Myr}$. Thermal (free-free) radiation from HII regions ionized by young, massive O/B stars also contributes to the radio continuum at $1$ -- $10$~GHz, typically accounting for $\approx 10$ -- $40$\% \citep[e.g.,][]{gioia1982_rc_spec,2001SSRv...99..243B,tabatabaei2017_spcidx_flat_sf}, tracing shorter timescale star formation ($\sim 10~{\rm Myr}$). Therefore, the total radio continuum is sensitive to both the overall magnetic field strength and star formation activity, while polarized emission particularly traces the ordered magnetic field\footnote{Following the convention in the literature (e.g., \citealt{beck2015}), we distinguish between the \textit{total} field, traced by the total continuum, and the \textit{ordered} field, which produces polarized continuum emission. In this paper, we use the term `ordered magnetic field' when discussing the polarized emission in general, and `regular magnetic field' when specifically referring to the large-scale, coherent component.}.
    
    The synchrotron emission spectrum is determined by the energy distribution of CRe, expressed as $j(E) \propto E^s$. The flux density $F_\nu$ is related to frequency $\nu$ by $F_\nu \propto \nu^\alpha$, where the spectral index $\alpha$ is derived from the CRe spectral slope $s$ via $\alpha = (s+1)/2$. Since the intrinsic CRe energy spectrum is often approximated by a power-law with a slope of $-3 \lesssim s \lesssim -2$ at $E = 1$ -- $10$~GeV (e.g., \citealt{orlando_2018_cre,padovani_2018_cre}), the resulting spectral index, $\alpha$, generally falls between $-0.5$ and $-1.0$. As CRe propagate through the ISM, they undergo frequency-dependent energy losses \citep{Longair_2011_cre}, such as synchrotron and inverse Compton losses. High-energy CRe are particularly susceptible to these processes, leading to a systematic steepening of the spectral index. In contrast, thermal (free-free) emission exhibits an almost flat spectral index ($\approx -0.1$) at frequencies of a few GHz, as expected from its emission mechanism. In general, isolated spiral galaxies exhibit axisymmetric total continuum morphologies with enhanced emission in spiral arm regions, while the strongest polarized emission and well-ordered spiral magnetic field patterns are found in the interarm regions \citep[e.g.,][]{beck2015}. Their spectral indices vary from about $-0.5$ in star-forming regions within the disk to $-1.3$ or steeper in the outer regions of galaxies \citep[e.g.,][]{klein_1984_spectral_index,hummel_1991_spectral_index,vargas_2018_specindx}.

    The external pressure from the ICM compresses the ISM on the galaxy's leading side, which can enhance magnetic field strengths, potentially triggering bursts of star formation \citep[e.g.,][]{crowl2006_ngc4522_sfr,abramson2011,vulcani_2018_sf_burst}. Meanwhile, in the stripped tail, the radio continuum traces displaced gas, cosmic rays, and magnetic fields as they are pulled away from the galactic disk and interact with the surrounding ICM. 
    Indeed, previous full-polarization radio continuum observations of ram pressure stripped Virgo spirals have revealed strongly asymmetric total emission and enhanced polarized emission along the leading side, with magnetic field vectors oriented roughly  perpendicular to the ICM wind direction \citep[e.g.,][]{chyzy2007_bfield,2004AJ....127.3375V, 2007A&A...464L..37V, 2010A&A...512A..36V, 2013A&A...553A.116V}. Although these studies revealed that the cluster environment can compress the ISM and order magnetic fields, they found little evidence that ram pressure directly affects the spectral index. Instead, the spectral index appears to be governed primarily by local star-forming activity. 
    
    More recently, high-sensitivity low-frequency surveys with LOFAR have significantly expanded the sample of known RPS galaxies. These observations have revealed extended radio tails and complex magnetic field structures not only in massive clusters but also in galaxy groups, demonstrating that nonthermal emission tails are frequent signatures of the stripping process \citep[e.g.,][]{Ignesti_2023,Roberts_2021,Roberts_2022,Roberts_2024}.
    However, the relatively coarse spatial resolutions ($\sim$~1.5 -- 3~kpc) of previous observations have limited our understanding of where, and to what extent, the ICM influences the ISM and its magnetic field in detail. This also hampers detailed multiwavelength comparisons. In addition, separating the thermal emission component has been challenging from MHz to GHz frequencies ($144$~MHz of LOFAR, $1.4$~GHz of L-band, and $5$~GHz of C-band) because of the low thermal fraction.    
    
    In this paper, we present high-resolution S-band ($3$~GHz) and X-band ($10$~GHz) continuum observations of NGC~4522 obtained with the Karl G. Jansky Very Large Array (VLA). These data provide spatial resolutions $2$ -- $5$ times higher than in previous studies. Our multi-band continuum observations enable us to construct detailed spectral index maps, which help disentangle the thermal emission from the dominant nonthermal synchrotron component and address several key questions:
    Are there signatures of cosmic-ray electron (CRe) re-acceleration driven by ICM shocks? How does the balance between thermal and nonthermal emission vary across the galaxy? How does the ICM influence the galactic magnetic field? Most importantly, where and how does the interaction between the ISM and the hot ICM occur?

    This paper is organized as follows. \autoref{sec:sec2} describes the target and observations. In \autoref{sec:sec3}, we present the overall morphology and spectral index distribution of the total continuum.  \autoref{sec:sec4_pol} focuses on the polarized emission and magnetic field structure. In \autoref{sec:discussion}, we discuss our results in the context of previous simulations and recent MUSE observations. In this section, we present a schematic overview of our main findings, which highlights the distinct physical processes identified across the galaxy. Finally, we summarize our main findings in \autoref{sec:sec_summary}.

    \section{Target and observations}\label{sec:sec2}
    
    \subsection{NGC 4522}
    
\begin{figure}
      \centering
      \includegraphics[width=\columnwidth]{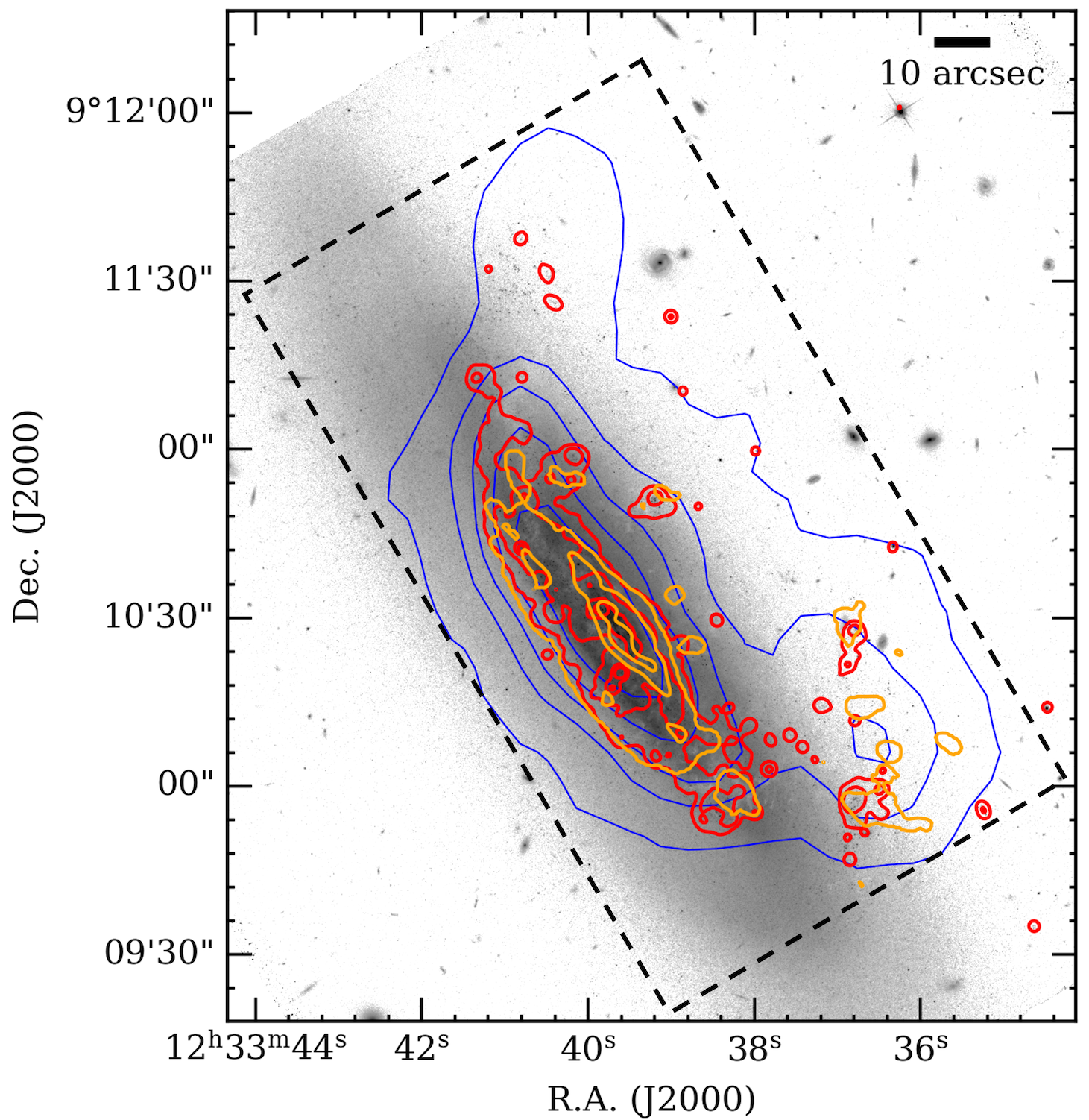}
      \caption{A composite map showing H$\alpha$ (red contours, \citealt{Boselli_2018_vestige}), ALMA \COO\, (yellow contours, \citealt{2018ApJ...866L..10L}), \ion{H}{1} (blue contours, \citealt{2009AJ....138.1741C}), and optical (background, \textit{HST} WFC F814W) of NGC~4522. This galaxy has a highly asymmetric \ion{H}{1} disk with extraplanar molecular gas and associated star-forming regions. The black dashed box indicates the region that is the focus of this study.}
      \label{fig:ngc4522}
    \end{figure}

    NGC~4522 is a Virgo Cluster spiral galaxy that shows clear evidence of ongoing \ion{H}{1} stripping by ram pressure from the ICM. Basic physical parameters of NGC 4522 are summarized in \autoref{tab:ngc4522}. Numerous observational studies have focused on this galaxy, including those based on \ion{H}{1}, H${\alpha}$, CO, and radio continuum data \citep[e.g.,][]{kenney2004_4522,2004AJ....127.3375V,2018ApJ...866L..10L,Boselli_2018_vestige}, as shown in \autoref{fig:ngc4522}.
    NGC~4522 is the nearest and clearest example in which multiple gas tracers, including molecular gas, are detected outside the stellar disk. It thus serves as an ideal laboratory to explore how ram pressure influences the multi-phase and magnetized ISM, the coupling among different components, and the stripping conditions that depend on the ISM density.
    
    \citealt{2004AJ....127.3375V} (hereafter \citetalias{2004AJ....127.3375V}) observed this galaxy in the L-band ($20$~cm) and C-band ($6$~cm) radio continuum at $50$ and $15$ arcsec resolutions, respectively, revealing an asymmetry similar to that seen in the \ion{H}{1} distribution at both frequencies. The C-band polarized continuum emission was also found to be highly asymmetric, predominantly located near the ICM wind front (the eastern side), with its peak offset from that of the total continuum. They further reported that the polarization fraction increases radially outward from the galaxy center and that the spectral index steepens from east to west. However, as noted earlier, their analysis was limited to the global distribution and trends of the (polarized) continuum because of the insufficient spatial resolution. To examine in greater detail how the ISM responds to the external ICM pressure, we have conducted new full-polarization VLA continuum observations of NGC~4522 with significantly improved resolution compared to previous studies.
    
\begin{table}
\centering
\caption{Basic properties of NGC~4522}
\label{tab:ngc4522}
\begin{tabular*}{\columnwidth}{@{\extracolsep{\fill}}lc}
\hline \hline
Quantity & Value \\
\hline
Type$^a$  &  SBc       \\
R.A. (J2000)$^a$ & $12^{\mathrm{h}}33^{\mathrm{m}}39^{s}.7$   \\
Dec. (J2000)$^a$ &   $+9^{\circ}10\arcmin30\farcs2$    \\
Distance (Mpc)$^b$ &   16.5   \\
Position Angle ($^{\circ}$)$^a$ &   35\\
Inclination ($^{\circ}$)$^a$ &   79\\
Systemic velocity (km s$^{-1}$)$^b$ &   2331 \\
$M_{\mathrm{H_I}}$ (M$_{\odot}$)$^b$  &   $5.64\times10^8$ \\
$M_{\mathrm{H_2}}$ (M$_{\odot}$)$^c$  &   $3.0\times10^8$ \\
$M_{\mathrm{stellar}}$ (M$_{\odot}$)$^d$  &  $2\times10^{9}$ \\
\hline
\end{tabular*}
\begin{minipage}{\columnwidth}
\vspace{0.1cm}
$^{a}$ HyperLeda \citep{Makarove2014_hyperleda}\\
$^{b}$ \citet{2009AJ....138.1741C}.\\
$^{c}$ \citet{2008A&A...491..455V}.\\
$^{d}$ \citet{cortese2012_ngc4522_mass}.\\
\end{minipage}
\end{table}

    \subsection{High-Resolution VLA Polarized Continuum Observations}
    
    We conducted VLA continuum observations of NGC~4522 at $3$~GHz (S-band) in February 2020 and at $10$~GHz (X-band) in April 2020, with bandwidths of $2$~GHz and $4$~GHz for the S- and X-bands, respectively, using the C-configuration (Project ID: 20A-310, PI: W. Choi). The full width at half power of the S- and X-bands is $15\arcmin$ and $4.5\arcmin$, respectively, which is sufficiently large to cover the entire stellar disk and extraplanar gas of NGC~4522 (diameter of $\approx4\arcmin$). To achieve a signal-to-noise ratio (SNR) of 3 for polarized emission, on-source integration times of $81$ and $280$ minutes were used for the S- and X-bands, respectively, assuming a polarization fraction of $\approx 7\%$ at $3$~GHz \citep{2004AJ....127.3375V,2013A&A...553A.116V}. This ensures a significantly higher SNR for the total continuum emission. To maximize the sensitivity of full-polarization observations, we used an 8-bit sampler for the S-band and a 3-bit sampler for the X-band. 3C~286 was used as the bandpass, flux, and polarization-angle calibrator, and J1239+0730 was used to calibrate the polarization leakage of the antennas. The observational setup is summarized in \autoref{tab:ngc4522_vla_obs}.

    The data were initially processed using the pipeline of the \textsc{Common Astronomy Software Applications package} (\textsc{CASA}; \citealt{mcmullin2007_casa}) 6.4.1 version, provided by NRAO. Additional radio frequency interference (RFI) flagging was manually performed using the task \texttt{flagdata}. To obtain both total and polarized continuum data, we followed the standard polarization calibration procedure provided by NRAO. Imaging was performed using the \texttt{tclean} task. We generated the mask interactively and cleaned the calibrated data to a depth of twice the root mean square (RMS) noise of the dirty image using this mask. Briggs weighting with robust of 0.5 and 1.5 was applied to the S- and X-band, respectively, during the cleaning process. We also employed the \textsc{Multi-scale Multi-frequency} synthesis (MT-MFS, \citealt{rau2011_mult}) algorithm to improve image quality, taking into account the relatively large bandwidth. Primary beam (PB) corrections are then applied to both bands using the \texttt{impbcor} task. The PB correction had a marginal effect, increasing the flux by only up to 1\% and 4\% at the $4\times\rm RMS$ noise level for the S- and X-bands, respectively, as the extent of our source is sufficiently small compared to the PB in both bands. Finally, we obtained Stokes I, Q, U, and V images for S- and X-bands with synthesized beams of $7\farcs8\times6\farcs0$ and $2\farcs8\times2\farcs2$\, achieving RMS noise levels of 6~$\mu$Jy beam$^{-1}$ and 1.5~$\mu$Jy beam$^{-1}$ for S and X-band, respectively. A detailed description of the calibration procedures and continuum imaging is provided in \autoref{app:sec_appA}.

\begin{table}[t]
    \centering
    \caption{NGC~4522 VLA Observation Parameters}
    \label{tab:ngc4522_vla_obs}   
    \begin{tabular*}{\columnwidth}{@{\extracolsep{\fill}}lcc}
    \hline \hline
    Parameter & S-band & X-band \\
     & (3~GHz) & (10~GHz) \\
    \hline
    Date of observations & 2020 Feb. 18 & 2020 Apr. 18 -- 21\\
    Configuration & C & C \\
    No. of antennas & 26 & 27 \\
    No. of spectral windows & 16 & 32 \\
    No. of channels per spectral & 64 & 64 \\ 
    window & & \\
    Total no. channels & 1024 & 2048 \\
    Channel separation (MHz) & 2.0 & 2.0 \\
    Total bandwidth (GHz) & 2.024 & 4.048 \\
    On-source time (min) & 81 & 280 \\
    Synthesized beam (arcsec)$^a$ & 6.9 & 2.5 \\
    RMS noise ($\mu$Jy beam$^{-1}$)$^a$ & 6.0 & 1.5 \\
    Flux calibrator$^b$ & 3C 286 & 3C 286 \\
    Phase calibrator & J1239+0730 & J1239+0730 \\
    Pol. leakage calibrator & J1407+2827 & J1407+2827 \\
    \hline
    \end{tabular*}
    \begin{minipage}{\columnwidth}
    \vspace{0.1cm}
    \footnotesize
    $^{a}$ Briggs weightings of 0.5 and 1.5 were applied to S-band and X-band, respectively.\\
    $^{b}$ This calibrator was also employed as the bandpass and polarization-angle calibrator.
    \end{minipage}
\end{table}

    \section{Total Continuum and Spectral Index}\label{sec:sec3}
    
    \subsection{Distribution of Continuum Emission}\label{sec:sec3_total_int}
    
    % Describe the overall total intensity of two bands and compare them with \ion{H}{1} and previous radio continuum observation 
    
    %\textcolor{red}{I'll revise this subsection. I'll make the description looks rich.}
    
\begin{figure*}
      \centering
      \includegraphics[width=1.7\columnwidth]{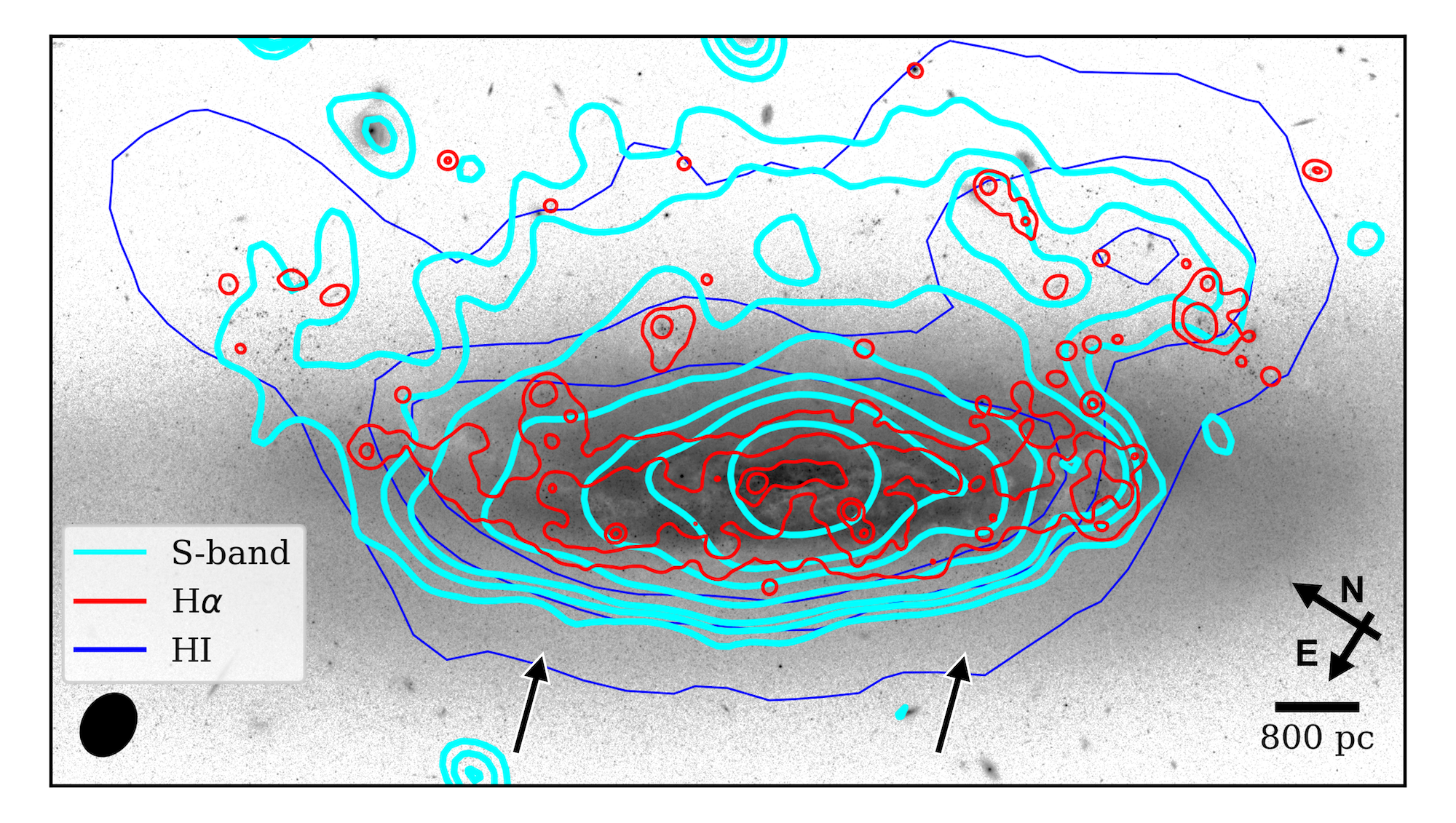}\\
      \vspace{-0.2cm}
      \includegraphics[width=1.7\columnwidth]{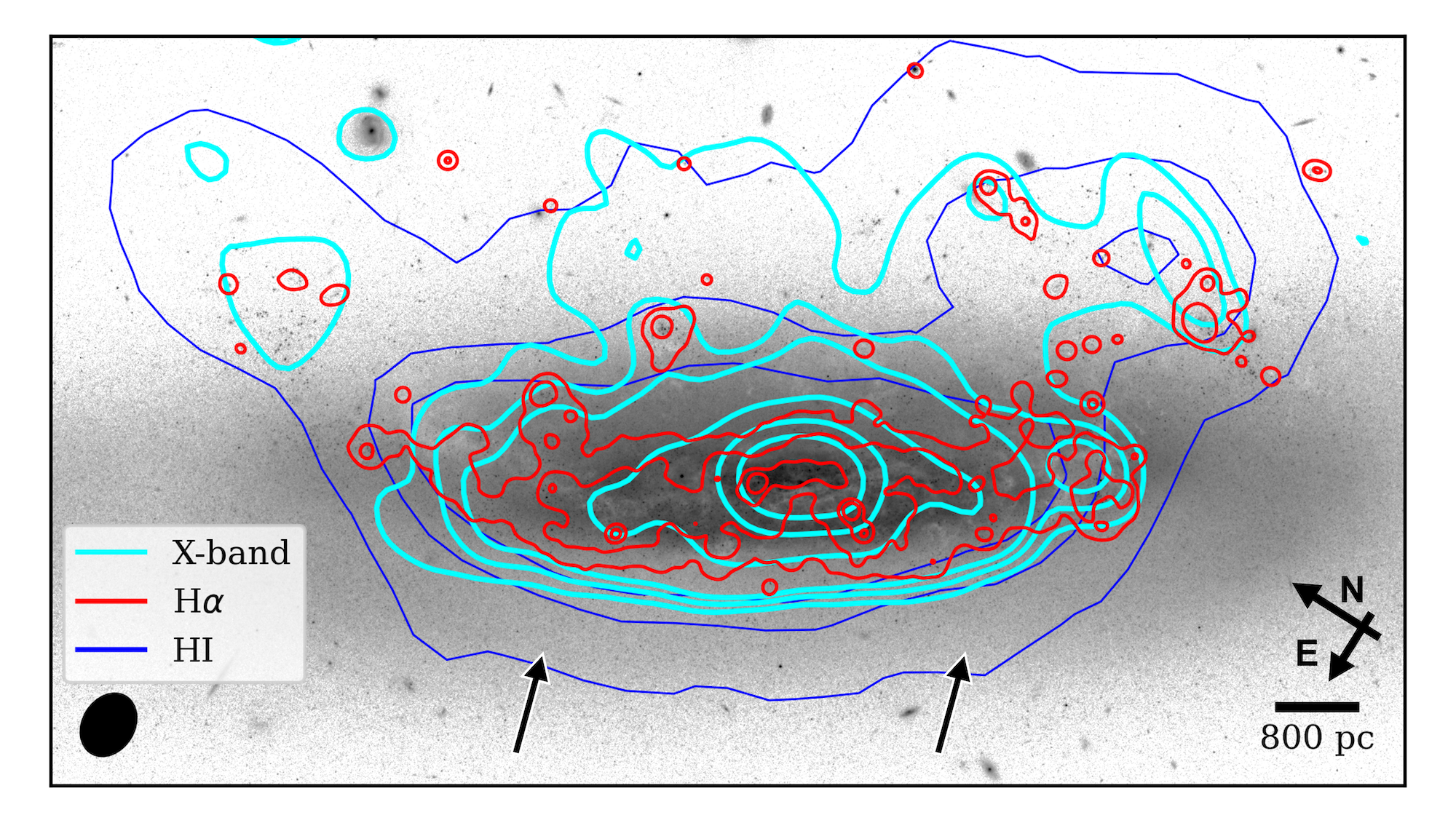}
      \caption{\textbf{Top:} A composite map of S-band total continuum intensity (cyan contours) overlaid with H$\alpha$ (red contours), %, ALMA \COO\, (yellow contours, \citealt{2018ApJ...866L..10L}), 
     \ion{H}{1} (blue contours), and optical (background, \textit{HST} WFC F814W) of NGC~4522. The contour levels of the S-band continuum are (4, 7, 10, 20, 35, 50, 90) $\times$ $6$~$\mu$Jy~beam$^{-1}$. The synthesized beam of $7\farcs8\times6\farcs0$ is shown in the bottom-left corner. Compared to the stellar disk, the S-band continuum emission appears asymmetric, extending toward the northwest, similar to the \ion{H}{1} gas distribution. \textbf{Bottom}: Same as the top panel but for the X-band total continuum convolved to the S-band resolution. The contour levels of the X-band continuum are (4, 6, 8, 20, 35, 50) $\times$ $5.5$~$\mu$Jy~beam$^{-1}$. The convolved synthesized beam of $7\farcs8\times6\farcs0$ is shown in the bottom-left corner. At this resolution and sensitivity, the X-band continuum is also extended toward the northwest. The expected direction of the ICM wind is indicated by the black arrow.}
      \label{fig:ngc4522_total_S}
    \end{figure*}

    The total continuum emission at both the S- and X-bands, shown in \autoref{fig:ngc4522_total_S}, exhibits a highly asymmetric morphology relative to the stellar disk. The emission is clearly compressed on the southeastern side (bottom of the figure) and extends into a broad tail toward the northwest (top of the figure). This morphology mirrors the stripped \ion{H}{1} gas distribution and aligns with the expected direction of the ICM wind, providing clear evidence that the radio continuum-emitting gas responds to ram pressure. At our sensitivity limit, there are regions where S-band emission is detected, but no X-band continuum is observed. The overall morphology of the continuum at these two bands is consistent with that seen in previous C-band ($6$~cm)  observations of this target (\citetalias{2004AJ....127.3375V}).

\begin{figure*}
      \centering
      \includegraphics[width=1.7\columnwidth]{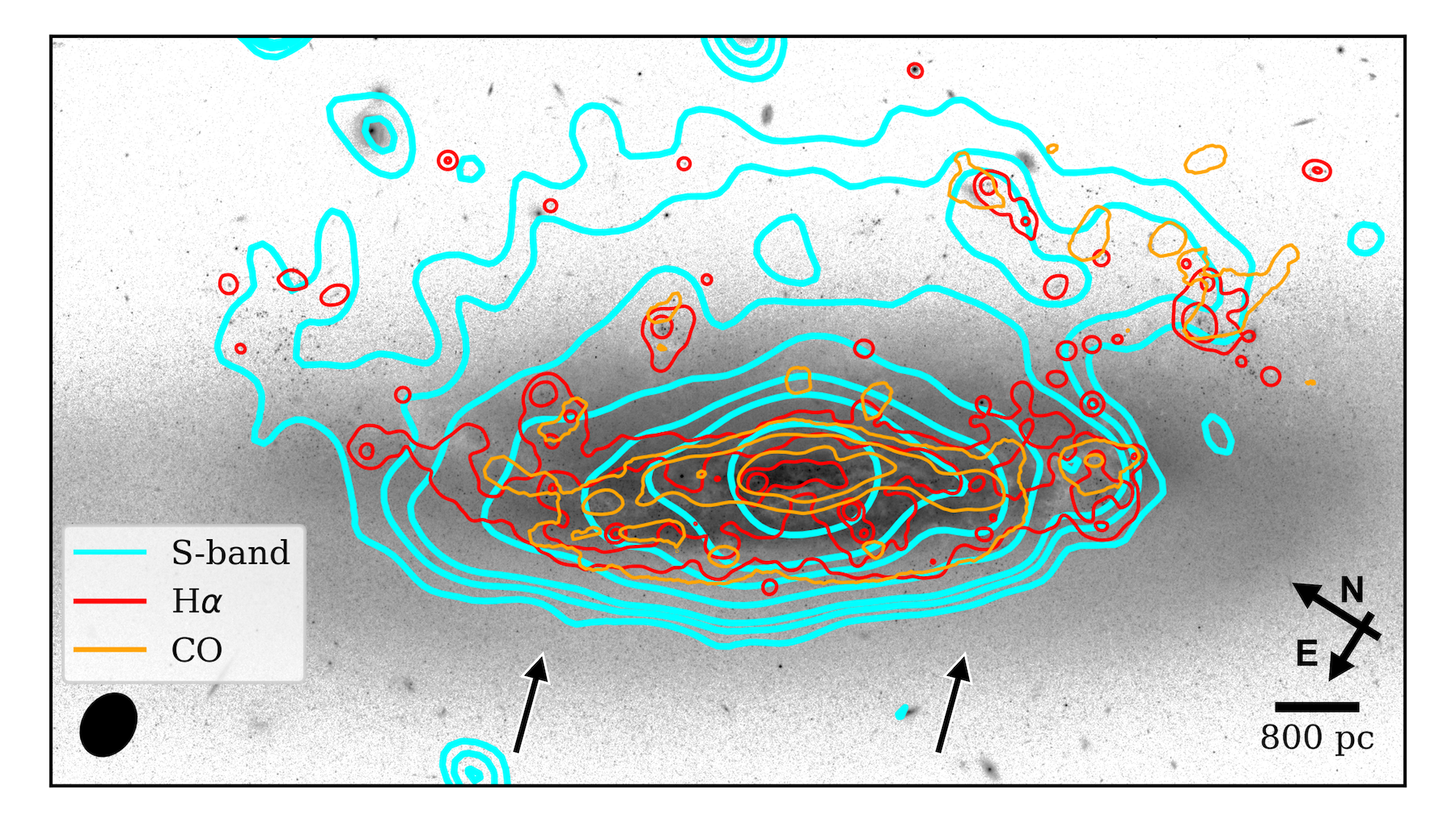}\\
      %\vspace{-0.2cm}
      \vspace{-0.2cm}
      \includegraphics[width=1.7\columnwidth]{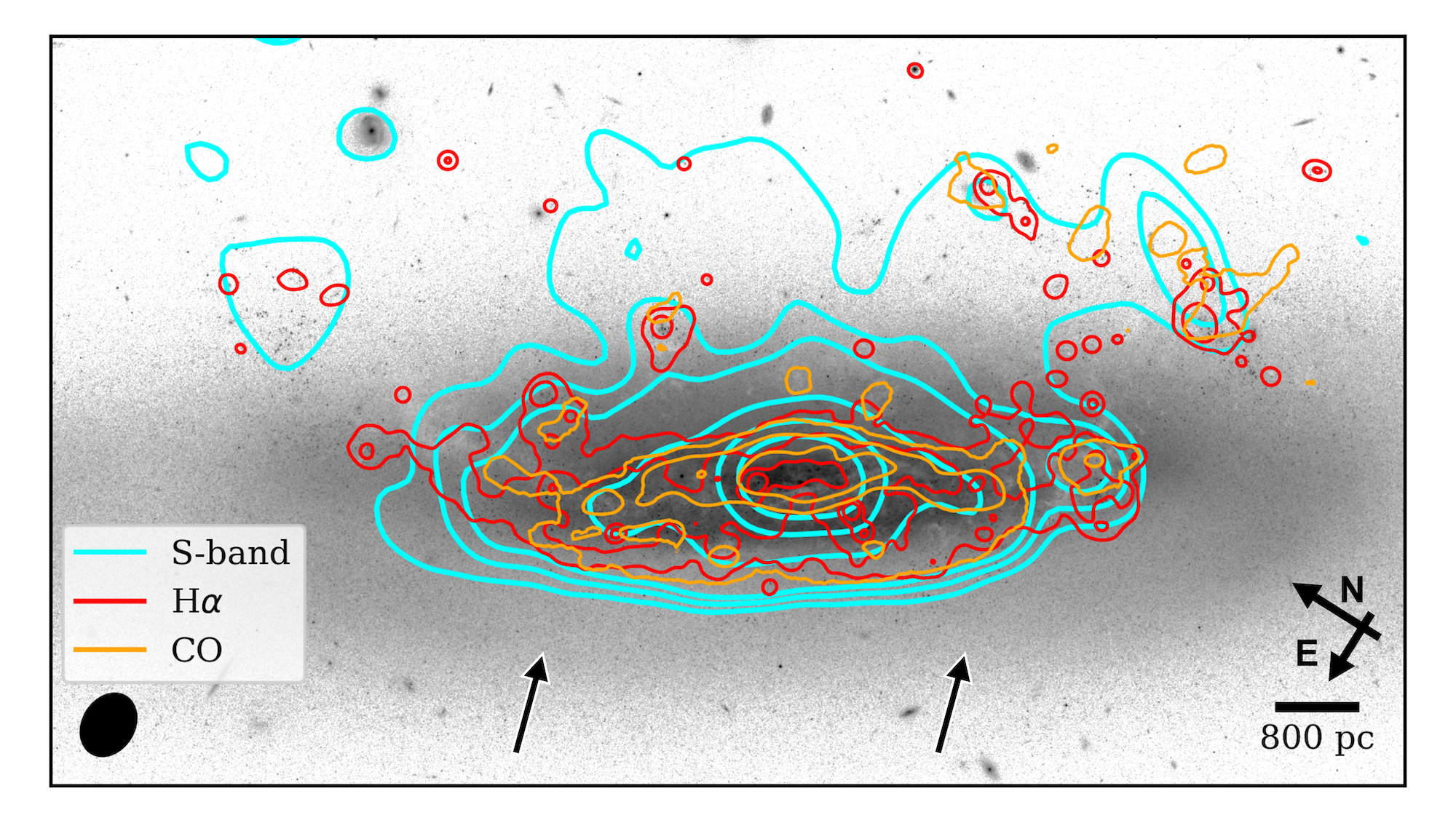}
      \caption{Same as \autoref{fig:ngc4522_total_S}, but overlaid with ALMA \COO\ emission, (yellow contours, \citealt{2018ApJ...866L..10L}) instead of \ion{H}{1}.}
      \label{fig:ngc4522_total_with_co}
    \end{figure*}

    The overall distributions within the stellar disk are similar across multi-wavelength observations (\autoref{fig:ngc4522_total_S} and \ref{fig:ngc4522_total_with_co}). At the western edge, where \ion{H}{1} is truncated, a sharp truncation is seen not only in the continuum but also in H$\alpha$\footnote{Due to the H$\alpha$ filter of the Canada-France-Hawaii Telescope, this H$\alpha$ data includes a contribution from [\text{\ion{N}{2}}] line; however, we will refer to it simply as H$\alpha$.} and molecular gas. In contrast, at the eastern edge, H$\alpha$ and molecular gas are less extended than the continuum and \ion{H}{1}. The continuum emission is clearly detected in the extraplanar region (upper-right), where H$\alpha$ and molecular gas are present. Interestingly, the continuum in this region at both bands is broadly correlated with H$\alpha$ and molecular gas, although spatial offsets exist between their local peaks. For example, in the northwest, H$\alpha$ and a CO clump are adjacent but only partially overlap, while the continuum extends across both. Moreover, some CO clumps lack corresponding H$\alpha$ but appear associated with local continuum peaks. For reference, the resolutions of the H$\alpha$ ($\approx 3''$) and CO ($\approx 2.3''$) observations are comparable.

    We measure the total flux of NGC~4522 to be $12.64\pm0.41$~mJy at S-band and $4.14\pm0.20$~mJy at X-band. We assumed a systematic uncertainty of $\sim3\%$ on the absolute flux scale for both bands \citep{perley2017}, which is accounted for in the spectral index calculation. These values are in good agreement with the overall galaxy spectral energy distribution implied by the L-band ($20$~cm) flux of $24.3$~mJy and the C-band ($6$~cm) flux of $7.6$~mJy reported by \citetalias{2004AJ....127.3375V}. These earlier fluxes yield a spectral index of $-0.97$. Our S- and X-band fluxes follow the power law extrapolated from the L- and C-band measurements, yielding a consistent spectral index of $-0.93\pm0.05$ across the entire L- to X-band range. The spatially-resolved spectral index distributions are further discussed in \autoref{sec:sec3_specind}.

    \subsection{Distribution of Spectral Indices}\label{sec:sec3_specind}
    
    The radio continuum emission at centimeter wavelengths consists of thermal bremsstrahlung (free-free) and nonthermal synchrotron emission. Each component has a distinct spectral index $\alpha$, which represents the slope of the power-law relation between flux density $S$ and frequency $\nu$ ($S \propto \nu^{\alpha}$). In star-forming galaxies, between $1$ and $30$~GHz \citep[e.g.,][]{condon1992_rc_review}, the nonthermal and thermal components typically have spectral indices of $\alpha_{\mathrm{NT}} \approx -0.8$ and $\alpha_{\mathrm{T}} \approx -0.1$, respectively. We derive a total spectral index of $-0.93\pm0.05$ from the integrated S- and X-band fluxes, indicating that nonthermal emission dominates over this frequency range in NGC~4522. This value is consistent with the result obtained from L- and C-band data by \citetalias{2004AJ....127.3375V} ($-0.97$) and with the typical indices found in field spiral galaxies \citep[e.g.,][]{gioia1982_rc_spec}, which also apply to Virgo spirals \citep[e.g.,][]{2013A&A...553A.116V}. However, since the global spectral index is largely dominated by the disk emission, it is therefore necessary to examine its spatial variation to investigate the effects of ram pressure. This is particularly crucial for the extraplanar regions, as the continuum emission there may arise from distinct physical processes.

    %C-band flux missing is large, so they did not show spectral index map of C-band.
    
    %L-band 5-sigma total intensity cutoff. Spectral index of average of the maps is around -0.83.
    
    %Halo spectral index is steeper than that of the disk. (-0.9 to -2)

    To generate the spectral index map between S- and X-bands, we first matched the resolution and pixel scale of the two images (as shown in \autoref{fig:ngc4522_total_S} and \ref{fig:ngc4522_total_with_co}). We then produced the spectral index map using the \textsc{CASA} task \texttt{immath} with mode=`spix'. Only pixels with convolved X-band emission above $4\times\sigma_{\mathrm{rms,convol}}$ ($\sigma_{\mathrm{rms,convol}}\approx5.5\mu$Jy~beam$^{-1}$) were included. In \autoref{app:sec_appB}, we present X-band images generated with different uv coverages, as well as spectral index maps derived from these images, to assess the effect of convolution. As discussed in \autoref{app:sec_appB}, we confirm that the convolution process does not affect the robustness of our results.
    
\begin{figure*}
      %\centering
      \includegraphics[width=\linewidth]{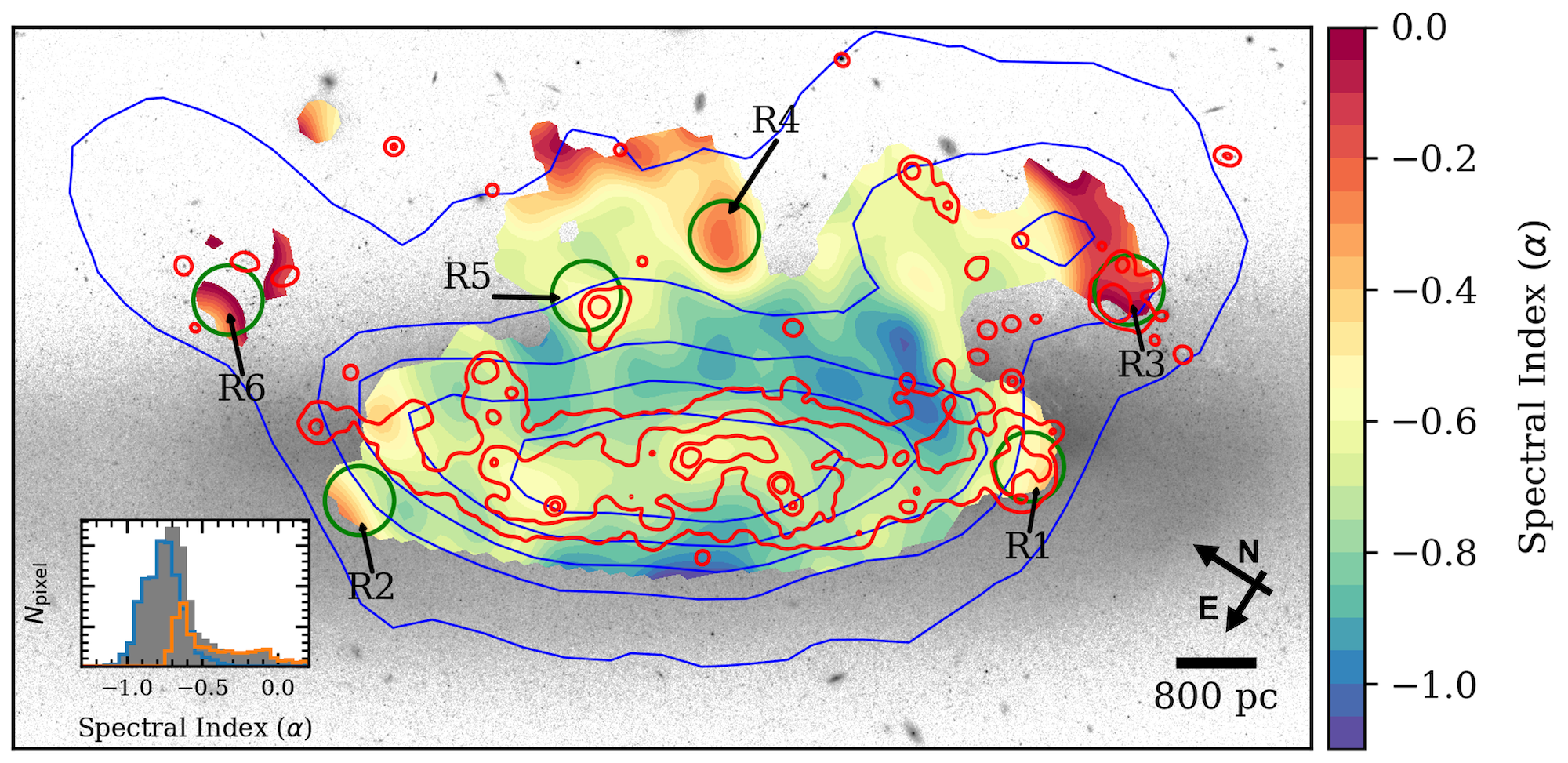}
      \caption{Spectral index map between the S-band and X-band total continuum. \ion{H}{1} and H$\alpha$ are shown as blue and red contours, respectively. Only pixels where the convolved X-band emission exceeds $4\times\sigma_{\mathrm{rms,convol}}$ ($\sigma_{\mathrm{rms,convol}}\approx5.5\mu$Jy~beam$^{-1}$) are displayed. Six regions of interest (R1 -- R6) are indicated by green circles. The spectral index generally steepens from the midplane to large scale heights, but increases again in the extraplanar regions, where very complex variations and extremely flat spectral indices are observed. The inset shows the spectral index distribution, with orange and blue histograms representing the extraplanar region (R3 -- R6 and their vicinity) and the rest of the galaxy, respectively. Overall, the spectral index is slightly flatter ($-1.2$ to $0.1$) than that reported by \citetalias{2004AJ....127.3375V} from L- and C-band data ($-2$ to $-0.7$).}
      \label{fig:ngc4522_specindx_sx}
    \end{figure*}

    \autoref{fig:ngc4522_specindx_sx} shows the spectral index map between the S- and X-band. Based on a pixel-by-pixel analysis, we divide the galaxy into four regions: (1) the main disk (within H$\alpha$ contour; $-0.8 \lesssim \alpha \lesssim -0.5$), (2) the main-disk edges (R1 and R2), (3) the outer disk (just outside the H$\alpha$ contour; $-1.1 \lesssim \alpha \lesssim -0.6$), and (4) the extraplanar region (R3 -- R6 and their vicinity; $-0.6 \lesssim \alpha \lesssim 0$).

    Across the main disk (defined by the H$\alpha$ contour), the spectral index gradually steepens from the galactic center toward the outskirts, with typical values ranging from $\alpha\approx-0.8$ to $-0.5$. Regions showing relatively flat spectral indices ($-0.6$ to $-0.5$) generally coincide with areas of bright H$\alpha$ emission.

    %implying that the contribution to the total continuum of thermal emission is larger than that of the rest of the target. The gradual outward steepening may arise from two effects: a decrease in the thermal emission contribution and "spectral aging" which is the natural result of the radiative energy loss of electrons that cause the highest energy electrons to lose energy first. 
    
    The main disk edges (R1 and R2), where \ion{H}{1} is truncated by RPS, tend to show a flattening of the spectral index from  $-0.6$ to $-0.3$. The flattening in R1 is likely due to a nearby strong H$\alpha$ region that enhances the thermal emission contribution, whereas no prominent H$\alpha$ is found in R2. Enhanced ICM-ISM interactions along the disk edges may also contribute to the spectral flattening, as discussed further in \autoref{sec:discussion}.

    The outer disk, defined as the region outside the outermost H${\alpha}$ contour but still within the stellar disk, shows a noticeably steeper spectral index than the main disk, with typical values of $\alpha\approx-1.1$ to $-0.6$. This region can be divided into two parts: the leading side, which corresponds to the near side of the outer disk \citep{2014AJ....147...63A} where the ICM wind compressed the continuum and \ion{H}{1}, and the northern region, corresponding to the far side of the outer disk. Hereafter, we refer to this northern region as the `far side'. However, due to the high inclination of the galaxy, the emission observed in this region is likely a superposition of the far side of the outer disk and the stripped gas locating at the near side (specifically, trailing gases projected along the line-of-sight). Consequently, the observed continuum here includes contributions from these near side stripped components, which are also likely to interact with the ICM wind. The steepening on the leading side mainly appears in the middle of the galaxy, whereas on the far side it is more widespread.

    %Since the leading side is directly affected by the ICM, steep spectral indices might be due to the re-acceleration of CRe from the ICM shocks. However, the far side is less likely to be severely affected by the ICM, implying that the steepening is rather due to other effects such as the decrease of contribution of thermal emission and the spectral aging or their combination. The origin of steep spectral indices in these regions will be discussed in more detail in \autoref{sec:discuss_spcidx}.
    
    %This indicates that the nonthermal emission dominates the total continuum. In addition,  
    
    %Intriguingly, the spectral index flattens significantly in the extraplanar region, which shows a mix of complex structures and changes. Some regions show spectral indices similar to those of the main disk (e.g., R5, $-0.7$ to $-0.5$), while others exhibit exceptionally flat spectral indices (e.g., R3, R4, R6, $-0.3$ to $0$). To our knowledge, such flat spectral indices have never been revealed in previous continuum observations for RPS galaxies especially in extraplanar regions. Near zero spectral indices imply complete dominance of thermal emission. However, star formation as traced by the H$\alpha$ emission is not always detected in these regions, which may require either the enhancement of thermal emission without corresponding star formation or the significant reduction in nonthermal emission. We conjecture that the mixing between the ISM and the ICM allows the ISM to emit thermal continuum emission. We will discuss this in more detail in \autoref{sec:discuss_spcidx}.    
    
    Most strikingly, the spectral index significantly flattens in the extraplanar region, displaying a complex mixture of structures and variations. While some regions show spectral indices comparable to those of the main disk (e.g., R5, $-0.7$ to $-0.5$), others exhibit exceptionally flat spectral values (e.g., R3, R4, R6, $-0.3$ to $0$). Among these regions, H$\alpha$ emission is detected in  R3 and R5, whereas it is nearly absent in R4 and R6. To our knowledge, such flat spectral indices have not been reported in previous continuum observations of RPS galaxies, particularly in extraplanar regions. The origin of these unusually flat spectra, especially in areas lacking clear signs of star formation, suggests the presence of complex physical processes, which we discuss in detail in \autoref{sec:discussion}.

    Compared to the spectral index map of \citetalias{2004AJ....127.3375V}, our results show a similar trend within the stellar disk, namely, a gradual steepening of the spectral index toward the outer disk. However, the overall spectral indices in our map are systematically flatter ($-1.1$ to $0$) than those reported by \citetalias{2004AJ....127.3375V} ($-2$ to $-0.6$). In addition, our high-resolution spectral index map reveals several intriguing features that were not identified in \citetalias{2004AJ....127.3375V}: (1) a steepening in the outer disk followed by a flattening in the extraplanar region, (2) a very steep spectral index on the leading side, and (3) locally flattened spectral indices along the main disk edges. We note that all of these newly detected features arise from differences in both spatial resolution and observing frequency.

    \section{Polarized Continuum Emission}\label{sec:sec4_pol}

    In this section, we present new high‐resolution S-band and X-band polarization maps of NGC~4522 to investigate in detail how the ICM wind shapes the galactic magnetic field. The polarized intensity, $P$, and the polarization angle, $\chi$, were derived from the Stokes $Q$ and $U$ maps, which were generated using the same imaging parameters as the total-intensity maps. We then constructed the polarized intensity and polarization angle maps as $P=\sqrt{Q^2+U^2}$ and $\chi=(1/2)\arctan(U/Q)$.
    
    \subsection{General Properties of the Polarized Emission}\label{sec:sec4_gen}
    
\begin{figure*}
      \centering
      \includegraphics[width=0.8\linewidth]{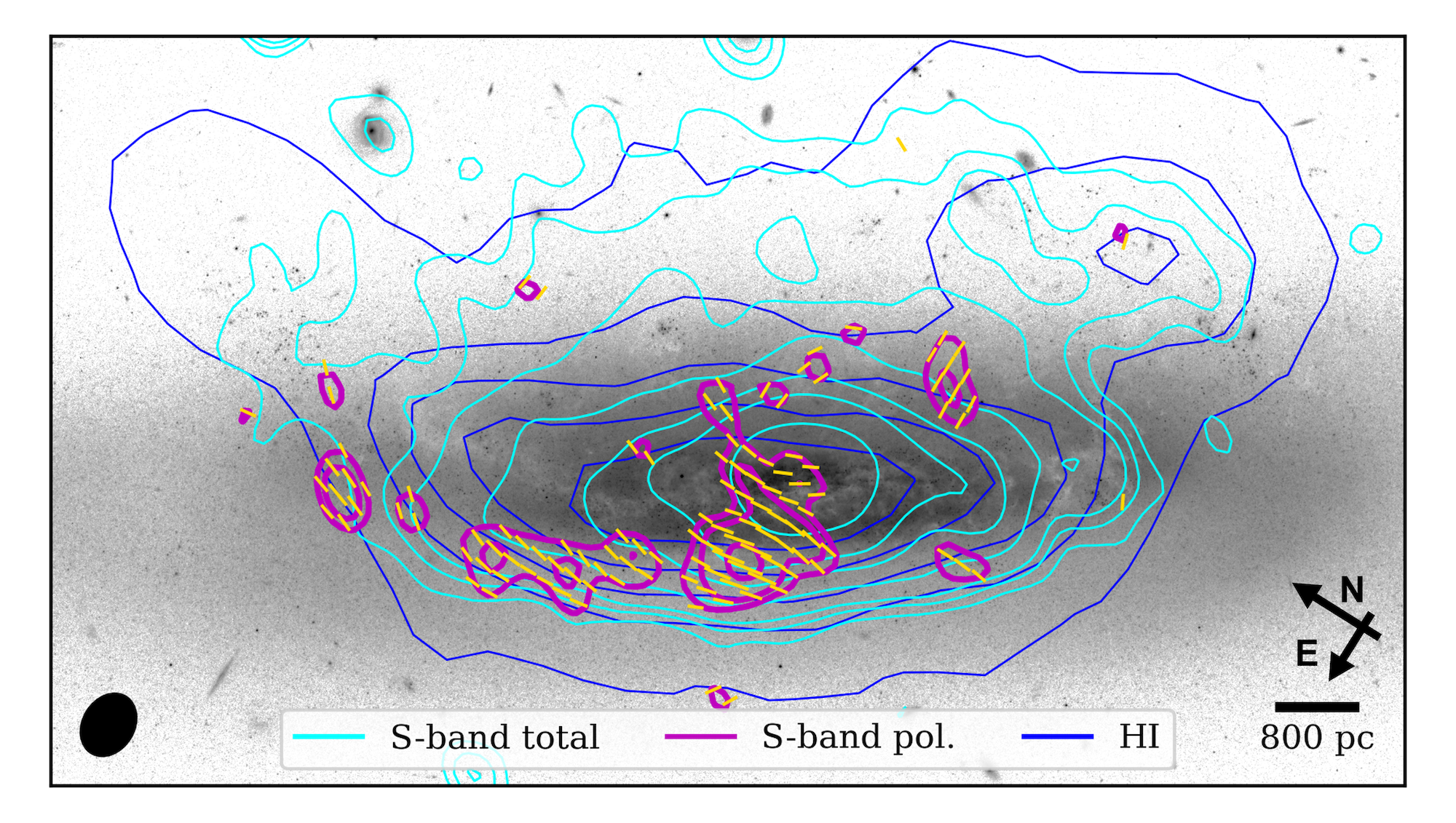}\\
      \vspace{-0.2cm}
      \includegraphics[width=0.8\linewidth]{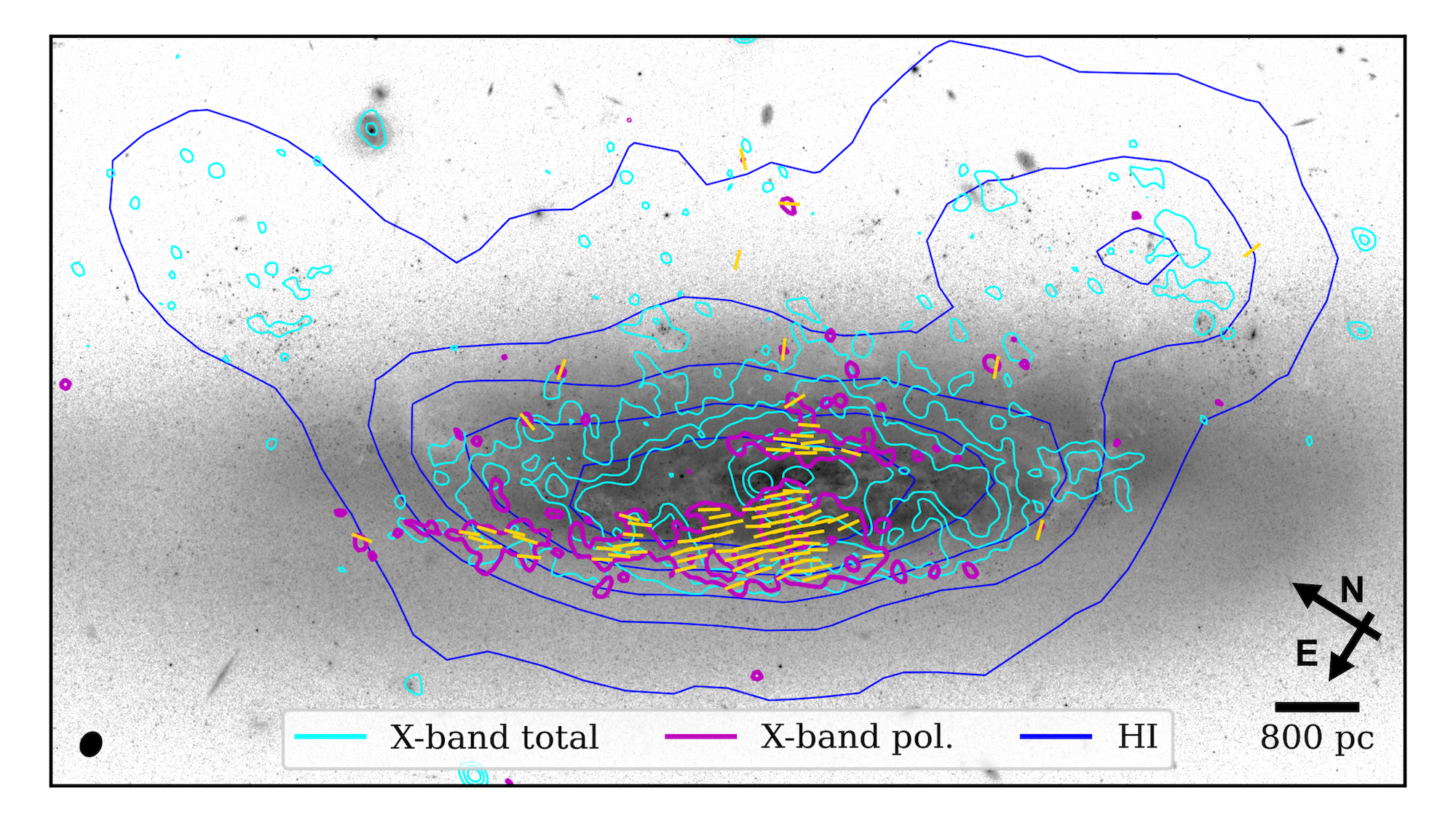}
      \caption{\textbf{Top}: S-band polarized continuum intensity (magenta contours) and apparent magnetic field vectors (yellow pseudo-vectors, uncorrected for Faraday rotation) overlaid on the optical image. Blue and cyan contours represent \ion{H}{1} and the S-band total continuum, respectively. The contour levels of the polarized emission are (3.5, 4.5, 7) $\times~6\mu$Jy~beam$^{-1}$. Polarized emission is highly asymmetric and mainly concentrated near the ICM wind front (southeast region). The polarization angles are roughly parallel to the galactic disk in this area. \textbf{Bottom}: Same as the top panel but for the X-band. The contour levels of the polarized emission are (4, 7) $\times~ 1.5\mu$Jy~beam$^{-1}$. As in the S-band, X-band polarized emission also appears near the ICM wind front and extends toward the northeast. The magnetic field vectors are nearly parallel to the \ion{H}{1} distribution, suggesting that the ISM in this region is strongly influenced by the ICM. Note that the X-band total-intensity map in this panel is shown at its native resolution ($2\farcs8\times2\farcs2$).}
      \label{fig:ngc4522_pol_s}
    \end{figure*}

    \autoref{fig:ngc4522_pol_s} shows the polarized continuum emission (magenta contours) and the apparent magnetic field vectors (yellow pseudo-vectors) of NGC~4522 at S-band (top panel) and X-band (bottom panel). The total-intensity map at X-band in this figure is shown at its native resolution ($2\farcs8\times2\farcs2$). The pseudo-vectors are rotated by $90$ degrees, such that they indicate the orientation of the ordered magnetic field, as the observed linear polarization traces the electric field, which is perpendicular to the magnetic field. We note again that no Faraday rotation correction has been applied to either band. 
    
    NGC~4522 exhibits highly asymmetric and localized polarized emission across the galaxy. %The most interesting and prominent point of the polarized emission is its highly asymmetric structures. 
    In the S-band, most of the polarized emissions is distributed along the northeastern edge of the main disk (i.e., the leading side), with local peaks displaced from the total continuum distribution (i.e., galactic midplane). Although noisier, the X-band polarized emission shows similar overall distribution along the leading side, with its peak also offset from the total continuum peak toward the leading edge. Despite this shared large-scale asymmetry, the detailed distributions of polarized emission differ between the two bands: the S-band reveals faint emission near the galactic center and in the outer disk, whereas the X-band emission is absent at the center but more concentrated around it. These characteristics stand in stark contrast to those of normal spiral galaxies, where polarization emission typically shows a symmetric pattern (see \citealt{beck2015} for a compilation of $6$~cm observations).

    In both bands, most of the polarized emissions do not coincide with optical galactic structures such as dust lanes or star-forming regions \citep[see ][]{2016AJ....152...32A}, but instead appear in the outer disk region. %This implies that the polarized emission near the ICM wind front is unlikely to originate from the galaxy's regular magnetic field. 
    Moreover, although both \ion{H}{1} and the total continuum clearly show that the ram pressure affects both the eastern and western sides of the ISM disk, it is unusual that polarized emission is detected only on one side \citep{2007A&A...464L..37V,2013A&A...553A.116V}. The enhanced and extended polarized emission toward the northeast is unlikely caused by an increased CRe density and/or magnetic field amplification due to compression, since the total continuum is not significantly enhanced in this region but merely shows compressed structures. %As suggested by \citetalias{2004AJ....127.3375V}, this feature may instead result from strong shear motions induced by the ICM-ISM interaction. However, it remains unclear why other regions do not show polarized emissions as seen in normal field spiral galaxies. One possible explanation for the lack of polarized emission across NGC~4522 is a recent ICM shock propagating through the ISM disk. Such a shock could have disrupted the regular magnetic field throughout the galaxy, leaving polarized emission only in regions where gas compression, shear motion, and magnetic draping are significant. This scenario, as well as the origin of the asymmetric polarized emission, will be discussed further in in \autoref{sec:disscuss_pol}.
    
    Meanwhile, \citetalias{2004AJ....127.3375V} reported that the flattest region of the spectral index coincides with the C-band polarized emission, but not with the H$\alpha$ or C-band total continuum peak. They suggested that this may result from a large-scale shock producing the polarized emission at the leading side. However, our observations reveal a clear offset between the peak of the S-band (or X-band) polarized emission and the flattest region of the spectral index. This discrepancy is likely due to the higher angular resolution of our data. The measured offset is less than $15$~arcsec, which could not have been distinguished in the previous observations with a resolution of $20$~arcsec. Alternatively, the difference may arise from the flux excesses in our X-band continuum data, leading to a slightly different spectral index distribution. 
    
    The overall magnetic field in the S-band appears slightly tilted with respect to the major axis of the galactic disk near the ICM wind front and gradually rotates clockwise toward the north. %The magnetic field on the opposite side of the ICM wind front is also parallel to the major axis, 
    Near the galactic center, some magnetic field vectors are oriented nearly perpendicular to the major axis. The magnetic field direction in the S-band shows an abrupt flip from approximately $-70$ $\sim$ $-80$ degrees near the peak of the polarized emission to $+70$ $\sim$ $+80$ degrees farther north, although this may be due to the 180 degree ambiguity inherent in polarization angles. The overall magnetic field morphology is consistent with that reported by \citetalias{2004AJ....127.3375V}, which also showed a magnetic field roughly parallel to the galactic disk with a slight bending toward the north. 
    
    The magnetic field in the X-band appears more coherent than that in the S-band. It is nearly parallel to the galactic disk, with orientations ranging from approximately $-40$ to $-55$, regardless of location. Considering that the Faraday rotation is negligible at X-band frequencies (\citealt{beck2015}; see also \autoref{sec:sec3_bfield_rotmeasure}), the X-band magnetic field directions are more reliable than those in the S-band. Our observations show only weak polarized emission and magnetic field signals in the extraplanar regions, indicating a lack of a well-aligned magnetic field there.

    %These are broadly in agreement with the magnetic field of the C-band continuum observation from \citetalias{2004AJ....127.3375V}.

    \subsection{Polarization Fraction}\label{sec:sec4_polfrac}
    
    As the polarized emission traces the ordered magnetic field, the polarization fraction ($p\equiv P/I$), defined as the ratio of polarized intensity to the total radio continuum intensity, can therefore be used to estimate the degree of magnetic field uniformity. The intrinsic polarization fraction $p_0$ of synchrotron emission in a perfectly ordered field is given by $p_0 = (1-\alpha)/(5/3-\alpha)$ \citep{beck2015}, yielding values of $60$ -- $75\%$ for $\alpha$ of $-0.5$ $\sim$ $-1.0$. However, the observed polarization fraction is generally smaller than $p_0$ due to thermal emission and Faraday depolarization \citep{beck2015}. In unperturbed spiral galaxies, the polarization fraction is typically higher in the interarm than along the spiral arms and shows a symmetric distribution with increasing values toward the outer disk (\citealt{beck2005_bfield}, and references therein), ranging from $10$ to $40\%$ in C-band ($6$~GHz) observations \citep{krause2020_bfield}. In our case, we expect a higher polarization fraction in the outer disk, particularly on the leading side, than in normal spiral galaxies, since polarized emission can be enhanced by ISM compression and/or shear motions induced by the ICM, or magnetic draping \citep{pfrommer_2010_mag_drap}.
    
\begin{figure*}
    \centering
    \includegraphics[width=1.4\columnwidth]{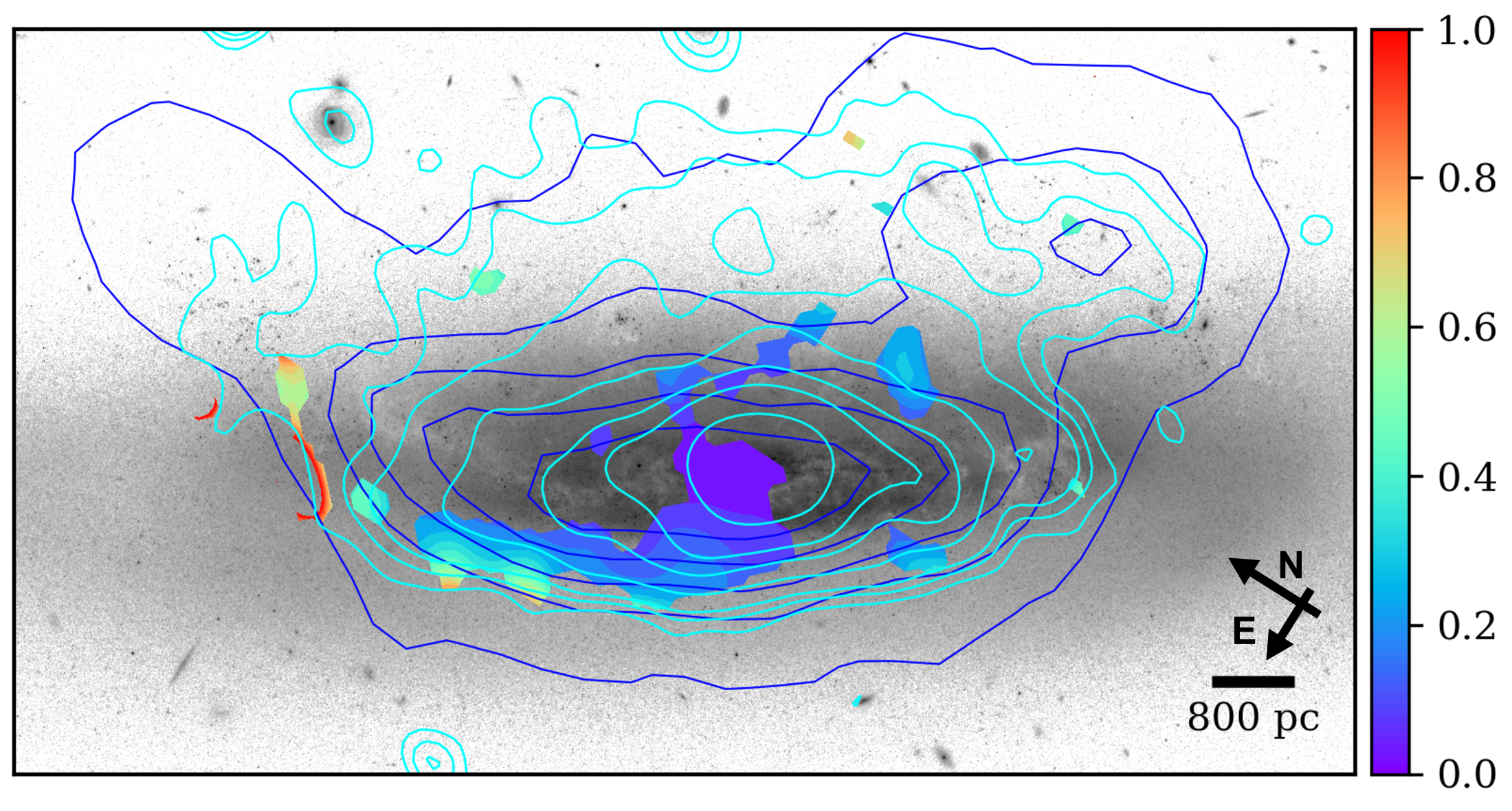}
    %\vspace{-0.4cm}
    \includegraphics[width=1.4\columnwidth]{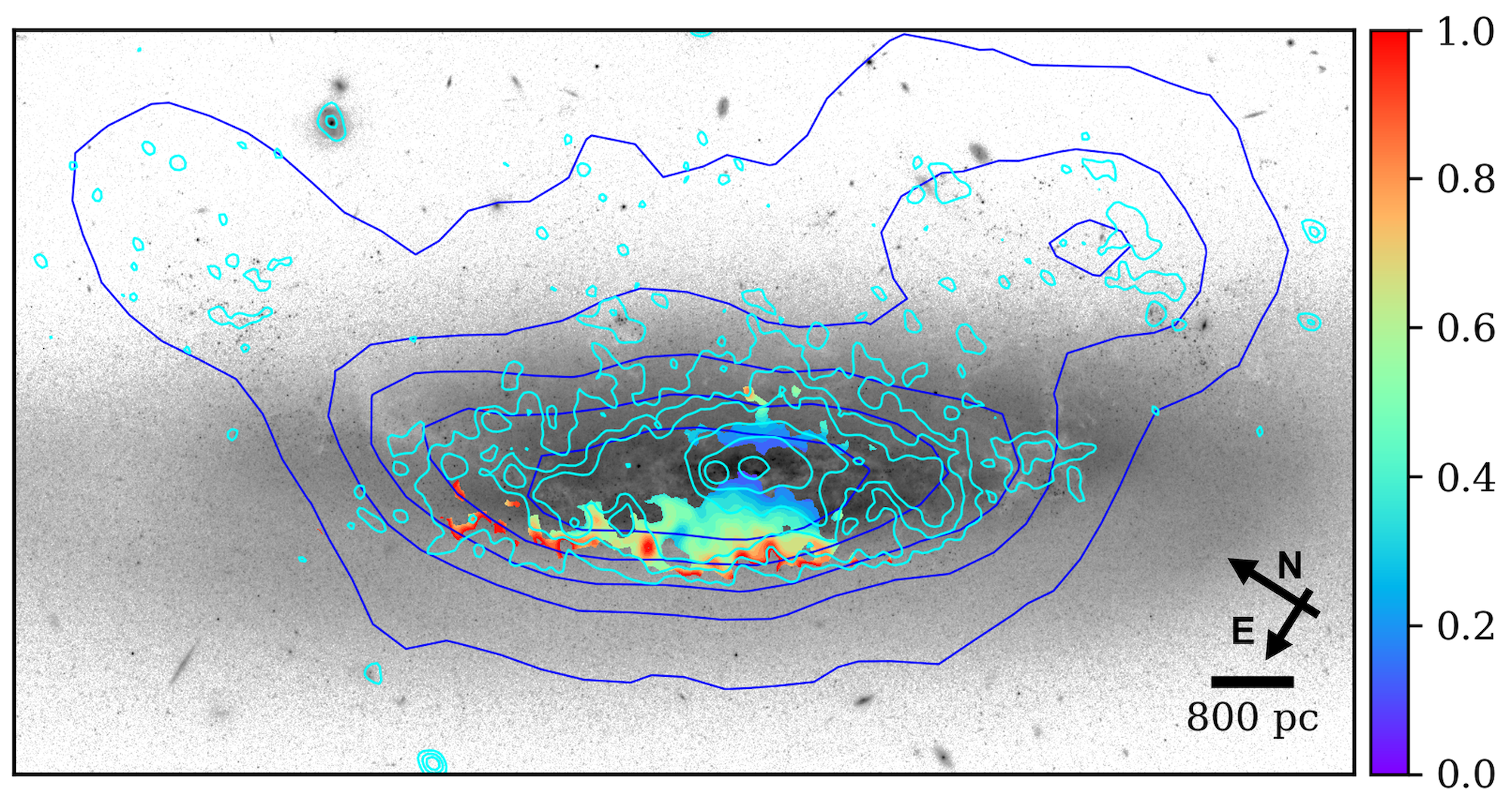}
    %\vspace{-0.4cm}
    %\includegraphics[width=0.9\columnwidth]{figures/NGC4522_degP_hist.png}
    \caption{\textbf{Top:} S-band polarization fraction map. Blue and cyan contours represent \ion{H}{1} and the S-band total continuum, respectively. The polarization fraction is generally low ($\lesssim10$\%), but increases gradually outward the outer regions, reaching its highest values near the ICM wind front. \textbf{Bottom:} Same as the top panel but for the X-band. The polarization fraction is very low near the galactic center and gradually increases outward. The distribution is roughly axisymmetric, with a few localized polarized regions in the northwest part of the galaxy.}
    \label{fig:ngc4522_degP_sx}
    \end{figure*}

    \autoref{fig:ngc4522_degP_sx} presents the polarization fraction maps of the S-band (top) and X-band (bottom). From the pixel-by-pixel analysis, the polarization fraction in the S-band ranges from $0$ to $80\%$, with an intensity-weighted mean of $\approx22\%$. The X-band shows a range, from $10$ to $100\%$, with an intensity-weighted mean of $\approx48\%$. When derived from the integrated total and polarized continuum intensities, the global polarization fractions are $12\%$ and $41\%$ for the S- and X-bands, respectively. The polarization fraction increases gradually from the galactic center to the outer disk and also rises toward the northeast direction. This trend closely matches that reported by \citetalias{2004AJ....127.3375V}, who found that the C-band polarization fraction increases from west to east, reaching its maximum in the northeast region.

    The localized polarized emission and high polarization fraction on the ICM wind front (northeastern direction) imply that RPS is responsible for enhancing the ordered magnetic field potentially via large-scale compression/stretching of the galaxy's field or magnetic draping. This aligns with previous studies that have also observed an enhanced polarization fraction on the leading side of RPS galaxies \citep[e.g.,][]{2005AJ....130...65C,vollmer2008_ngc4501,2013A&A...553A.116V}
    
    %The polarization fraction clearly peaks at the ICM wind front and decreases toward the galactic center, indicating a diminishing influence of the ICM. The localized polarized emission, which appears only on the side of directly interacting with the ICM, suggests that it does not originate from the galaxy's regular magnetic field but rather from other mechanisms such as gas compression or stretching, shear motions, and magnetic draping. This interpretation is consistent with previous studies that have reported enhanced polarization fractions on the leading sides of RPS galaxies and attributed them to ICM-induced compression and shear \citep[e.g.,][]{2005AJ....130...65C,vollmer2008_ngc4501,2013A&A...553A.116V}.

    As shown in the \autoref{fig:ngc4522_degP_sx}, the overall polarization fraction is higher in the X-band than in the S-band. %It has been suggested that the contribution of thermal emission and Faraday depolarization can lower the polarization fraction \citep[e.g.,][]{beck2015}.
    Since Faraday depolarization is proportional to the square of the wavelength ($\propto\lambda^2$), the S-band continuum suffers stronger Faraday depolarization than the X-band. This can therefore be a major cause of the lower polarization fraction in the S-band. Meanwhile, we suggest that thermal continuum emission is enhanced near the \ion{H}{1} truncation radius and the extraplanar region based on the flat spectral indices in \autoref{fig:ngc4522_specindx_sx}. The enhancement of thermal continuum emission may account for the non-detection of polarized emission in these regions and some regions within the stellar disk. However, because the thermal emission has a flat spectrum, the overall difference in the polarization fraction between the S-band and X-band is not due to the enhanced thermal contribution.

    %Meanwhile, as discussed earlier, thermal continuum emission is likely enhanced near the \ion{H}{1} truncation radius and in the extraplanar region, based on the flat spectral indices shown in \autoref{fig:ngc4522_specindx_sx}. Although such enhancement of thermal emission may explain the absence of polarized emission in these regions and in parts of the stellar disk, it is unlikely to account for the overall difference in polarization fraction between the two bands, since the thermal component cannot selectively increase in the S-band.

    \subsection{Rotation Measure and Magnetic Field Strength}\label{sec:sec3_bfield_rotmeasure}
    
    %Related references: \citet{chyzy2008_rotation_measure_ngc4254,heald2009_rotation_measure,fletcher2011_rotation_measure_m51}
    
    The polarization angle is rotated as the emission propagates through a magnetized medium, a phenomenon known as Faraday rotation. The change in the polarization angle ($\Delta\Psi$) depends on the observed wavelength ($\lambda$) and the rotation measure (RM) according to the following relation:
    \begin{equation}
    \Delta\Psi = \mathrm{RM}\times\lambda^2,
    \end{equation}
    where $\Delta\Psi=\chi - \chi_0$ represents the change in the polarization angle, $\chi$ and $\chi_0$ are the observed and intrinsic polarization angles, respectively, and $\lambda$ is the wavelength in meters. RM is proportional to the line-of-sight component of the magnetic field ($B_{\parallel}$, in $\mu$G) and the electron density ($n_e$, in cm$^{-3}$),
    \begin{equation}\label{eq:RM}
    \mathrm{RM}=0.812\int n_e B_{\parallel} dl.
    \end{equation}
    Thus, by assuming reasonable values for the electron density and the path length ($dl$), we can estimate the strength of the line-of-sight magnetic field.
    
    Since the intrinsic polarization angle is unknown, we derived the RM from the difference in the observed polarization angles between the S-band and X-band. While RM synthesis analysis \citep[e.g.,][]{brentjens_2005_rm_synthesis} is ideal for disentangling multiple emitting sources along the line-of-sight, our dataset lacks the wide $\lambda^2$ coverage required to achieve sufficient resolution in Faraday depth space. Therefore, we derive the RM assuming a simple foreground screen model, calculating it as $\Delta(\Delta\Psi) = \mathrm{RM}\times(\lambda_{\rm s-band}^2-\lambda_{\rm x-band}^2)$. To calculate $\Delta(\Delta\Psi)$, we matched the resolution and pixel scale of the two bands by convolving the X-band Stokes $Q$ and $U$ maps to the synthesized beam of the S-band and then regridding them to the same pixel size. We then generated polarization angle vectors from the convolved Stokes $Q$ and $U$ maps. The differences between the original and convolved X-band polarization angles are typically less than $10$ degrees, and we clipped all pixels below four times the rms noise of the convolved X-band map ($\sigma_{\rm rms,convol}\approx5.5$~$\mu$Jy beam$^{-1}$).
    
\begin{figure*}
      \centering
      \includegraphics[width=1.4\columnwidth]{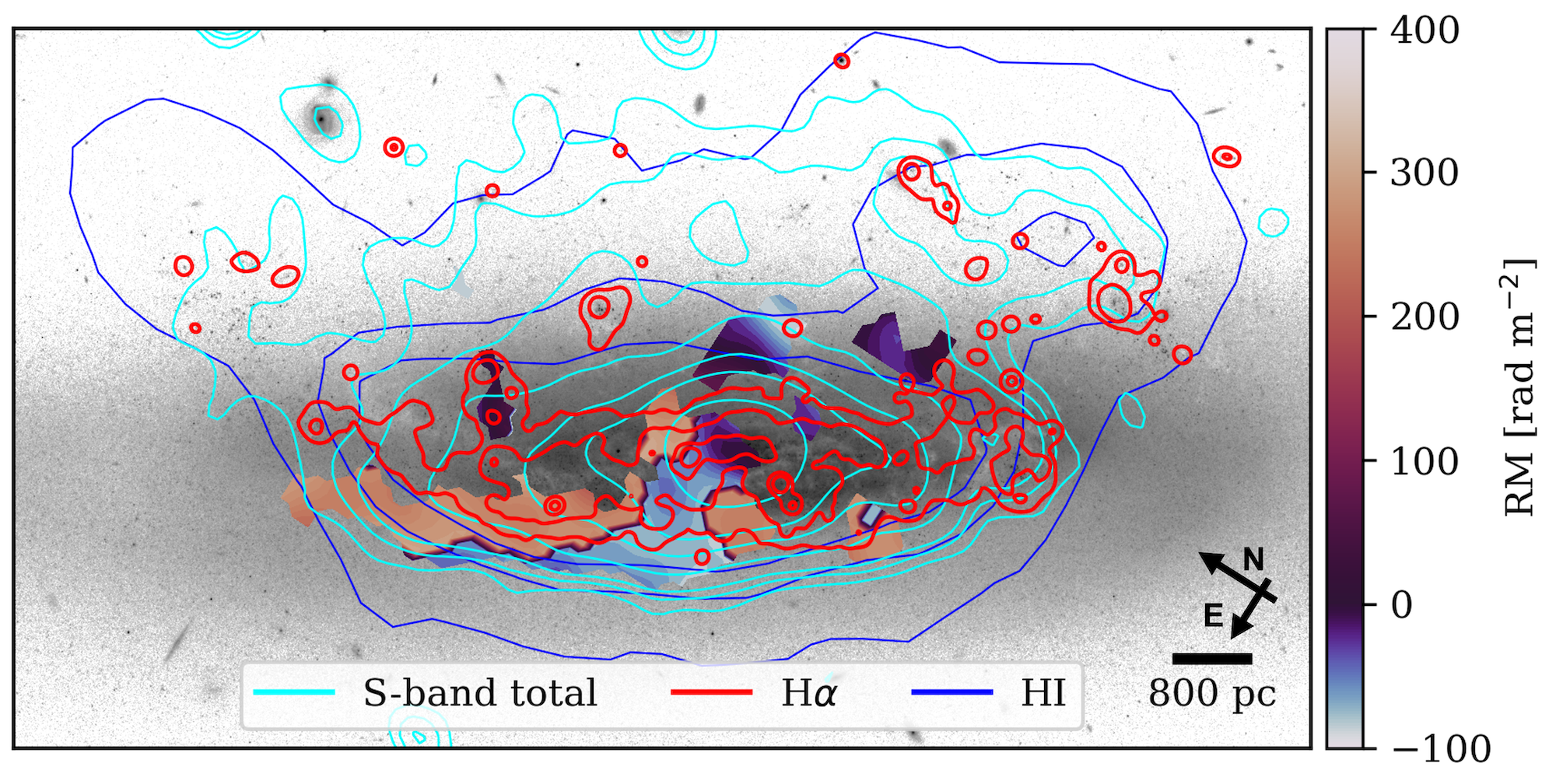}
      %\vspace{-0.5cm}
      \includegraphics[width=1.4\columnwidth]{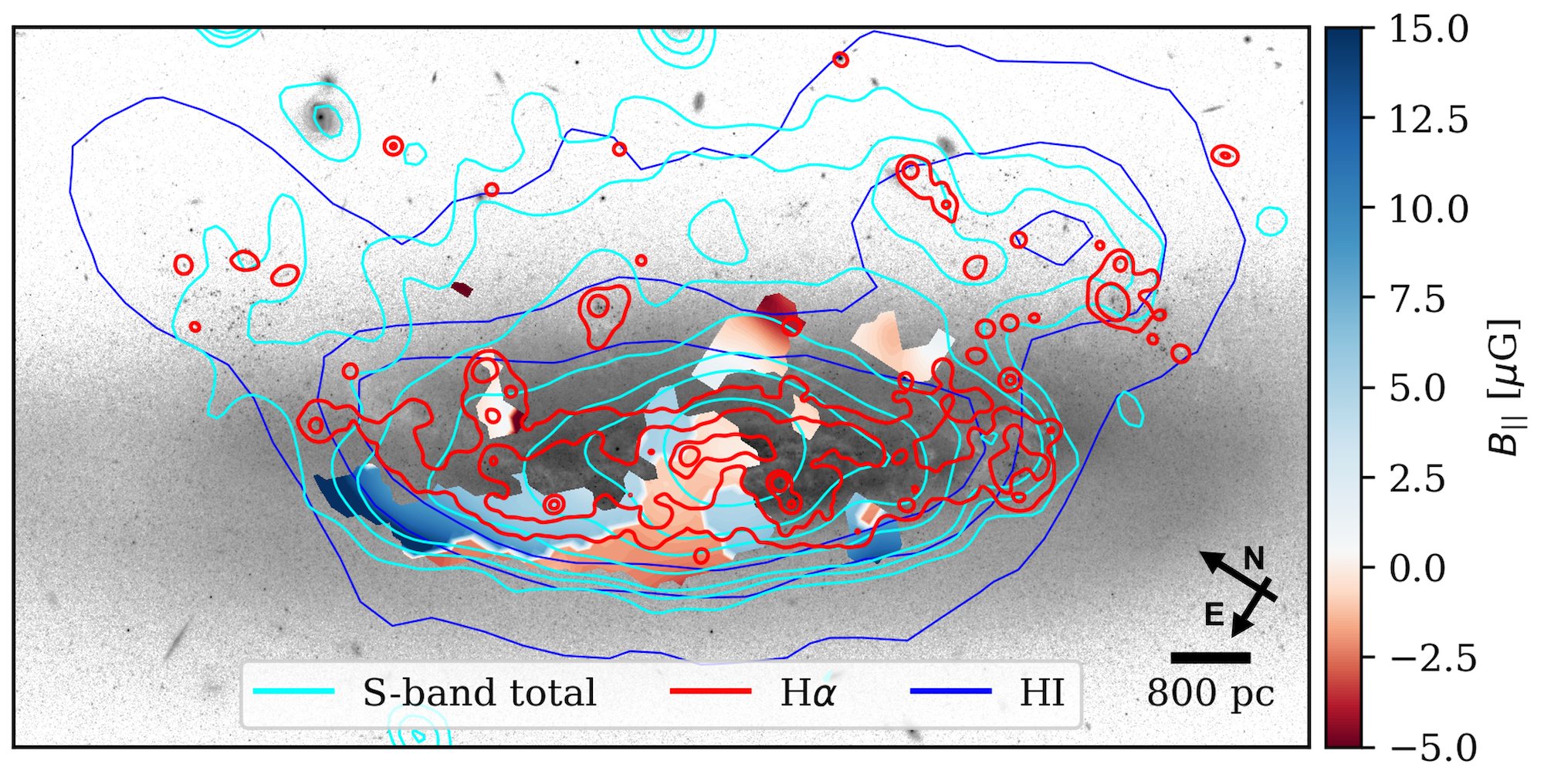}
      %\vspace{-0.4cm}
      \caption{\textbf{Top:} A map of the rotation measure between the S-band and X-band, shown using a cyclic color scale. \textbf{Bottom:} A map of the line-of-sight magnetic field strength derived from \autoref{eq:RM}, shown with a red-blue color scale. In both panels, the blue, cyan, and red contours represent \ion{H}{1}, the S-band total continuum, and H$\alpha$, respectively.}
      \label{fig:ngc4522_RM}
    \end{figure*}

    The top panel of \autoref{fig:ngc4522_RM} shows the RM map derived from the S-band and X-band data, overlaid with \ion{H}{1}, H$\alpha$, and S-band total continuum emission on the optical image. RM values range from $-100$ to $350$~rad m$^{-2}$, clustering mainly around two ranges $-100$ -- $0$~rad m$^{-2}$ (bluish regions) and $200$ -- $300$~rad m$^{-2}$ (orange regions). These absolute values are slightly higher than those typically found in external galaxies - $|\mathrm{RM}| \lesssim 150$~rad m$^{-2}$ (\citealt{klein_Fletcher_2015_bfield}, and references therein). Due to large variations in polarization angle seen in the S-band (\autoref{fig:ngc4522_pol_s}), RM also shows strong spatial variations near the ICM wind front and around the galactic center. According to the definition of RM, these sign reversals (from positive to negative and vice versa) imply reversals of the line-of-sight magnetic field between those regions.

    We assessed the potential contribution of foreground Faraday rotation from both the Milky Way and the Virgo cluster ICM. The Galactic contribution in the direction of NGC~4522 is estimated to be less than $20$~rad~m$^{-2}$ \citep{taylor2009_rm_mw}. This value is an order of magnitude smaller than the RM variations observed in NGC~4522 (up to $\sim 300$~rad~m$^{-2}$). Regarding the ICM, previous studies have revealed that the RM reaches $10$~rad~m$^{-2}$ in Virgo Cluster \citep[e.g.,][]{vallee_1990_virgo_rm}, which is also much smaller than that of NGC~4522. Therefore, we did not apply any correction for the foreground media, as it does not affect our interpretation of the RM gradients and magnetic field orientations.

    The bottom panel of \autoref{fig:ngc4522_RM} shows the electron-density-weighted magnetic field strength along the line-of-sight ($B_{\parallel, {\rm avg}}$), derived by inverting \autoref{eq:RM}, i.e.,
    \begin{equation}\label{eq:bpara}
    B_{\parallel, \text{avg}} \equiv \frac{\text{RM}}{0.812 N_e} \approx \frac{\text{RM}}{0.812 N_{\rm H~\textsc{I}} x_e}.
    \end{equation}
    Here, $N_e\approx N_{\rm H~\textsc{I}}x_e$ represents the electron column density, neglecting the contribution from fully ionized gas. This approximation is reasonable for the leading side of the disk, where polarization emission is strong, since these regions lie outside the H$\alpha$ contour. To construct this map, we adopt the \ion{H}{1} column density of NGC~4522 from \citet{2009AJ....138.1741C} and assume a typical electron fraction in the warm neutral medium, $x_e\sim0.1$ \citep[e.g.,][]{linzer_2024}.
    
    As a result, the strength of $|B_{\parallel}|$ ranges from $0$ -- $15$~$\mu$G with a median value of $\approx6$~$\mu$G. Regions with large RM values (e.g., the outer disk) tend to exhibit stronger magnetic fields than those with small RM values (e.g., the galactic center). The values of $|B_{\parallel}|$ are comparable to typical magnetic field strength of normal spiral galaxies ($10$ -- $15$~$\mu$G, \citealt{beck2015}). However, the distribution of magnetic field strength within NGC~4522 remain inconclusive, as the current results only reflect the ordered, line-of-sight component of the total magnetic field.
    
    %Since NGC~4522 is a highly inclined system, if regular magnetic fields are dominant, the line-of-sight magnetic field strength should be comparable to the typical regular magnetic field strength in spiral galaxies ($10$ -- $15$~$\mu$G, \citealt{beck2015}). However, our observations reveal weaker $|B_{\parallel}|$ values, predominantly distributed in the outer regions of the stellar disk. 
    
    %Nevertheless, the magnetic field distribution within NGC~4522 remains uncertain, as the present results only traces the line-of-sight component of the total magnetic field. %In the near future, to assess the reliability of the line-of-sight magnetic field derived from RM and to explore the role of magnetic fields at the ICM wind front, we plan to estimate the total magnetic field strength and its spatial distribution using the revised minimum-energy approach of \citet{beck_krause_2005_bfield}. %(\textcolor{red}{Total magnetic field map can be added.}).

    We note that the electron fraction is likely higher than $0.1$ inside the H$\alpha$ contour due to the presence of ionized gas. Therefore, the $|B_{\parallel}|$ values within the H$\alpha$-emitting region in \autoref{fig:ngc4522_RM} should be regarded as upper limits. We also emphasize that this estimate is simplified, as it assumes no correlation between $|B_{\parallel}|$ and $n_e$. However, such a correlation may exist in reality; thus independent observations and direct measurements of $n_e$ are required for a more accurate determination of $|B_{\parallel}|$.
    
    % Lifetime of synchrotron-emitting electrons
    
    % $t_{\mathrm{syn}}\approx1~\mathrm{Gyr} B_{\perp}^{-1.5} \nu_{\mathrm{syn}^{-0.5}}$

    \section{Discussion}\label{sec:discussion}

    \subsection{Possible Origins of Complex Spectral Indices}\label{sec:discuss_spcidx}
    
    % CGK: possible source of spectral index steepening (discussion with Nora)
    % CR electron energy spectrum can steepen in dense environments due to energy dependent energy loss (synchrotron) -- this is seen in post-processing of the R2 model
    % energy spectrum can also get shallower due to energy dependent transport if loss is not strong. High energy diffuses fast making more uniform distribution while low energy diffuses more slowly -- this is seen in post-processing of the R8 model
    % Then, synchrotron emmision picks up different range of energy spectrum depending on magnetic field strength. -- Nora cannot create corresponding synchrotron emission from the R2 model as it has much stronger magnetic fields, which map her modeled energy range to much higher frequency (>10 GHz)
    % It is non-trivial what to expect just from spectral aging.
    
    We observe systematic variations in the spectral index, characterized by $\alpha\sim-0.6$ in the main disk (enclosed by the innermost \ion{H}{1} and H$\alpha$ contours), which steepens to $\alpha\sim -1$ in the outer disk and then significantly flattens to $\alpha\sim 0$ in the ram-pressure-stripped gas and clouds. A gradual steepening of the spectral index away from the main disk is commonly observed in other galaxies and is often interpreted in the context of CRe aging. For example, \citet{vargas_2018_specindx} observed a sample of edge-on galaxies in the L (1.5~GHz) and C (6~GHz) bands, decomposing the total radio emission into the thermal and nonthermal components by estimating the thermal fraction from the star formation rate inferred from H$\alpha$. They concluded that the outward steepening is primarily due to energy losses of CRe as they propagate away from their acceleration sites \citep{condon1992_rc_review}. In addition, because the thermal contribution near the midplane is generally higher, the steepening of the total emission with increasing height is more pronounced than that of the nonthermal component alone. 
    
    To interpret the spectral index variations across NGC~4522, we employed a simple two-component model. The total flux density $S_{\nu}$ at each frequency $\nu$ was modeled as the sum of thermal and nonthermal component, $S_{\nu} = S_{\nu, \rm th} + S_{\nu, \rm nt}$. The thermal emission is represented by a power law with a fixed spectral index of $\alpha_{\rm th} = -0.1$ ($S_{\nu, \rm th} \propto \nu^{-0.1}$). The nonthermal emission was also modeled as a power law, $S_{\nu, \rm nt} \propto \nu^{\alpha_{\rm nt}}$, but both its spectral index $\alpha_{\rm nt}$ and normalization were treated as free parameters. We note that modeling the nonthermal emission as a single power law is an approximation, as an aged CRe population is physically expected to exhibit a curved radio spectrum due to energy losses, potentially introducing uncertainties in regions dominated by CRe aging.

\begin{figure}
        \centering
        \includegraphics[width=1\linewidth]{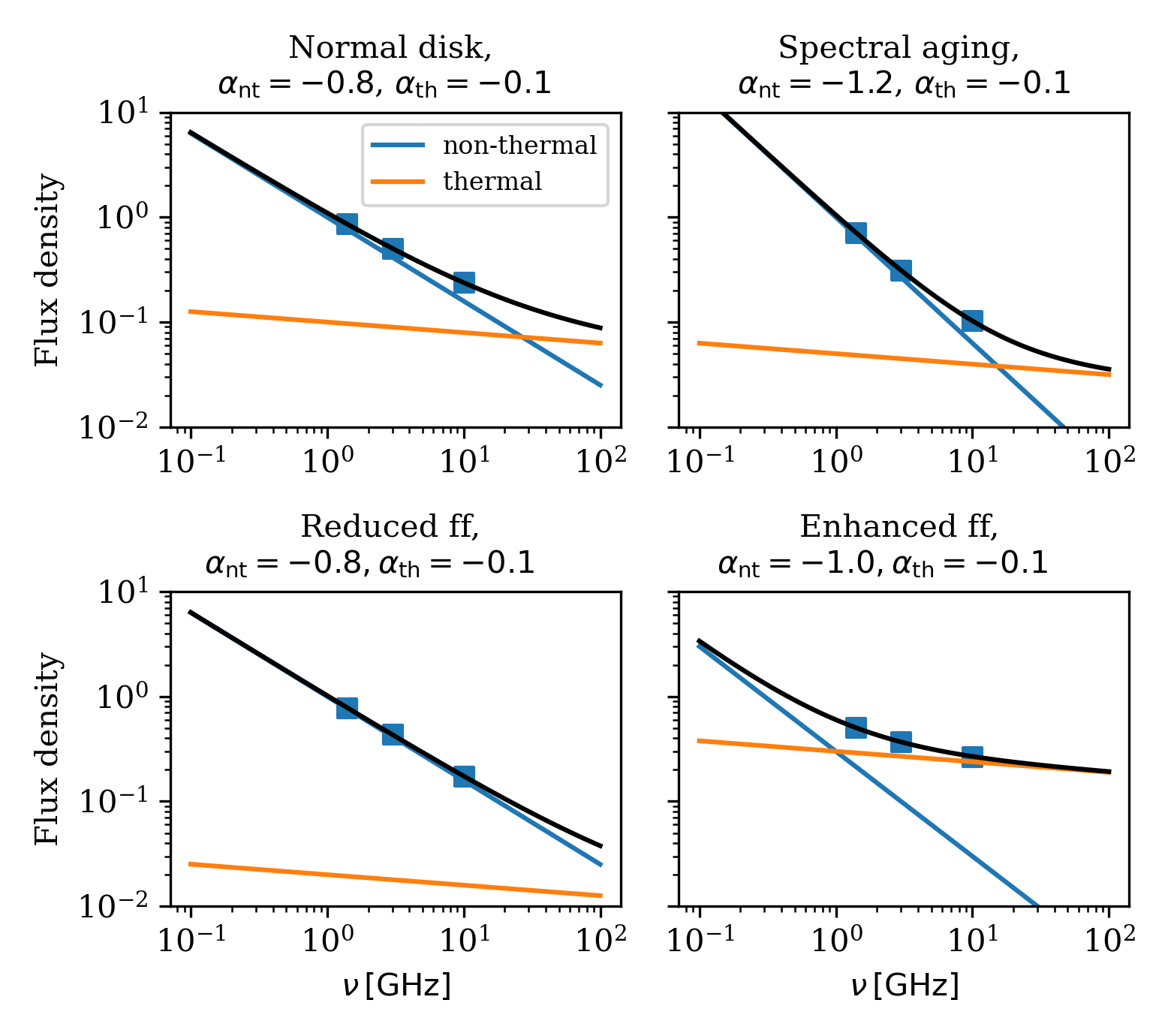}
        \caption{Schematic diagram illustrating four different cases of the two-component radio continuum. The total continuum (black curve) is composed of synchrotron (blue line) and free-free (orange line) emissions. The blue squares indicate the three data points used to constrain the three free parameters of the model. \textbf{Top left}: The nonthermal emission dominates over the thermal emission at all frequencies, as typically expected in the main disk. \textbf{Top right}: A steeper spectral index results from the reduced high-frequency nonthermal emission (i.e., CRe aging), commonly seen in outer regions. \textbf{Bottom left}: Purely nonthermal spectrum caused by a significant reduction of thermal emission. \textbf{Bottom right}: Flattened spectral index due to enhanced thermal emission. }
        \label{fig:schematic}
    \end{figure}

    The possible outcomes of the interplay between these two components are illustrated in \autoref{fig:schematic}. Depending on the relative contributions of each component and the degree of CRe aging, which steepens $\alpha_{\rm nt}$, the observed spectrum may appear steep, flat, or curved. For instance, in the main disk, we generally expect a larger fraction of nonthermal emission compared to the thermal component within the frequency range of interest (top left). If CRe aging occurs, typically found in the outer regions of galaxies \citep[e.g.,][]{vargas_2018_specindx,Heesen_2019_spix,stein_2023_changes}, the higher-frequency nonthermal emission is suppressed, resulting in a steeper spectral index (top right). When the free-free emission is significantly reduced, a purely nonthermal spectrum emerges (bottom left). Conversely, if the thermal emission dominates over the synchrotron component, the resulting spectral index becomes flatter (bottom right).

\begin{figure*}
        \centering
        \includegraphics[width=0.7\linewidth]{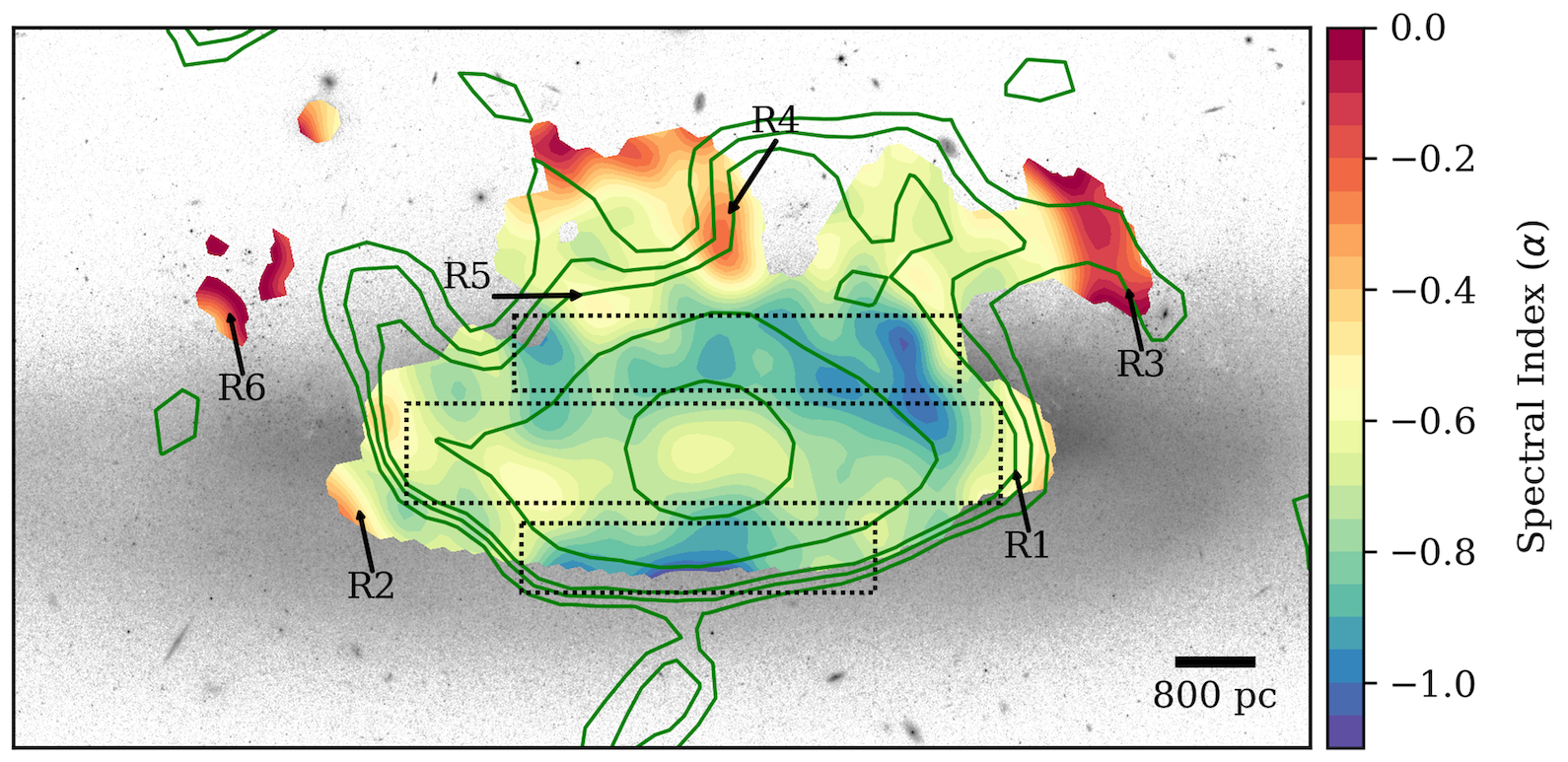}
        \includegraphics[width=1\linewidth]{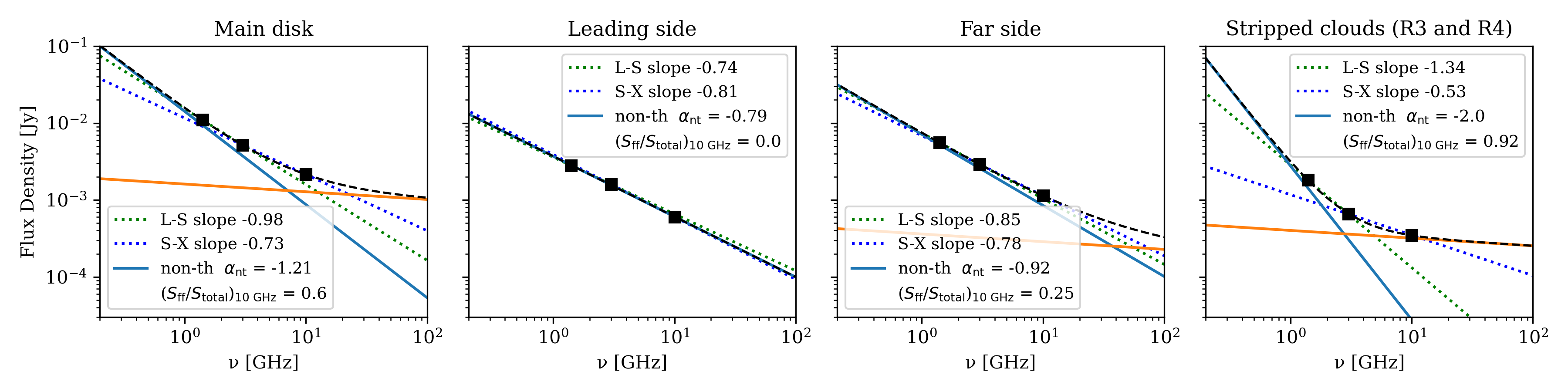}
        \caption{\textbf{Top:} L-band continuum intensity contours (green) overlaid on the optical and spectral index map. The contour levels of the L-band continuum emission are (3, 4, 5, 10, 30)$\times 120~\mu$Jy~beam$^{-1}$. The black dotted boxes indicate, from bottom to top, the leading, main disk, and far side, respectively. \textbf{Bottom:} Integrated flux density spectra and two-component fits for four regions in NGC~4522. The stripped clouds corresponds to regions R3 and R4. The black squares represent the total flux densities measured in three bands (L, S, and X). The black dashed line shows the best-fit total model, composed of nonthermal (blue) and thermal (orange). components. The green and blue dotted lines indicate the spectral slopes obtained from two-band fits between the L/S and S/X bands, respectively. ($S_{\rm ff}$/$S_{\rm total}$)$_{\rm 10~GHz}$ denotes the flux density ratio of the thermal to total emission at $10$~GHz. All measured flux densities exceed at least four times the RMS noise level.}
        \label{fig:two-comp-fitting}
    \end{figure*}

    To perform this fitting, we incorporated archival lower-resolution L-band ($1.4$~GHz) continuum observations from \citealt{2009AJ....138.1741C} ($\sim14$~arcsec of synthesized beam) and NRAO VLA Sky Survey ($45$~arcsec of synthesized beam, \citealt{condon_1998_nvss}), together with our data. We ran the \textsc{CASA} task \texttt{feather} to combine two L-band continuum images, which help to recover more of the faint extended emission. The S- and X-band maps were convolved and reprojected to match the resolution and pixel grid of the L-band map. The top panel of \autoref{fig:two-comp-fitting} shows the L-band continuum intensity map, with the regions used for our analysis indicated by black dotted boxes. This three-bands approach enables a direct decomposition of the radio continuum by simultaneously solving for the three free parameters: thermal flux, nonthermal flux, and $\alpha_{\rm nt}$. We note that this method provides a more robust decomposition than those based on H$\alpha$ scaling, especially facilitated by our $10$~GHz data, where the thermal contribution becomes significant.
    
    The bottom panels of \autoref{fig:two-comp-fitting} show the results of two-component fits for the regions of interest. Note that the spectral indices between the convolved S- and X-band images are slightly different from those in \autoref{fig:ngc4522_specindx_sx} due to additional convolution, although the overall trend remains consistent. The main disk exhibits a typical combination of nonthermal and thermal components, with a thermal fraction of $\approx 14\%$ at $1.4$~GHz, consistent with that of nearby galaxies at the same frequency \citep[e.g.,][]{condon1992_rc_review,tabatabaei2017_spcidx_flat_sf}. At the X-band ($10$~GHz), the thermal fraction is considerably higher, exceeding the nonthermal contribution. This value is significantly higher than that typically found in nearby galaxies, where the mean thermal fraction is reported to be around $23\%$ at a comparable frequency of $5$~GHz \citep{tabatabaei2017_spcidx_flat_sf}.
    
    In the outer disk (both the leading and far sides), the emission is dominated by nonthermal emission with $\alpha_{\rm nt}$ of $-0.8$ $\sim$ $-0.9$ which is flatter compared to the main disk. On the leading side, the galaxy likely experiences a supersonic encounter with the ICM. Considering the relative velocity of NGC~4522 to the ICM ($\sim 1500$ -- $3000$~\kms) and the sound speed of the ICM ($c_{s, \rm ICM}$), the estimated Mach number is $\mathcal{M} \sim 2$ -- $3$, implying the formation of a global bow shock encompassing the galaxy. In the downstream of the bow shock, the ICM is thermalized and slows down to the speed subsonic compared to $c_{s, \rm ICM}$. However, the postshock velocity remains highly supersonic with respect to the cold and warm ISM ($c_{s, \rm ISM} \approx 1 - 10~\mathrm{km/s} \ll c_{s, \rm ICM}$).
    
    Consequently, local shocks are likely driven into the ISM at the interfaces, re-accelerating CRe and enhancing the nonthermal emission. On the far side, although significant spectral aging is expected given its distance from the galactic midplane, clear evidence of such aging is barely visible. This region also exhibits a spectral shape close to a pure power-law that is slightly flatter than that of the main disk, similar to the leading edge. Given the galaxy's high inclination and the 3D geometry of the interaction between the ICM and the galaxy \citep{2006A&A...453..883V}, the global bow shock and the postshock flow envelop the galactic disk rather than being confined to the immediate leading side. Thus, the far side is also subject to these local shocks driven by the interactions between the ICM and the ISM, where CRe aging is mitigated by re-acceleration and/or the injection of fresh CRe.

    This interpretation is consistent with the high [\ion{S}{2}]/H$\alpha$ ratios (indicating shock-excited gas) observed in these regions (see \autoref{sec:discuss_mauve}). Furthermore, the derived spectral index ($\approx-0.8$) is broadly consistent with that of radio continuum emission at cluster merger shocks, such as those observed in the merging cluster Abell 3411-3412, where electron re-acceleration also occurs \citep{vanWeeren2017_cr_reaccel}. We note that while the leading side, where localized polarized emission is observed, likely experiences magnetic draping, this mechanism alone cannot explain the flattened non-thermal spectral index. An enhanced magnetic field accelerates synchrotron cooling ($t_{\rm cool,syn} \propto B^{-2}$), which would lead to even steeper spectral indices. At the same time, it leads us to observe electrons with lower energy at a fixed observing frequency ($E \propto B^{-0.5}$; e.g., \citealt{padovani_2021_cre}), which could potentially result in a flatter spectral index, especially if the emission originates from an aged CRe population. Consequently, the net impact of magnetic draping on the spectral index remains inconclusive due to the complex interplay between accelerated synchrotron cooling and the shift in the sampled electron energy range.

    The stripped clouds (R3 and R4) exhibit a dominance of thermal emission at higher frequencies and a very steep nonthermal spectral index, indicating strong CRe aging. The inferred thermal emission levels are higher than those in the outer disk. These results highlight the need for an additional source of thermal emission in the extraplanar region. It should be noted that,  owing to the shallower sensitivity of the L-band continuum observation compared to the S- and X-band data, only two regions were analyzed.
    
    Meanwhile, the slope between the L- and S-bands (green dotted line) steepens from $-0.98$ to $\approx-1.3$ between the main disk and the extraplanar region, whereas the slope between the S- and X-bands (blue dotted line) flattens from $-0.73$ to $\approx-0.5$. This difference explains why our spectral index variations differ from those reported by \citetalias{2004AJ....127.3375V} and underscores the importance of high-frequency bands (e.g., X-band), which can more reliably trace thermal contribution in the continuum analysis of RPS galaxies.

    %We suggest that the mixing between the ISM and the ICM can be a possible origin, as we will explain step by step. The flat spectral indices near $\alpha\sim 0$ require the dominance of thermal emission as the nonthermal emission intrinsically have spectral index steeper than $\sim-0.6$. Therefore, the increase of the thermal component fraction in the total continuum emission should happen either reduction of the nonthermal component of the continuum emission at $10$~GHz (and $3$~GHz), or an increment of the thermal component of the continuum emission should happen in these regions. A combination of the two effects is also possible.
    
    We now focus on the pronounced flattening of the spectral index observed in the ram-pressure-stripped clouds. In particular, R3 and R6 exhibit spectral indices of $\alpha\sim 0$, which can only be explained by a complete dominance of thermal emission, as shown in \autoref{fig:two-comp-fitting}. This implies that either the thermal emission is strongly enhanced or the nonthermal emission is significantly suppressed. What could cause such a relative enhancement of thermal emission? Recently formed, young and massive stars emit copious ionizing radiation that creates ionized regions with high electron densities, producing strong free-free emission. Consequently, a flatter radio spectrum is naturally expected where H$\alpha$ emission is strong. This may apply to the star-forming molecular cloud R3, although it also shows somewhat extended emission toward the top of the map, which the spectral index remains very flat despite no detectable H$\alpha$ emission. Another extraplanar molecular cloud, R5, exhibits visible H$\alpha$ emission with a moderately flat spectral index ($\alpha\sim-0.5$). In contrast, R6 shows a very flat spectrum despite weak H$\alpha$ emission, while R4 displays a moderately flat spectrum $(\alpha\sim-0.3)$ with no indication of ongoing star formation.%(also without CO detection). 
    
    \subsection{Insights from Previous Simulations}\label{sec:disscuss_simul}
    
    While many of the flat spectra in the extraplanar clouds can be explained by enhanced thermal emission associated with star formation, some regions (R4, R6, and the region above R3) may require an alternative mechanism capable of increasing thermal emission without star formation. To gain further insight, we analyze the simulations presented in \citet{choi2022}. The simulations employ a local wind-tunnel model, in which the ICM wind is continuously driven into a local patch of the ISM disk. The ISM disk in the absence of an ICM wind is adopted from the solar-neighborhood model constructed with the TIGRESS framework \citep{kim2017_tigress}, which self-consistently models a turbulent, magnetized ISM including radiative cooling, gaseous self-gravity, fixed stellar and dark matter gravity, and prescriptions for star formation and feedback. The simulated multiphase ISM evolves to a self-regulated, quasi-steady state, after which ICM winds of different strengths are injected from the bottom boundary of the simulation domain to model a face-on interaction. We use three models: one without an ICM wind (\noicm{}) and two with weak and strong ICM winds (\icmpw{} and \icmps{}, respectively). The ram pressure of the weak ICM model is comparable to the anchoring pressure ($\pi G \Sigma_{\rm gas} \Sigma_{\rm star}$; \citealt{1972ApJ...176....1G}), whereas that of the strong ICM model exceeds it. Full details of the simulation setup and model parameters are provided in \citet{choi2022}. Note that the ICM wind is unmagnetized.
    
    In these simulations, ionizing radiation and CRe are not included, preventing us from producing direct synthetic observations of thermal and nonthermal emission at the radio frequencies of interest. Instead, we estimate the free–free emission using the electron density under the assumption of collisional ionization equilibrium. The volume emissivity of the free–free emission is calculated by
    \begin{equation}
        j_{\nu,ff} \propto n_e n_i T^{-0.5} e^{-h\nu/k_BT},
    \end{equation}
    which is integrated along the line-of-sight ($x$-axis of the simulation domain) to produce a brightness map. 
    For the nonthermal component, we present the magnetic field strength perpendicular to the line-of-sight (i.e., $B_{\perp}=(B_y^2+B_z^2)^{1/2}$), since the synchrotron emission strength is approximately proportional to $B_{\perp}$.

\begin{figure}
        \centering
        \includegraphics[width=\linewidth]{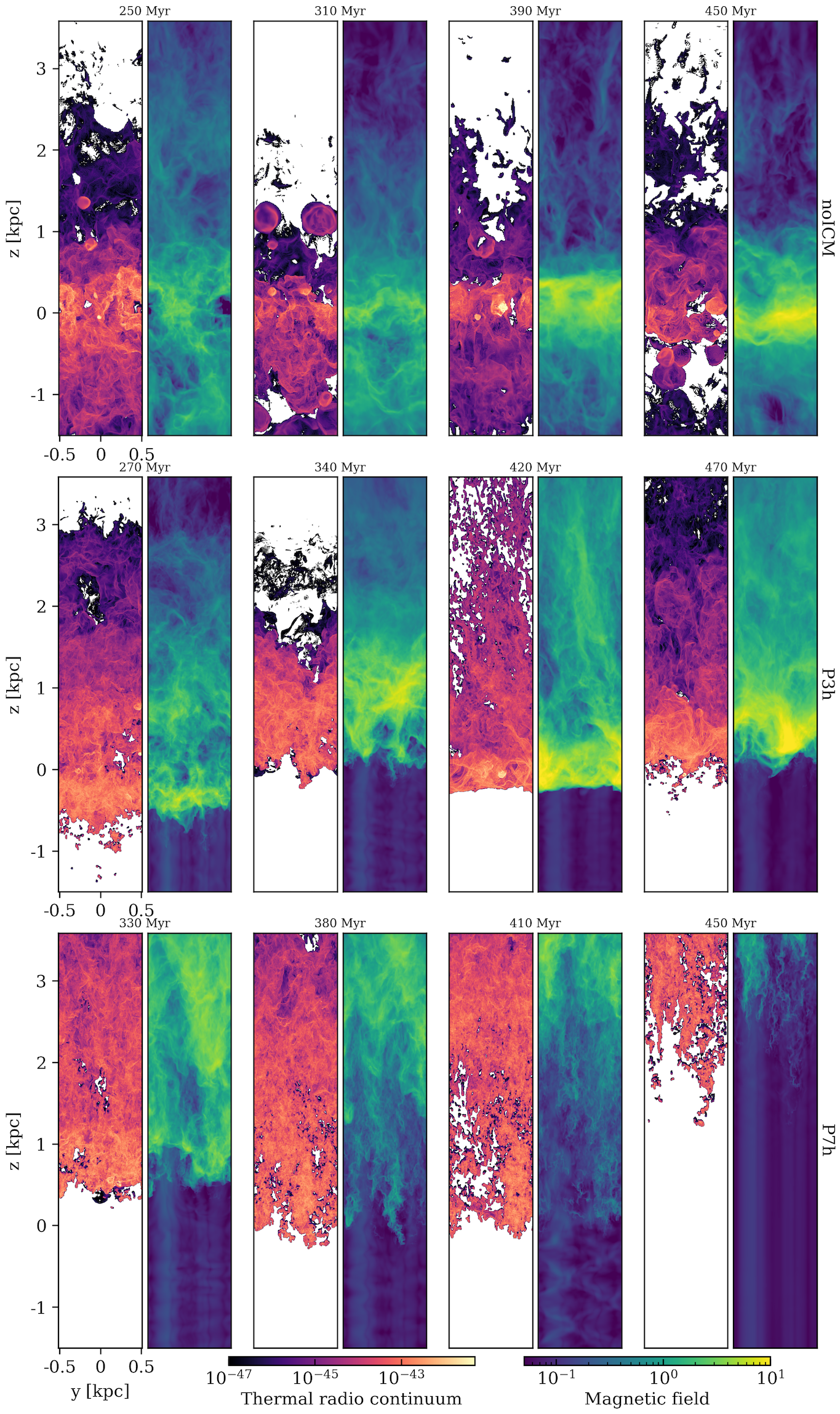}
        \caption{Two-dimensional maps of thermal emission and magnetic field strength for the \noicm{} (top), \icmpw{} (middle), and \icmps{} (bottom) models. The ICM wind begins to flow in at $250$~Myr from the bottom boundary of the simulation box. Only the regions with z $>$ -1.5 kpc is shown to focus on the upper disk. The color bars at the bottom indicated the thermal volume emissivity in erg~s$^{-1}$~cm$^{-3}$~Hz$^{-1}$~sr$^{-1}$ and magnetic field strength in $\mu$G.}
        \label{fig:2dmap-tigress}
    \end{figure}

    \autoref{fig:2dmap-tigress} shows the two-dimensional maps of the thermal emission and magnetic field for the \noicm{}, \icmpw{}, and \icmps{}. We select four snapshots for each model that clearly represent their distinct features. In the \noicm{} model, both the free-free emission and magnetic field strength are roughly symmetric, although they undergo compression and expansion over time due to bursty star formation and feedback. In general, the vertical extent (or scale height) of the free-free emission is shorter than that of the perpendicular magnetic field strength. For example, the free-free emission decreases by a factor of $100$ from the midplane to $z=1~\kpc$, whereas $B_\perp$ decreases by a factor of $10$. If the thermal contribution to the total emission at high frequencies (e.g., X-band) is substantial, this difference in the vertical scale heights of the two components alone can lead to a steepening of the spectral index from the midplane toward the extraplanar region. We note that the absence of ionizing radiation may reduce the overall level of free–free emission, but its scale height is expected to remain similar. The extraplanar gas density profile primarily determines the emission profile, as ionizing radiation can escape efficiently when supernovae create bubbles and fountain flows \citep{2020ApJ...897..143K, linzer_2024}. 
    
    In the \icmpw{} model, after introduction of the ICM wind, the ISM disk quickly becomes asymmetric, with gas compressed on the windward side. The compression of this leading edge of the ISM enhances the overall gas density and amplifies the magnetic field strength in the same region. At the same time, the interaction with the hot ICM, including mixing between the two media, increases the overall ISM temperature, leading to an overall enhancement of the thermal emission compared to the \noicm{} case. However, the extended gas on the tail side is accompanied by extended magnetic field, causing the relative contribution of the thermal and nonthermal components to remain similar. The corresponding change in the spectral index is therefore nontrivial.
    
    Finally, the \icmps{} model exhibits distinct behavior. The gas is gradually stripped away as the ICM ram pressure exceeds the anchoring pressure of the ISM. However, the stripping process is not impulsive. The main ISM disk is initially displaced upward by $\sim 1~\kpc$, then falls back before being completely blown away.  During this evolution, the develops an offset between the peaks of the thermal emission and magnetic field strength. This occurs primarily because the unmagnetized ICM mixes with and dilutes the magnetic field in the ISM, while the same mixing tends to maintain, or even enhance, the thermal emission, despite the absence of direct ionizing radiation. When focusing on the peak of the thermal emission (which also appears bright in other gas tracers such as \ion{H}{1} or CO), the stripped cloud region can become almost completely dominated by thermal emission and exhibit a very flat spectral index of $\alpha\sim-0.1$. 
    
    We note that if the ICM is magnetized, a magnetic draping effect can occur, amplifying the ICM field in front of the leading side of the disk. \citet{pfrommer_2010_mag_drap} suggested that this mechanism could explain the enhanced synchrotron emission observed on the leading side of NGC~4522, which coincides with the steepened spectral index seen in \autoref{fig:ngc4522_specindx_sx}. If this magnetic draping effect similarly influences the stripped clouds, one might expect an enhanced magnetic field near the thermal emission peak, unlike what is seen in the \icmps{} model. 
    
    In summary, our analysis indicates that multiple mechanisms drive the observed spectral index variations: (1) a combination of typical nonthermal emission and a moderate thermal fraction in the main disk; (2) ICM shock-induced CRe re-acceleration accompanied by suppressed thermal emission in the outer disk; and (3) strong CRe aging together with significantly enhanced thermal emission in the extraplanar clouds. For the latter, while star formation contributes where present, we propose that thermal enhancement through ICM–ISM mixing may play a critical role in regions lacking active star formation.

    \subsection{Comparison with ionized gas properties from MAUVE-MUSE}\label{sec:discuss_mauve}
    
    %Recent MUSE observations for Virgo cluster galaxies have been conducted through Multiphase Astrophysics to Unveil the Virgo Environment (MAUVE) survey. They have revealed interesting features in the ionized gas distribution (Brown et al. in prep.). One of their analysis is the 2D BPT diagram \citep{bpt1981} using [\ion{O}{III}]/H$\beta$, [\ion{N}{2}]/H$\alpha$ and [\ion{S}{2}]/H$\alpha$. 
    
    In this section, we compare our radio continuum results with recent MUSE observations of NGC~4522, obtained as part of the Multiphase Astrophysics to Unveil the Virgo Environment (MAUVE) project \citep{watts_2024_mauve,cainella_2025_mauve}. In particular, \citet{brown_2025} conducted a detailed analysis of the ionization sources in Virgo galaxies, highlighting the diverse processes influencing the warm phase of the ISM. Here, we use the same MUSE dataset for NGC~4522, with the only difference being that our line maps are generated without pre-masking and by binning the data cube to achieve a SNR of 10 in the 6760 -- 6790 \AA\ wavelength range (corresponding to the [\ion{S}{2}] doublet), in order to maximize sensitivity to diffuse extraplanar emission. This differs from the approach adopted by Brown et al., who used a SNR threshold based on the 4800 -- 7000 \AA\ range and pre-masked individual spaxels with S/N $<$ 1.5.
    
\begin{figure}
      \centering
      \includegraphics[width=\columnwidth]{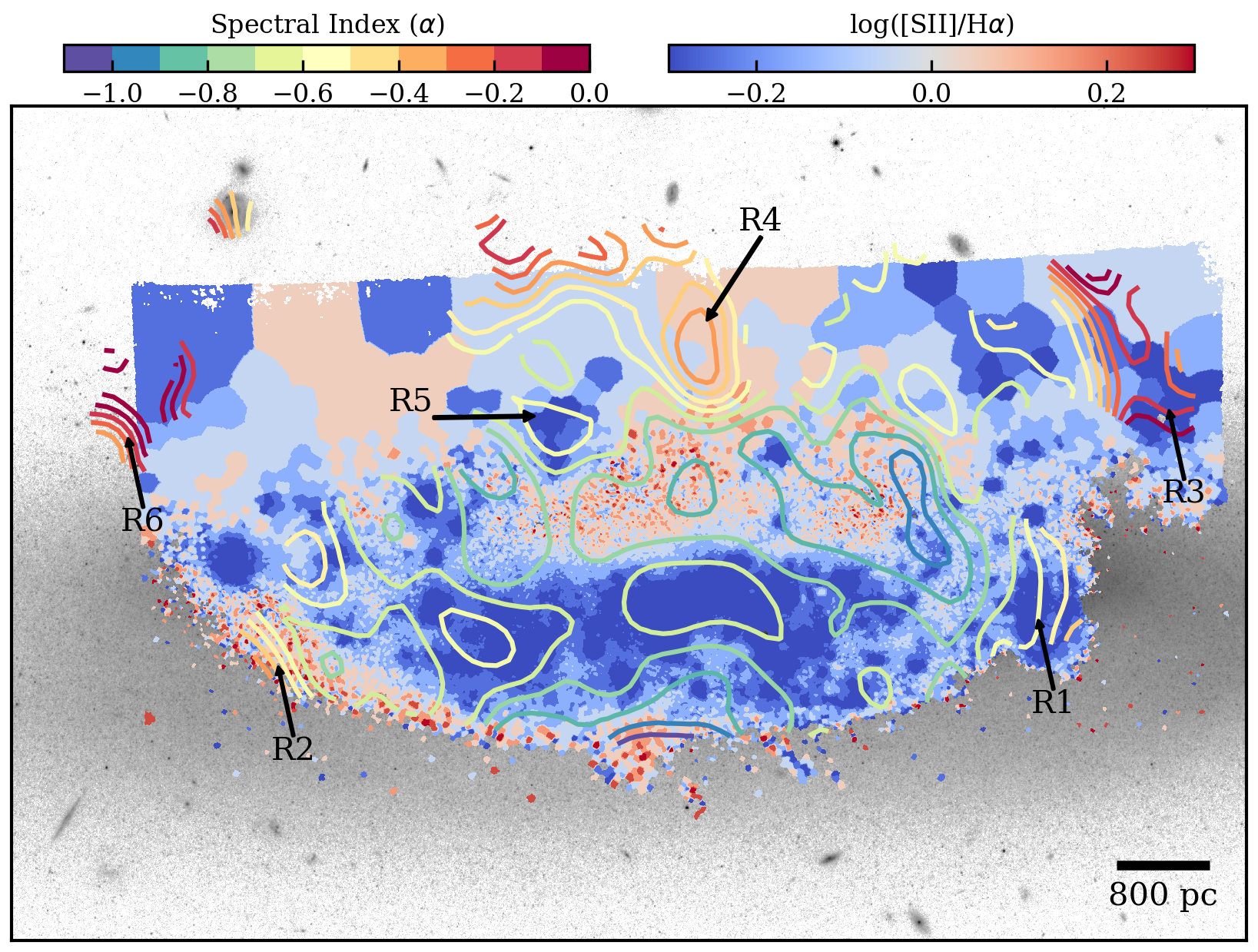}
      \caption{Map of [\ion{S}{2}]/H$\alpha$ from MUSE observations (MAUVE-MUSE; \citealt{brown_2025}), overlaid with the contours of the spectral index between S and X bands.}
      \label{fig:ngc4522_spec_mauve}
    \end{figure}

    \autoref{fig:ngc4522_spec_mauve} shows the two-dimensional distribution of the [\ion{S}{2}] $\lambda$6716,6731/H$\alpha$ (the corresponding [\ion{N}{2}] $\lambda$ 6583/H$\alpha$ map is shown in \autoref{fig:mauve-nii}). We include only spaxels with S/N greater than 3 and flux densities higher than 1$\times$10$^{-19}$ erg cm$^{-2}$ s$^{-1}$. We are unable to correct these ratios for dust attenuation because of the low S/N of the H$\beta$ line in the extraplanar region. However, dust has only a minor effect on the [\ion{S}{2}]/H$\alpha$ ratio and a negligible effect on the  [\ion{N}{2}]/H$\alpha$ \citep{Osterbrock_2006_eline}. These maps reveal that most of the main disk is characterized by line ratios consistent with ionization by star formation (i.e., low [\ion{S}{2}]/H$\alpha$ and [\ion{N}{2}]/H$\alpha$), whereas several regions in the outer disk and extraplanar area exhibit elevated  [\ion{S}{2}]/H$\alpha$ ratios, likely tracing ICM interaction and shock excitation in NGC~4522.

    %This analysis reveals whether the gas is primarily affected by star formation (indicated by low [\ion{S}{2}]/H$\alpha$) or by other mechanisms, such as ICM shocks (indicated by high [\ion{S}{2}]/H$\alpha$).

    %Here, we make use of the same MUSE data products for NGC~4522, with the only difference being that the line maps are obtained by binning the cube to a signal-to-noise ratio of 10 in the $6760$ -- $6790$ \AA\ wavelength range (corresponding to the [\ion{S}{2}] doublet) in order to focus on line emission, differing from the binning used in Brown et al. which was based on the S/N across the $4800$ -- $7000$ \AA\ range.

    %\autoref{fig:ngc4522_spec_mauve} shows the 2D distribution of the distance from Ke01 line \citep{Kewley_2001_ke01} of [\ion{S}{2}]/H$\alpha$ (corresponding map for [\ion{N}{2}]/H$\alpha$ is shown in \autoref{fig:mauve-nii}). This analysis reveals whether the gas is primarily affected by star formation (indicated by low [\ion{S}{2}]/H$\alpha$) or by other mechanisms, such as ICM shocks (indicated by high [\ion{S}{2}]/H$\alpha$). 
    
     %\textcolor{red}{WR: this will be revised when I have a new map.} Note that their 2D-BPT analysis is limited to the height of R5 from the main disk due to a sensitivity of line emissions ([\ion{O}{III}], [\ion{N}{2}], and [\ion{S}{2}]), although H$\alpha$ distributes similarly to the S-band continuum. Thus, the comparison with the MAUVE data can be made only for the main and outer disk and a small region of R3, not for R4 and R6.
    
    %Thus, R4 and R6 cannot be compared with our results, while a small region of R3 could be.
    
    First, they identify prominent shock-induced features (very high [\ion{S}{2}]/H$\alpha$) in the outer disk (i.e., the leading and far sides), where steep spectral indices ($\alpha < -0.7$) are observed in \autoref{fig:ngc4522_specindx_sx}. This is expected, as shocked gas emits a synchrotron-dominated continuum due to CRe re-acceleration. This interpretation is also supported by the two-component analysis in this region (\autoref{fig:two-comp-fitting}), which reveals a nearly pure power-law spectrum without the curvature characteristic of radiative cooling. Moreover, the nonthermal spectral index ($\alpha_{\rm nt}$) in these regions is found to be flatter than that of the main disk. The coincidence of these spectral features and high [\ion{S}{2}]/H$\alpha$ ratios provides a coherent picture, confirming that the ICM shock likely governs the emission properties on both the leading and far sides of the outer disk.

    In the main disk, both [\ion{S}{2}]/H$\alpha$ and [\ion{N}{2}]/H$\alpha$ are low, clearly indicating that star formation is the dominant source of ionization. This is consistent with the relatively flat spectral index ($-0.6$) observed in this region. Although R5 is located in the extraplanar region, its behavior is similar to that of the main disk, showing low [\ion{S}{2}]/H$\alpha$ and [\ion{N}{2}]/H$\alpha$ ratios and a relatively flat spectral index. Intriguingly, the vicinity of R5 exhibits similarly flat spectral indices despite the absence of detectable H$\alpha$ emission. This suggests that the ICM-ISM mixing affects the extraplanar gas, or that effective diffusion of free electrons enhances the thermal emission without triggering star formation, therefore without producing detectable H$\alpha$ emission.
    
    Other regions, however, exhibit more complex features. R3 shows clear H$\alpha$ emission with a low [\ion{S}{2}]/H$\alpha$ ratio, and R4  exhibits a high [\ion{S}{2}]/H$\alpha$ ratio but no detectable H$\alpha$ emission. Despite these differences in ionized gas properties, both regions display a very flat spectral index ($\alpha > -0.3$). R2 shows a high [\ion{S}{2}]/H$\alpha$ radio without strong H$\alpha$ emission, similar to the leading side of the galaxy. However, its spectral index is much flatter than that of the leading side, suggesting that both the ICM shock and ICM-ISM mixing influence R2. In contrast, R1 exhibits strong H$\alpha$ emission with low [\ion{S}{2}]/H$\alpha$ and [\ion{N}{2}]/H$\alpha$ ratios, indicating that it is primarily affected by star formation. Nevertheless, we speculate that R1 may also experience some influence from the ICM–ISM mixing, as its spectral index tentatively flattens outward in a manner similar to R2.

    \begin{table*}
    \centering
    %\hline
        \begin{tabular}{ccccll}
            \hline 
            Location & Spectral Index & H$\alpha$ & log([\ion{S}{2}]/H$\alpha$) & Thermal & Nonthermal \\ \hline \hline
            Main disk & $-0.8 \lesssim \alpha \lesssim -0.5$ & Strong & -0.6 -- -0.2 & SF & Sync \\ \hline
            Leading side & $-1.1 \lesssim \alpha \lesssim -0.7$ & Weak & 0 -- 0.4 & -- & Shock + Sync\\ \hline
            Far side & $-1.1 \lesssim \alpha \lesssim -0.6$ & Weak & 0 -- 0.4 & -- & Shock + Sync \\ \hline
            R1 & $-0.61 \pm 0.22$ & Strong & -0.4 -- 0 & SF + Mixing? & Sync \\ \hline
            %R2 & $\approx -0.5$ & Strong & Very low & SF + Mixing? & Sync \\ \hline
            R2 & $-0.56 \pm 0.25$ & Weak & 0.1 -- 0.4  & Mixing & Sync + Shock? \\ \hline
            R3 & $-0.16 \pm 0.21$ & Moderate & -0.2 -- 0 & SF + Mixing & CRe aging \\ \hline
            R4 & $-0.36 \pm 0.05$ & Weak & -0.1 -- 0.1 & Mixing & CRe aging \\ \hline
            R5 & $-0.63 \pm 0.11$ & Moderate & -0.3 -- 0 & SF + Mixing & Sync + Shock? \\ \hline
            R6 & $-0.13 \pm 0.08$ & None & None & Mixing & CRe aging \\ \hline
        \end{tabular}
    
        \caption{Summary of regional characteristics. The spectral indices and log([\ion{S}{2}/H$\alpha$]) for the main disk, leading side, and far side are measured across each respective region, whereas those for R1 -- R6 are measured within a region corresponding to two synthesized beam sizes.}
        \label{tab:region_summary}
    \end{table*}

    \begin{figure*}
        \centering
        \includegraphics[width=1.6\columnwidth]{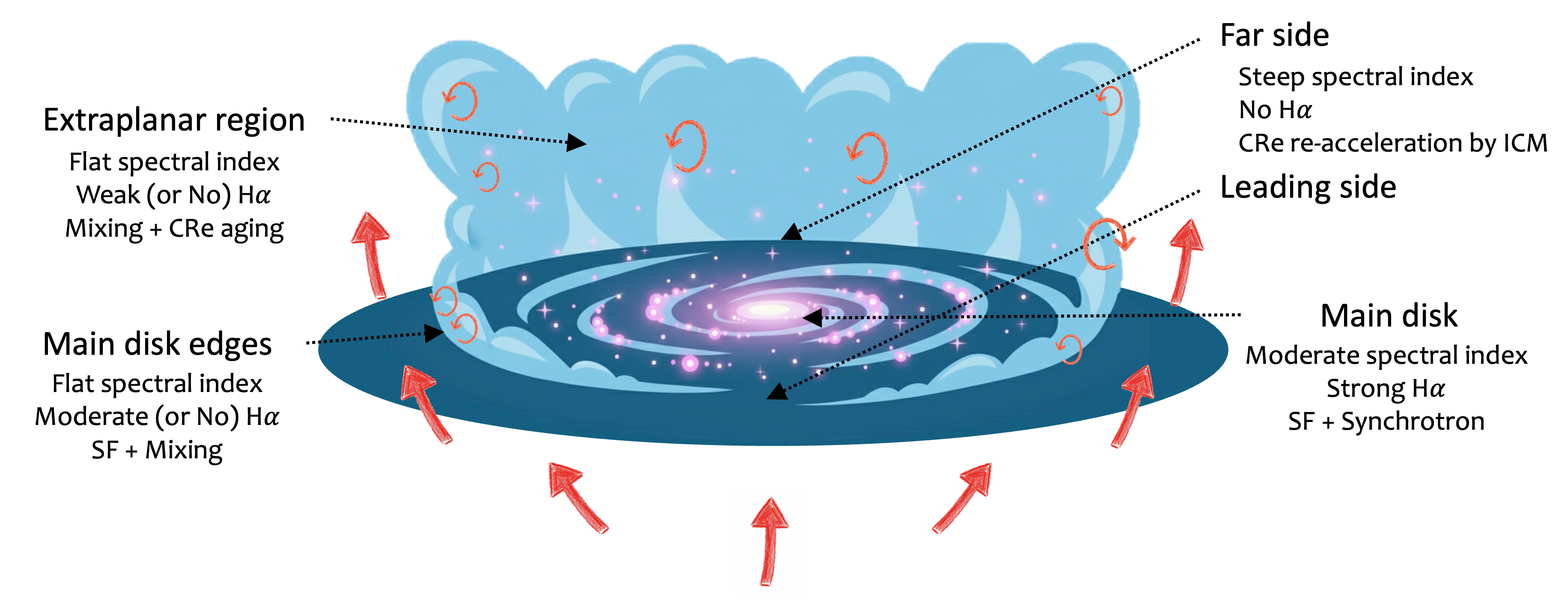}
        \includegraphics[width=1.2\columnwidth]{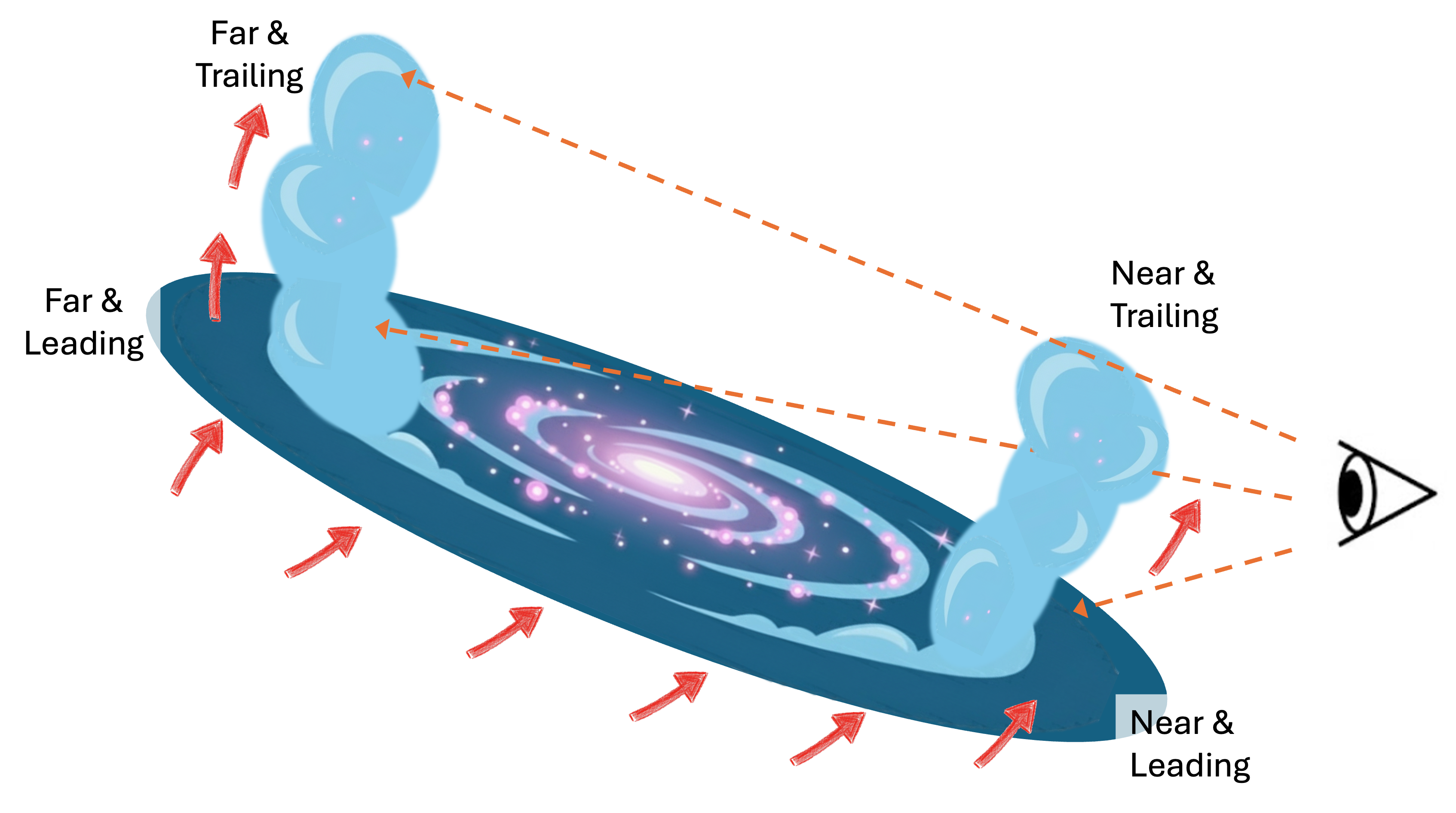}
        \caption{Schematic overview and geometry of NGC~4522. The red arrows show the ICM wind direction. \textbf{Top}: A face-on view highlighting regions of interest (the main disk, the outer disk, and the extraplanar regions) and their characteristics. \textbf{Bottom}: A profile view illustrating the relative geometry between the observer and the galaxy. The diagram defines the far/near and leading/trailing sides and show the potential line-of-sight confusion between different regions. Note that, for visual clarity, stripped gas components in the central part of the diagrams are omitted in both panels.}
        \label{fig:4522_schematic}
    \end{figure*}
    %R1 is the result of the combination of star formation and ICM thermalization.
    
    %The outer regions where the ICM is presumed to be acting directly show a generally flat spectral index.
    
    Consequently, we can classify NGC~4522 into four categories: (1) Strong SF and synchrotron emission (main disk), (2) Strong SF with mixing and synchrotron emission (R1 and R5), (3) Weak or no SF with mixing and shock-induced synchrotron emission (outer disk and R2), and (4) Mixing and CRe aging (R3, R4, and R6). The characteristics of each region are summarized in \autoref{tab:region_summary}. To facilitate the interpretation of these properties, \autoref{fig:4522_schematic} shows a schematic overview and a profile view of NGC~4522, illustrating the geometry and spatial distribution of different emission components relative to the ICM wind.
    
    \subsection{Highly Asymmetric Polarized Emissions}\label{sec:disscuss_pol}
    
    We detected highly asymmetric polarized emission in both S- and X-bands, as described in \autoref{sec:sec4_pol}. Such features are rarely observed in nearby field galaxies, even among highly inclined galaxies \citep[e.g.,][]{beck2005_bfield,krause2020_bfield}, whereas 16 out of 29 observed Virgo spiral galaxies exhibit strongly asymmetric polarized emission \citep{2004AJ....127.3375V,2007A&A...464L..37V,2013A&A...553A.116V}. This suggests that the observed asymmetry is not solely due to the viewing-angle effect but is instead driven by environmental influences within the galaxy cluster, including ram pressure by the ICM. Compression and shear motions, caused by ram pressure, tidal interactions, gas accretion, or magnetic draping, have been proposed as the most likely mechanisms responsible for the enhanced polarized emission on the leading sides of galaxies \citep[e.g.,][]{2006A&A...453..883V,pfrommer_2010_mag_drap}. While these mechanisms can explain the aligned magnetic field and enhanced polarization on the leading side, they do not account for the absence of polarized emission in the remaining regions of our target, which ultimately results in the strongly asymmetric morphology.
    
    We conjecture that the highly asymmetric polarized emission arises from a combination of enhanced polarization on the leading side and reduced polarized emission across the main disk. The ICM wind sweeping across the disk can disrupts the regular magnetic field while simultaneously enhancing (or maintaining) the turbulent magnetic component in the same region. This process likely decreases the polarized emission within the disk, even though the total continuum emission remains largely unchanged. To test this hypothesis, we estimate the evolution of the regular and turbulent magnetic fields using the simulations described in \autoref{sec:discuss_spcidx}. We calculate regular and turbulent magnetic fields as $B_{\rm reg} = (\bar{B}_x^2+\bar{B}_y^2+\bar{B}_z^2)^{0.5}$ and $B_{\rm turb} = ((B_x - \bar{B}_x)^2+(B_y - \bar{B}_y)^2+(B_z - \bar{B}_z)^2)^{0.5}$, respectively, where $B_{x,y,z}$ are the magnetic field components in each simulation cell along the $x$, $y$, $z$ directions, and the mean magnetic field, $\bar{B}_{x,y,z}$, are their mean values, computed by averaging each component over the horizontal ($x-y$) plane at a given vertical height $z$. To quantify the ratio between the regular and turbulent fields ($B_{\rm reg}/B_{\rm turb}$) within the disk, we consider the region within $\pm300$~pc of the midplane. In the \noicm{} model, this ratio ranges between $0.2$ and $0.8$ (mean $\approx 0.5$) over 250 Myr, with $B_{\rm reg}=0.3$ -- $3.5$~$\mu$G and $B_{\rm turb}=1$ -- $5$~$\mu$G. In the \icmpw{} model, the ratio decreases to $0.1$ -- $0.4$ (mean $\approx 0.3$) over the same period, with $B_{\rm reg}=0.2$ -- $2.5$~$\mu$G and $B_{\rm turb}=1$ -- $6.5$~$\mu$G. These results indicate that ram pressure can suppress the regular magnetic field while enhancing the turbulent component, thereby reducing the polarized emission from the disk.
    
    In addition to the increased turbulent-to-regular magnetic field ratio, Faraday depolarization is also likely to contribute to the suppression of polarized emission. \citet{arshakian_2011_pol} demonstrated that the intensity of polarized radio continuum emission decreases as a function of the rotation measure (RM) at a given wavelength. Since the RM is proportional to both the electron density and the line-of-sight magnetic field component (\autoref{eq:RM}), enhancements in either quantity will increase the RM and, consequently, reduce the polarized emission. Although we could not directly measure the magnetic field strength in the regions lacking polarized emission, the regions where polarization is detected exhibit relatively weak magnetic fields. This suggests that an increase in electron density is the more plausible cause of the depolarization. This interpretation is consistent with our earlier suggestion that ICM–ISM mixing, which can heat the ISM and produce additional free electrons, plays an important role in shaping the overall ISM properties of NGC~4522. 
    
    Consequently, the combination of these suppression effects with mechanisms that enhance polarized emission on the leading side, such as compression, shear motions, and magnetic draping, can naturally explain the highly asymmetric polarized emission observed in NGC~4522.

    %We note that NGC~4522 is a highly inclined galaxy and is therefore likely to be subject to the line-of-sight confusion and projection. However, in the C-band continuum observations, relatively symmetric polarized emission distributions have been observed in normal edge-on galaxies \citep[e.g.,][]{krause2020_bfield}. Thus, although we cannot rule out the impact of projection, it is reasonable to attribute the asymmetric polarized emission of NGC~4522 to environmental effects.

    \section{Summary}\label{sec:sec_summary}
    
    We present S-band (3 GHz) and X-band (10 GHz) VLA continuum observations of the ram pressure–stripped galaxy NGC 4522 at spatial resolutions of $\sim7\arcsec$ and $\sim2.5\arcsec$, respectively. The VLA data were calibrated to obtain both total and polarized radio continuum maps. We examine the overall distribution of the total continuum in comparison with multi-wavelength observations and analyze the spectral index distribution and investigate the physical mechanisms driving its variation. Using the polarized continuum data, we investigate the spatial distribution of polarized emission, magnetic field orientations, and polarization fraction variations, and discuss the physical origins of the highly asymmetric polarized morphology. We further derive the rotation measure from the polarization angles in both bands, estimate magnetic field strengths, and discuss their implications.
    
    The main findings are summarized as follows:
    
    \begin{enumerate}
        \item The total radio continuum emission at S and X bands is asymmetric, extending toward the northwest, broadly resembling the \ion{H}{1} gas distribution. Within the stellar disk, continuum emissions spatially correlate with CO and H$\alpha$, whereas the extraplanar regions exhibit local offsets, suggesting a complex interplay between the ISM and the ICM. The integrated spectral index between the S and X bands ($\alpha = -0.93\pm0.05$) is consistent with previous L- and C-band measurements ($\alpha = -0.97$, \citealt{2004AJ....127.3375V}). However, our 2D spectral index map (\autoref{fig:ngc4522_specindx_sx}) reveals, for the first time, significantly flatter indices in the extraplanar regions.
    
        \item The spectral index steepens systematically from $\alpha \sim -0.6$ in the main disk to $\alpha \sim -1.1$ in the outer disk. The galaxy’s leading side, directly impacted by the ICM wind, shows moderately steep indices ($\alpha \approx -0.8$), and the far side also shows similar spectral slopes. Two component fitting shows that the outer disk exhibits a near pure power-law spectrum without clear evidence of CRe aging and their non-thermal spectral index is flatter compared to the main disk. These features suggest recent CRe re-acceleration by the ICM-induced shock. A significant reduction in the contribution from thermal emission is observed across the outer disk.
    
        \item Extraplanar clouds (R3 -- R6) exhibit unusually flat spectral indices ($\alpha \sim -0.3$ to $0$). While some regions coincide with strong H$\alpha$ emission (e.g., R3 and R5), others show little or no detectable H$\alpha$ (e.g., R4, R6). These flat indices result from a combination of strongly aged, steep nonthermal spectra and significantly enhanced thermal emission. We propose that ICM–ISM mixing serves as the primary mechanism enhancing thermal emission independent of star formation. Our previous simulations support this scenario, demonstrating that stripped clouds mixed with the ICM can sustain thermal emission while the magnetic field becomes diluted. The mild flattening observed at the edges of the main disk (R1 with H$\alpha$ and R2 without H$\alpha$) may also partly arise from such mixing processes.
    
        \item Polarized continuum emission in both bands is highly asymmetric, although spatial offsets exist between the locations and distributions of their peaks. The magnetic field orientation in the X-band is nearly parallel to the galactic disk, whereas in the S-band it tends to be rotated toward the north. In both bands, the polarization fraction increases radially outward from the galaxy center, suggesting that the polarized continuum emission may originate from gas compression by the ICM wind and/or the magnetic draping effect. The overall polarization fraction is higher in the X-band, likely because the S-band is more strongly affected by Faraday depolarization.
        
        \item Using polarization angles from both bands, we derive rotation measures (RM) and estimate the line-of-sight magnetic field strength. Due to the limited wavelength coverage, we adopt a foreground screen model rather than performing RM synthesis analysis. The RMs range from $-100$ to $350$~rad~m$^{-2}$, clustering around two main intervals: $-100$ to $0$ and $200$ to $300$~rad~m$^{-2}$. The estimated line-of-sight magnetic field strengths range from $0$ to $15$~$\mu$G, with a median value of $\approx 6$~$\mu$G, while those within the main disk remain relatively low ($\lesssim 5$~$\mu$G).
    
        \item We suggest that the highly asymmetric polarized emission arises from a combination of two effects: enhanced polarized emission at the leading side, likely due to the compression and the magnetic draping via RPS; and suppressed polarized emission within the disk, where the interaction with the ICM increases turbulent magnetic fields and CR electrons, while regular magnetic fields are reduced by ICM shock propagation.
    
    \end{enumerate}
    
    Although we showed \autoref{app:sec_appB} that convolving the X-band images has only a minor impact on the overall results, it would be preferable to obtain images with similar spatial resolution to minimize uncertainties introduced during post-processing. Therefore, we plan to propose additional VLA observations of NGC~4522 at lower-resolution X-band (or high-resolution S-band) to improved both the image consistency and sensitivity. Moreover, since our current analysis is based on a single galaxy, we will also propose VLA observations of other Virgo galaxies  exhibiting clear extraplanar gas to test our hypothesis. These future observations will help determine whether NGC~4522 represents a unique case and provide a deeper understanding of ram-pressure effects on the general ISM. Finally, while our simulation analysis has provided valuable insights, quantitative interpretation was limited by the absence of CRe and the restricted simulation volume. A galaxy-scale simulation incorporating both magnetic fields and CRe would enable a more complete investigation of the evolution of thermal and nonthermal radio emission.

    %% Please use the acknowledgment and contribution environments. This will 
    %% be anonomyized when the "anonymous" style option is used. 
    \begin{acknowledgments}
    We thank the anonymous referee for helpful and constructive comments. The work of WC was supported by Basic Science Research Program through the National Research Foundation of Korea (NRF) funded by the Ministry of Education (RS-2024-00413394). AC acknowledges support by the NRF, No. RS-2022-NR070872 and RS-2022-NR069020. 
    
    This work has used the Karl G. Jansky Very Large Array
    operated by the National Radio Astronomy Observatory
    (NRAO). The NRAO and Green Bank Observatory are facilities of the U.S. National Science Foundation operated under cooperative agreement by Associated Universities, Inc. All the {\it HST} data presented in this paper were obtained from the Mikulski Archive for Space Telescopes (MAST) at the Space Telescope Science Institute. The specific observations analyzed can be accessed via MAST:\dataset[10.17909/pahn-3r19]{http://dx.doi.org/10.17909/pahn-3r19}. STScI is operated by the Association of Universities for Research in Astronomy, Inc., under NASA contract NAS5–26555. Support to MAST for these data is provided by the NASA Office of Space Science via grant NAG5–7584 and by other grants and contracts.
    
    \end{acknowledgments}
    
    % \begin{contribution}
    % %%This section gives authors the space to recognize author contributions. The text inside this environment is NOT counted towards the total word quanta. At a minimum, manuscripts are expected to include this text:
    
    % All authors contributed equally to the Terra Mater collaboration.
    
    %% But authors are expected to provide more specific details, e.g. 
    %%
    %%SC was responsible for writing and submitting the manuscript.
    %%WWM came up with the initial research concept and edited the manuscript.
    %%OTS obtained the funding and edited the manuscript.
    %%EBF provided the formal analysis and validation. He also edited the manuscript.
    %%GEH Supervised the undergraduates, wrote the software and administers the project github and Zenodo repositories.
    %%
    %% Authors can use the Contributor Role Taxonomy (CRediT) at
    %% for ideas on how write a good statement tailored to their needs.
    
    %\end{contribution}
    
    \facilities{VLA}
    
    \software{astropy \citep{astropy2013,astropy2018},
              matplotlib \citep{Matplotlib}, 
              numpy \citep{numpy}, 
              scipy \citep{scipy}
              APLpy \citep{APLpy}
              }
    
    
    \appendix

    \section{Data Reduction}\label{app:sec_appA}
    
    The data were reduced using the \textsc{Common Astronomy Software Applications} (\textsc{casa}; \citealt{mcmullin2007_casa}), version 6.4.1. Each observation was processed in an identical manner. To obtain the polarization information from the VLA continuum observations, we calibrated the data as follows\footnote{Note that these reductions are consistent with the standard continuum reduction tutorials (see \url{https://casaguides.nrao.edu/index.php?title=CASA_Guides:Polarization_Calibration_based_on_CASA_pipeline_standard_reduction:_The_radio_galaxy_3C75-CASA6.4.1)}. The parameters and options were adjusted to reflect the characteristics of our observations.}:

    \begin{enumerate}
        \item Run the pipeline, which is provided by the NRAO, using CASA. By doing so, we obtained the calibrated parallel-hand (RR/LL)\footnote{Note that the VLA antennas detect right- (R) and left-hand (L) circularly polarized emission.}  cross-correlation visibilities, which allow us to derive only the total intensity of the continuum rather than the polarization information. Therefore, additional calibration steps were required to obtain the polarized continuum.
        
        \item To continue the polarization calibration, we needed to remove the parallactic-angle correction that had been applied by the standard pipeline. Therefore, we ran the task \texttt{applycal}, reapplying the calibration to the corrected column with parang=False, thereby disabling the parallactic-angle corrections. Subsequently, the parallel-hand (RR and LL) data for all calibrators and for the target were flagged to remove severe radio-frequency interference (RFI) using the task \texttt{flagdata} with the mode set to "rflag".
    
        \item Another round of RFI flagging was performed for the parallel-hand data of all calibrators using the mode 'tfcrop', and for the cross-hand data (RL and LR) of all calibrators using the mode "rflag", with the \texttt{flagdata} task. In addition, manual flagging was carried out on the calibrators to remove low-level RFI and bad data that were not eliminated during the previous flagging steps. During this process, spectral window 3 (spw3) was completely flagged for both the parallel- and cross-hand data of all calibrators and the target source due to severe RFI contamination.
    
        \item We set the full polarization models for the flux-density, phase, and polarization-leakage calibrators using the polarization fractions and position angles as functions of frequency derived by \citet{perley2017}. These models were then used as reference values for the subsequent calibration steps.
        
        \item To determine the instrumental delay between the two polarization outputs, we derived the cross-hand delay ("kcross") solutions using a calibrator with known polarization. For this purpose, 3C 286 was used to solve for the cross-hand delays with the task \texttt{gaincal}.
        
        \item To determine the frequency-dependent polarization leakage between the R and L polarization channels (the so-called D-terms), which arise from instrumental effects, we solved for the polarization leakage per channel using the unpolarized calibrator J1407+2827 with the task  \texttt{polcal} and poltype="Df".
        
        \item To determine the absolute polarization angle (Xf), we ran the task \texttt{polcal} with poltype="Xf", using a calibrator with a known polarization position angle (3C 286).
    
        \item We then applied all calibration solutions to the raw data, including all three calibrators and the target source (NGC 4522), using the \texttt{applycal}  task with parang=True.
            
    \end{enumerate}
    
    \begin{figure}
      \centering
      \includegraphics[width=0.45\columnwidth]{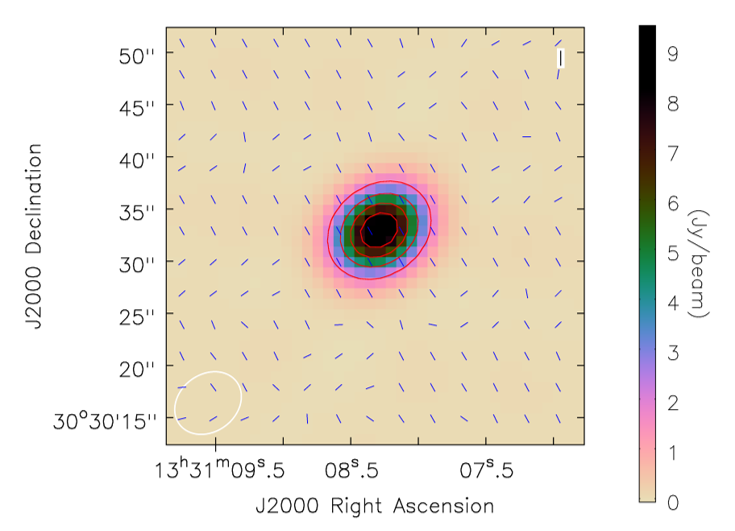}
      \includegraphics[width=0.45\columnwidth]{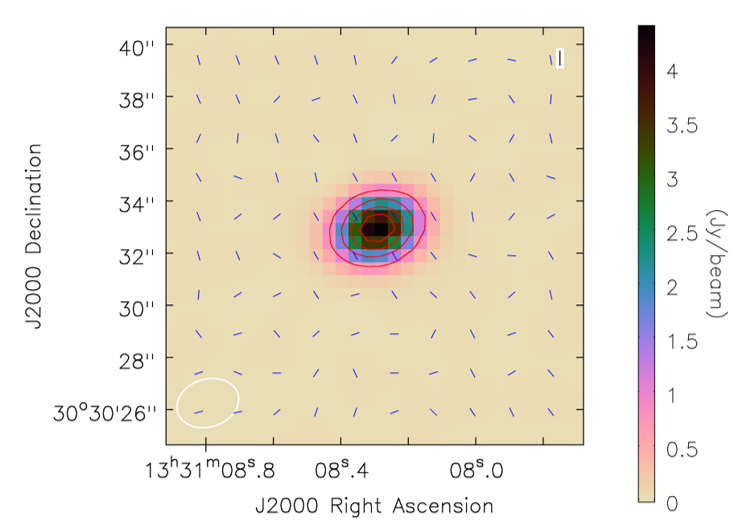}
      \caption{Total continuum intensity maps of 3C~286 overlaid with the polarized intensity (red contours) and polarization angle (blue vectors) for the S-band (left) and X-band (right). Contour levels are (0.2, 0.4, 0.6, 0.8) $\times$ 1.02 (0.53)~Jy~beam$^{-1}$ for the S-band (X-band). The synthesized beams of $7\farcs9\times6\farcs0$ and $2\farcs2\times1\farcs9$, respectively, and are shown in the bottom-left corner of the maps. These results confirm the accuracy of our data calibration. }
      \label{fig:3c286}
    \end{figure}
    
    To assess the robustness of our calibration, we first imaged the well-known calibrator 3C~286 and measured its total continuum flux, polarized flux, and polarization angle. The same imaging parameters and pixel scales as described above were used. \autoref{fig:3c286}  presents the total-intensity maps overlaid with the polarized continuum (red contours) and polarization angles (blue vectors) for the S-band (left) and X-band (right) of 3C~286. The measured total flux densities of 3C~286 are 10.0~Jy at S-band and 4.6~Jy at X-band; the polarized flux densities are 1.07~Jy and 0.55~Jy, respectively; and the polarization angles are $33.4^{\circ}$ and $33.9^{\circ}$, respectively. The corresponding reference values reported by \citep{perley2017} are 9.92~Jy at 3 GHz and 4.53~Jy at 10~GHz, 1.09~Jy at 3 GHz and 0.55~Jy at 10~GHz, and $33.0^{\circ}$ at 3 GHz and $33.4^{\circ}$ at 10 GHz, respectively. These results confirm that our polarization calibration was performed successfully.
    
    To produce a clean image of our target, we first generated a mask interactively using the \texttt{tclean} task, defining the regions where the algorithm was allowed to search for emission and encompassing all bright sources. We then deconvolved the calibrated data with \texttt{tclean} using the generated mask, cleaning down to a threshold of twice the RMS noise of the dirty image. The cleaned components were imaged with Briggs weighting, adopting a robust parameter of 0.5 for the S band and 1.5 for the X band. During imaging, we employed the \textit{Multi-scale Multi-frequency} (MT-MFS, \citealt{rau2011_mult}) algorithm, which improves upon the traditional CLEAN method for wide-band VLA observations. In this scheme, \texttt{tclean} was set to converge at spatial scales corresponding to the synthesized beam, 5$\times$ the synthesized beam, and 15$\times$ the synthesized beam. Compared with traditional CLEAN algorithms, MT-MFS models source structures as a collection of truncated paraboloids rather than point sources and simultaneously accounts for spectral variations of the source flux across the observing bandwidth. These features make MT-MFS particularly well suited for our data, as NGC~4522 exhibits extended radio continuum emission and the VLA observations cover a wide frequency range.
    
    Consequently, we achieved synthesized beam sizes of $7\farcs9\times6\farcs0$ for the S-band and $2\farcs6\times2\farcs1$ for the X-band in the final images of NGC~4522. To balance spatial sampling against image size, pixel sizes of $1.2$ arcsec and $0.5$ arcsec were adopted for the S- and X-bands, respectively, corresponding to $\approx 5.8$ and $\approx 4.7$ pixels across the synthesized beams.
    
    %Note that we did not correct the Faraday rotation since we would like to show the observed polarization angle explicitly, although the rotation measure ranges between $-100$ and $300$~rad m$^{-1}$, resulting in the Faraday rotation of $-57$ $\sim$ $180$ deg in the S-band (see \autoref{sec:sec3_bfield_rotmeasure}). Moreover, since Faraday rotation $\propto \lambda^2$, Faraday rotation in the X-band ($3$~cm) is expected to be negligible. We also note that the rotation measure of our Milky Way toward NGC~4522 is less than $20$~rad m$^{-1}$ \citep{taylor2009_rm_mw}, so we also did not correct for the Milky Way effect.
    
    We did not apply any correction for Faraday rotation, as we intended to present the observed polarization angles directly. The rotation measure (RM) ranges from $-100$ to $300$~rad m$^{-1}$, corresponding to a Faraday rotation of approximately $-57$ to $180$ deg in the S band (see \autoref{sec:sec3_bfield_rotmeasure}). Because Faraday rotation scales as $\lambda^2$, its effect at X-band (3~cm) is expected to be negligible. In addition, the Galactic rotation measure toward NGC~4522 is less than $20$~rad m$^{-2}$ \citep{taylor2009_rm_mw}, and thus no correction for the Milky Way contribution was applied.
    
    % find that the differential ionospheric Faraday rotation within S-band ranges between -40 and 30~rad m$^{-1}$, resulting in $-23$ $\sim$ $17$ deg at $10$~cm (see \autoref{sec:sec3_bfield_rotmeasure}). This will not significantly affect our interpretations of the results. Moreover, since Faraday rotation $\propto \lambda^2$, Faraday rotation at $3$~cm (X-band) is expected to be negligible. 

    %%%%%%%%%%%%%%%%%%%%%%%%%%%%%%%%%%%%%%%%%%%%%%%%%%

    %\appendix
    
    \section{Test of X-band imaging and smoothing}\label{app:sec_appB}
    
    \begin{figure}
        \centering
        \includegraphics[width=0.5\linewidth]{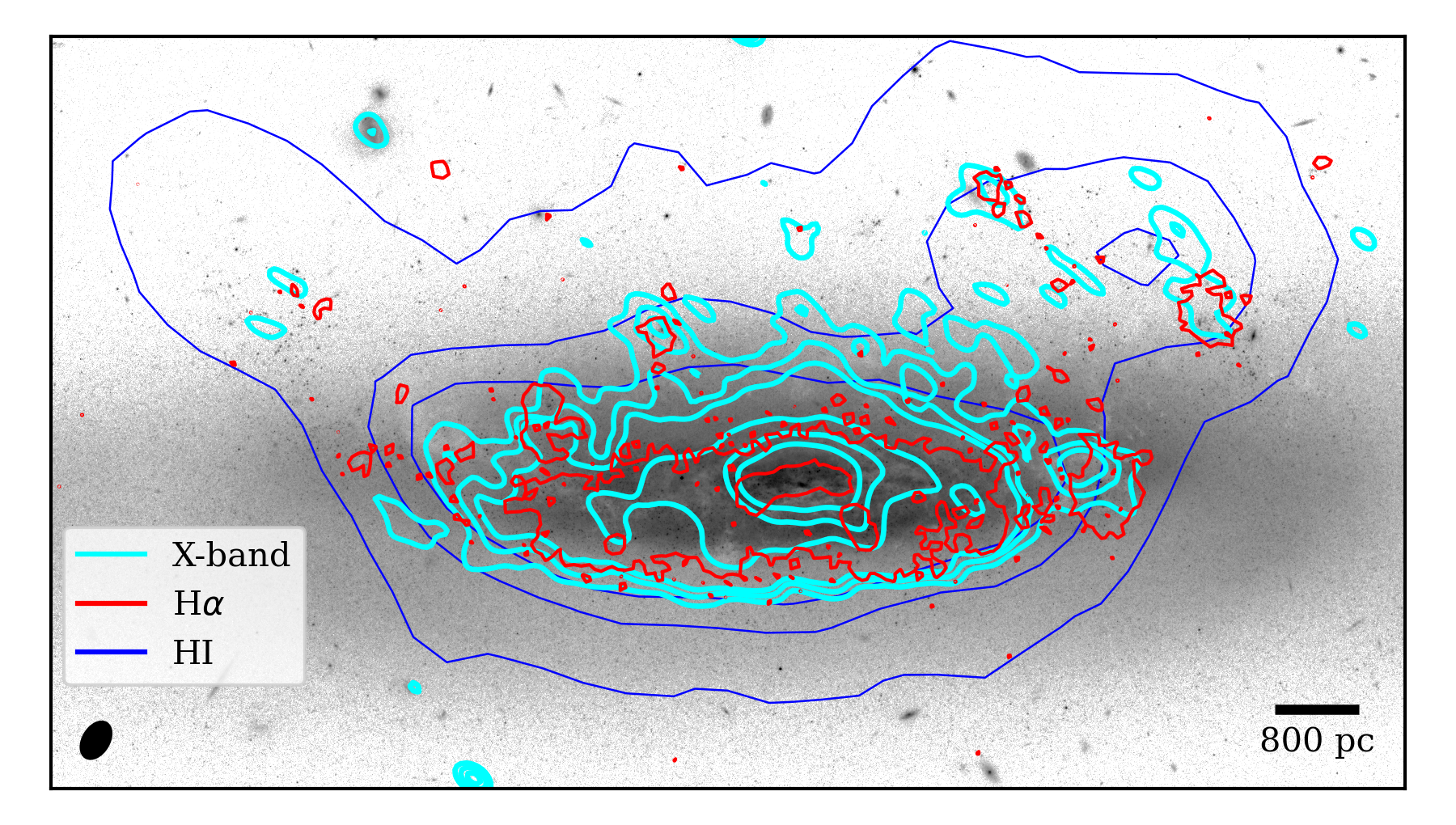}
        \includegraphics[width=0.5\linewidth]{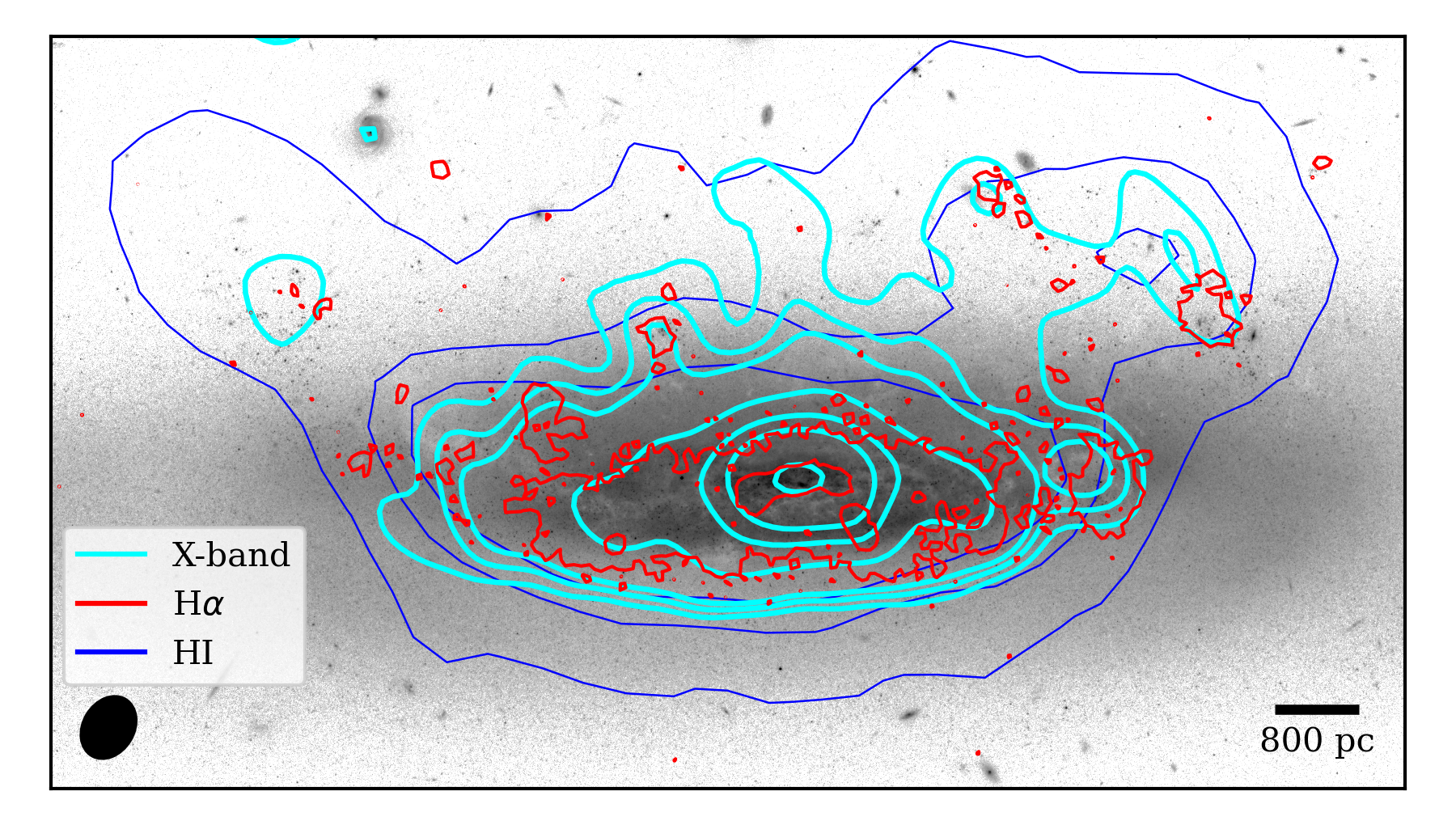}
        \includegraphics[width=0.5\linewidth]{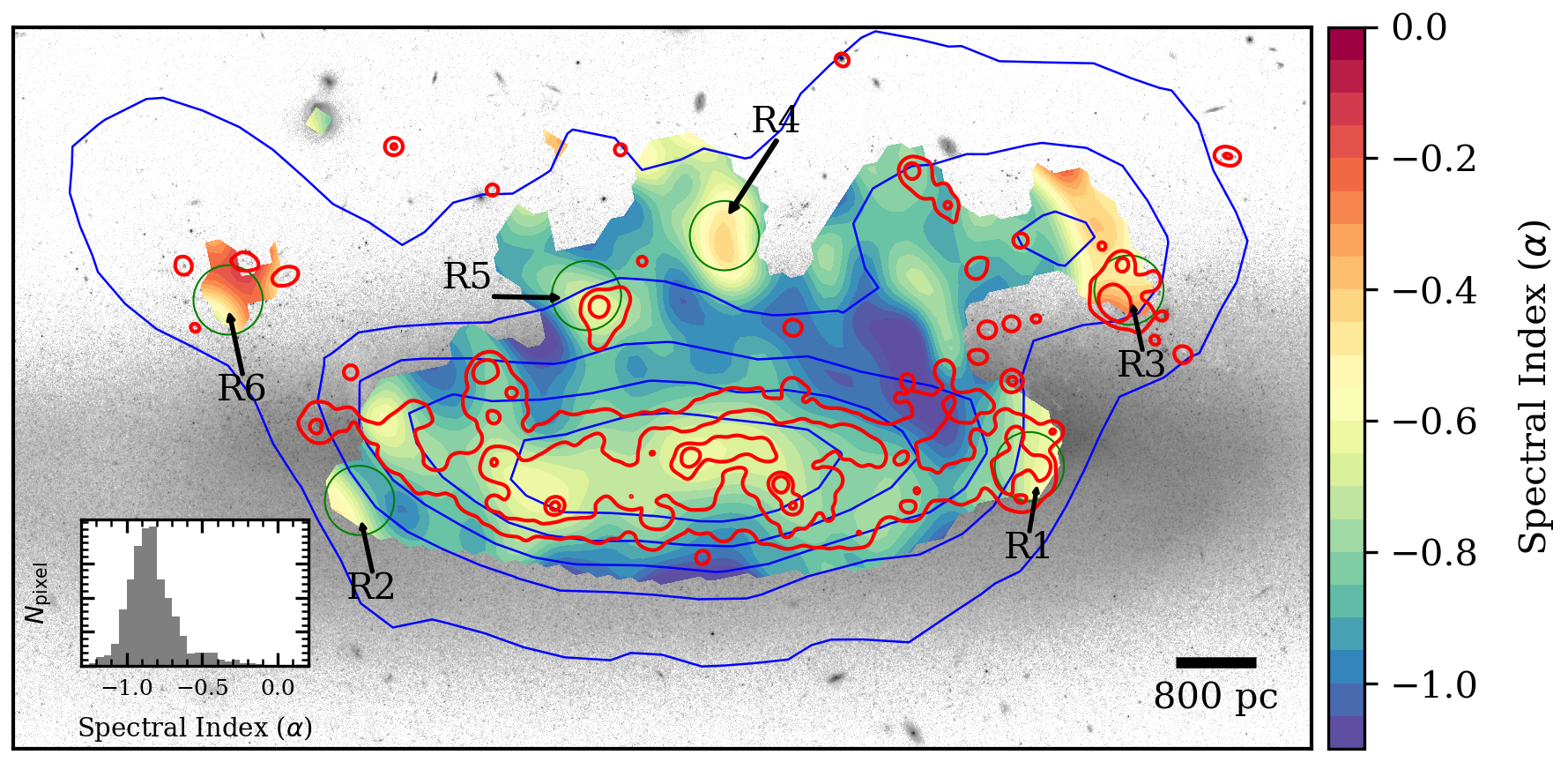}
        \caption{Same as \autoref{fig:ngc4522_total_S}, but for the UV-tapered X-band continuum map (top) and the X-band map convolved to the S-band resolution (middle). The UV range above $30$k$\lambda_{3\rm cm}$ was tapered, resulting in a synthesized beam of $4\farcs7\times3\farcs0$. The bottom panel shows the spectral index map derived from the UV-tapered X-band and S-band maps. }
        \label{fig:xband_uvtap}
    \end{figure}

    In \autoref{fig:xband_uvtap}, we present the UV-tapered X-band continuum map (top) and the X-band map convolved to the S-band resolution (middle). The UV range above $30$k$\lambda_{3\rm cm}$, resulting in a synthesized beam of $4\farcs7\times3\farcs0$ and an RMS noise of 2~$\mu$Jy~beam$^{-1}$.  The bottom panel shows the spectral index map derived from the UV-tapered X-band and S-band maps. The overall trend of the spectral index is similar to that in  \autoref{fig:ngc4522_specindx_sx}, although the values are systematically lower by approximately 0.15.
    
    \section{[\ion{N}{2}]/H$\alpha$ map}\label{sec:sec_appC}
    
    \begin{figure}
        \centering
        \includegraphics[width=0.5\linewidth]{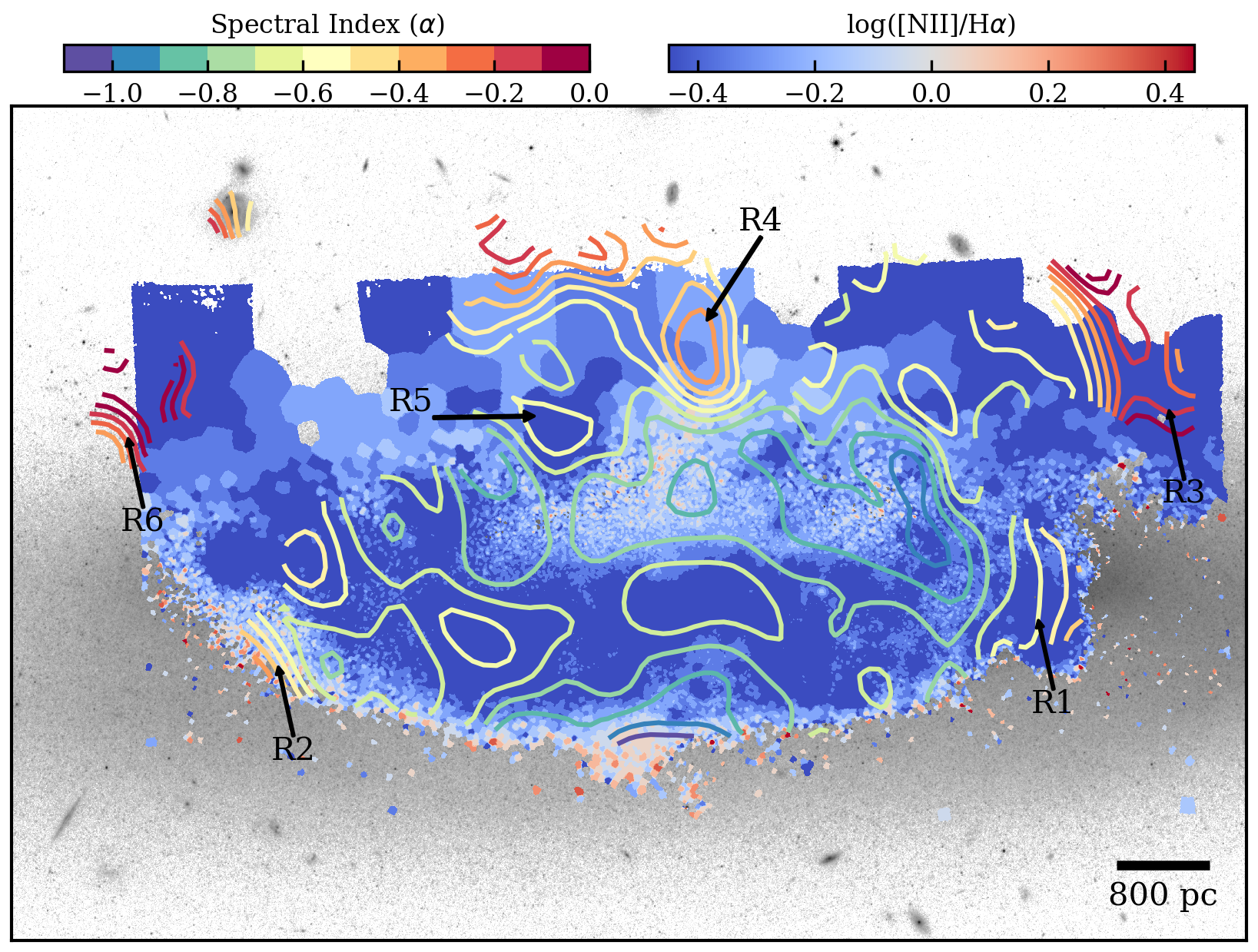}
        \caption{A map [\ion{N}{2}]/H$\alpha$ from MUSE observations, overlaid with contours of the spectral index map.}
        \label{fig:mauve-nii}
    \end{figure}

    \autoref{fig:mauve-nii} is the same as \autoref{fig:ngc4522_spec_mauve}, but for [\ion{N}{2}]/H$\alpha$. Although the overall values are lower than those of [\ion{S}{2}]/H$\alpha$, the general trend remains very similar.
    
    %% For this sample we use BibTeX plus aasjournalv7.bst to generate the
    
    \bibliography{ref.bib}{}

@INPROCEEDINGS{mcmullin2007_casa,
       author = {{McMullin}, J.~P. and {Waters}, B. and {Schiebel}, D. and {Young}, W. and {Golap}, K.},
        title = "{CASA Architecture and Applications}",
    booktitle = {Astronomical Data Analysis Software and Systems XVI},
         year = 2007,
       editor = {{Shaw}, R.~A. and {Hill}, F. and {Bell}, D.~J.},
       series = {Astronomical Society of the Pacific Conference Series},
       volume = {376},
        month = oct,
        pages = {127},
       adsurl = {https://ui.adsabs.harvard.edu/abs/2007ASPC..376..127M}
}

@article{1972ApJ...176....1G,
 adsurl = {https://ui.adsabs.harvard.edu/abs/1972ApJ...176....1G},
 author = {{Gunn}, James E. and {Gott}, J. Richard, III},
 doi = {10.1086/151605},
 journal = {\apj},
 month = aug,
 pages = {1},
 title = {{On the Infall of Matter Into Clusters of Galaxies and Some Effects on Their Evolution}},
 volume = {176},
 year = {1972}
}

@article{1990AJ....100..604C,
 adsurl = {https://ui.adsabs.harvard.edu/abs/1990AJ....100..604C},
 author = {{Cayatte}, V. and {van Gorkom}, J.~H. and {Balkowski}, C. and
{Kotanyi}, C.},
 doi = {10.1086/115545},
 journal = {\aj},
 month = sep,
 pages = {604},
 title = {{VLA Observations of Neutral Hydrogen in Virgo Cluster Galaxies. I. The Atlas}},
 volume = {100},
 year = {1990}
}

@article{2001SSRv...99..243B,
 adsurl = {https://ui.adsabs.harvard.edu/abs/2001SSRv...99..243B},
 archiveprefix = {arXiv},
 author = {{Beck}, Rainer},
 eprint = {astro-ph/0012402},
 journal = {\ssr},
 month = oct,
 pages = {243-260},
 primaryclass = {astro-ph},
 title = {{Galactic and Extragalactic Magnetic Fields}},
 volume = {99},
 year = {2001}
}

@article{kenney2004_4522,
 adsurl = {https://ui.adsabs.harvard.edu/abs/2004AJ....127.3361K},
 archiveprefix = {arXiv},
 author = {{Kenney}, Jeffrey D.~P. and {van Gorkom}, J.~H. and {Vollmer}, B.},
 doi = {10.1086/420805},
 eprint = {astro-ph/0403103},
 journal = {\aj},
 month = jun,
 number = {6},
 pages = {3361-3374},
 primaryclass = {astro-ph},
 title = {{VLA H I Observations of Gas Stripping in the Virgo Cluster Spiral NGC 4522}},
 volume = {127},
 year = {2004}
}

@article{2004AJ....127.3375V,
 adsurl = {https://ui.adsabs.harvard.edu/abs/2004AJ....127.3375V},
 archiveprefix = {arXiv},
 author = {{Vollmer}, B. and {Beck}, R. and {Kenney}, Jeffrey D.~P. and
{van Gorkom}, J.~H.},
 doi = {10.1086/420802},
 eprint = {astro-ph/0403054},
 journal = {\aj},
 month = jun,
 number = {6},
 pages = {3375-3381},
 primaryclass = {astro-ph},
 title = {{Radio Continuum Observations of the Virgo Cluster Spiral NGC 4522: The Signature of Ram Pressure}},
 volume = {127},
 year = {2004}
}

@article{2004ApJ...613..851K,
 adsurl = {https://ui.adsabs.harvard.edu/abs/2004ApJ...613..851K},
 archiveprefix = {arXiv},
 author = {{Koopmann}, Rebecca A. and {Kenney}, Jeffrey D.~P.},
 doi = {10.1086/423190},
 eprint = {astro-ph/0209547},
 journal = {\apj},
 month = oct,
 number = {2},
 pages = {851-865},
 primaryclass = {astro-ph},
 title = {{Massive Star Formation Rates and Radial Distributions from H{\ensuremath{\alpha}} Imaging of 84 Virgo Cluster and Isolated Spiral Galaxies}},
 volume = {613},
 year = {2004}
}

@article{2005AJ....130...65C,
 adsurl = {https://ui.adsabs.harvard.edu/abs/2005AJ....130...65C},
 archiveprefix = {arXiv},
 author = {{Crowl}, Hugh H. and {Kenney}, Jeffrey D.~P. and {van Gorkom}, J.~H. and
{Vollmer}, Bernd},
 doi = {10.1086/430526},
 eprint = {astro-ph/0503422},
 journal = {\aj},
 month = jul,
 number = {1},
 pages = {65-72},
 primaryclass = {astro-ph},
 title = {{Dense Cloud Ablation and Ram Pressure Stripping of the Virgo Spiral NGC 4402}},
 volume = {130},
 year = {2005}
}

@article{2006A&A...453..883V,
 adsurl = {https://ui.adsabs.harvard.edu/abs/2006A&A...453..883V},
 archiveprefix = {arXiv},
 author = {{Vollmer}, B. and {Soida}, M. and {Otmianowska-Mazur}, K. and
{Kenney}, J.~D.~P. and {van Gorkom}, J.~H. and {Beck}, R.},
 doi = {10.1051/0004-6361:20064954},
 eprint = {astro-ph/0603854},
 journal = {\aap},
 month = jul,
 number = {3},
 pages = {883-893},
 primaryclass = {astro-ph},
 title = {{A dynamical model for the heavily ram pressure stripped Virgo spiral galaxy NGC 4522}},
 volume = {453},
 year = {2006}
}

@article{2007A&A...464L..37V,
 adsurl = {https://ui.adsabs.harvard.edu/abs/2007A&A...464L..37V},
 archiveprefix = {arXiv},
 author = {{Vollmer}, B. and {Soida}, M. and {Beck}, R. and {Urbanik}, M. and
{Chy{\.z}y}, K.~T. and {Otmianowska-Mazur}, K. and {Kenney}, J.~D.~P. and
{van Gorkom}, J.~H.},
 doi = {10.1051/0004-6361:20066980},
 eprint = {astro-ph/0701610},
 journal = {\aap},
 month = mar,
 number = {3},
 pages = {L37-L40},
 primaryclass = {astro-ph},
 title = {{The characteristic polarized radio continuum distribution of cluster spiral galaxies}},
 volume = {464},
 year = {2007}
}

@article{2008A&A...491..455V,
 adsurl = {https://ui.adsabs.harvard.edu/abs/2008A&A...491..455V},
 archiveprefix = {arXiv},
 author = {{Vollmer}, B. and {Braine}, J. and {Pappalardo}, C. and {Hily-Blant}, P.},
 doi = {10.1051/0004-6361:200810432},
 eprint = {0809.5178},
 journal = {\aap},
 month = nov,
 number = {2},
 pages = {455-464},
 primaryclass = {astro-ph},
 title = {{Ram-pressure stripped molecular gas in the Virgo spiral galaxy NGC 4522}},
 volume = {491},
 year = {2008}
}

@article{crowl2008,
 adsurl = {https://ui.adsabs.harvard.edu/abs/2008AJ....136.1623C},
 archiveprefix = {arXiv},
 author = {{Crowl}, Hugh H. and {Kenney}, Jeffrey D.~P.},
 doi = {10.1088/0004-6256/136/4/1623},
 eprint = {0807.3747},
 journal = {\aj},
 month = oct,
 number = {4},
 pages = {1623-1644},
 primaryclass = {astro-ph},
 title = {{The Stellar Populations of Stripped Spiral Galaxies in the Virgo Cluster}},
 volume = {136},
 year = {2008}
}

@article{2009AJ....138.1741C,
 adsurl = {https://ui.adsabs.harvard.edu/\#abs/2009AJ....138.1741C},
 author = {{Chung}, Aeree and {van Gorkom}, J.~H. and {Kenney}, Jeffrey D.~P. and
{Crowl}, Hugh and {Vollmer}, Bernd},
 doi = {10.1088/0004-6256/138/6/1741},
 journal = {\aj},
 month = dec,
 pages = {1741-1816},
 title = {{VLA Imaging of Virgo Spirals in Atomic Gas (VIVA). I. The Atlas and the H I Properties}},
 volume = {138},
 year = {2009}
}

@article{2010A&A...512A..36V,
 adsurl = {https://ui.adsabs.harvard.edu/abs/2010A&A...512A..36V},
 archiveprefix = {arXiv},
 author = {{Vollmer}, B. and {Soida}, M. and {Chung}, A. and {Beck}, R. and
{Urbanik}, M. and {Chy{\.z}y}, K.~T. and {Otmianowska-Mazur}, K. and
{van Gorkom}, J.~H.},
 doi = {10.1051/0004-6361/200913591},
 eid = {A36},
 eprint = {1001.3597},
 journal = {\aap},
 month = mar,
 pages = {A36},
 primaryclass = {astro-ph.CO},
 title = {{The influence of the cluster environment on the large-scale radio continuum emission of 8 Virgo cluster spirals}},
 volume = {512},
 year = {2010}
}

@article{2013A&A...553A.116V,
 adsurl = {https://ui.adsabs.harvard.edu/abs/2013A&A...553A.116V},
 archiveprefix = {arXiv},
 author = {{Vollmer}, B. and {Soida}, M. and {Beck}, R. and {Chung}, A. and
{Urbanik}, M. and {Chy{\.z}y}, K.~T. and {Otmianowska-Mazur}, K. and
{Kenney}, J.~D.~P.},
 doi = {10.1051/0004-6361/201321163},
 eid = {A116},
 eprint = {1304.1279},
 journal = {\aap},
 month = may,
 pages = {A116},
 primaryclass = {astro-ph.CO},
 title = {{Large-scale radio continuum properties of 19 Virgo cluster galaxies. The influence of tidal interactions, ram pressure stripping, and accreting gas envelopes}},
 volume = {553},
 year = {2013}
}

@article{2014AJ....147...63A,
 adsurl = {https://ui.adsabs.harvard.edu/abs/2014AJ....147...63A},
 archiveprefix = {arXiv},
 author = {{Abramson}, Anne and {Kenney}, Jeffrey D.~P.},
 doi = {10.1088/0004-6256/147/3/63},
 eid = {63},
 eprint = {1401.0023},
 journal = {\aj},
 month = mar,
 number = {3},
 pages = {63},
 primaryclass = {astro-ph.GA},
 title = {{Hubble Space Telescope Imaging of Decoupled Dust Clouds in the Ram Pressure Stripped Virgo Spirals NGC 4402 and NGC 4522}},
 volume = {147},
 year = {2014}
}

@article{2014ApJ...780..119K,
 adsurl = {https://ui.adsabs.harvard.edu/abs/2014ApJ...780..119K},
 archiveprefix = {arXiv},
 author = {{Kenney}, Jeffrey D.~P. and {Geha}, Marla and {J{\'a}chym}, Pavel and
{Crowl}, Hugh H. and {Dague}, William and {Chung}, Aeree and
{van Gorkom}, Jacqueline and {Vollmer}, Bernd},
 doi = {10.1088/0004-637X/780/2/119},
 eid = {119},
 eprint = {1311.5501},
 journal = {\apj},
 month = jan,
 number = {2},
 pages = {119},
 primaryclass = {astro-ph.CO},
 title = {{Transformation of a Virgo Cluster Dwarf Irregular Galaxy by Ram Pressure Stripping: IC3418 and Its Fireballs}},
 volume = {780},
 year = {2014}
}

@article{2016AJ....152...32A,
 adsurl = {https://ui.adsabs.harvard.edu/abs/2016AJ....152...32A},
 archiveprefix = {arXiv},
 author = {{Abramson}, A. and {Kenney}, J. and {Crowl}, H. and {Tal}, T.},
 doi = {10.3847/0004-6256/152/2/32},
 eid = {32},
 eprint = {1604.01883},
 journal = {\aj},
 month = aug,
 number = {2},
 pages = {32},
 primaryclass = {astro-ph.GA},
 title = {{HST Imaging of Dust Structures and Stars in the Ram Pressure Stripped Virgo Spirals NGC 4402 and NGC 4522: Stripped from the Outside In with Dense Cloud Decoupling}},
 volume = {152},
 year = {2016}
}

@article{2018ApJ...866L..10L,
 adsurl = {https://ui.adsabs.harvard.edu/abs/2018ApJ...866L..10L},
 archiveprefix = {arXiv},
 author = {{Lee}, Bumhyun and {Chung}, Aeree},
 doi = {10.3847/2041-8213/aae4d9},
 eid = {L10},
 eprint = {1810.04188},
 journal = {\apj},
 month = oct,
 number = {1},
 pages = {L10},
 primaryclass = {astro-ph.GA},
 title = {{The ALMA Detection of Extraplanar $^{13}$CO in a Ram-pressure-stripped Galaxy and Its Implication}},
 volume = {866},
 year = {2018}
}

@article{2019MNRAS.487.4580R,
 adsurl = {https://ui.adsabs.harvard.edu/abs/2019MNRAS.487.4580R},
 archiveprefix = {arXiv},
 author = {{Ramatsoku}, M. and {Serra}, P. and {Poggianti}, B.~M. and
{Moretti}, A. and {Gullieuszik}, M. and {Bettoni}, D. and {Deb}, T. and
{Fritz}, J. and {van Gorkom}, J.~H. and {Jaff{\'e}}, Y.~L. and
{Tonnesen}, S. and {Verheijen}, M.~A.~W. and {Vulcani}, B. and
{Hugo}, B. and {J{\'o}zsa}, G.~I.~G. and {Maccagni}, F.~M. and
{Makhathini}, S. and {Ramaila}, A. and {Smirnov}, O. and {Thorat}, K.},
 doi = {10.1093/mnras/stz1609},
 eprint = {1906.03686},
 journal = {\mnras},
 month = aug,
 number = {4},
 pages = {4580-4591},
 primaryclass = {astro-ph.GA},
 title = {{GASP - XVII. H I imaging of the jellyfish galaxy JO206: gas stripping and enhanced star formation}},
 volume = {487},
 year = {2019}
}

@article{2020ApJ...897..143K,
 adsurl = {https://ui.adsabs.harvard.edu/abs/2020ApJ...897..143K},
 archiveprefix = {arXiv},
 author = {{Kado-Fong}, Erin and {Kim}, Jeong-Gyu and {Ostriker}, Eve C. and {Kim}, Chang-Goo},
 doi = {10.3847/1538-4357/ab9abd},
 eid = {143},
 eprint = {2006.06697},
 journal = {\apj},
 month = jul,
 number = {2},
 pages = {143},
 primaryclass = {astro-ph.GA},
 title = {{Diffuse Ionized Gas in Simulations of Multiphase, Star-forming Galactic Disks}},
 volume = {897},
 year = {2020}
}

@ARTICLE{vulcani_2020_sfr,
       author = {{Vulcani}, Benedetta and {Poggianti}, Bianca M. and {Tonnesen}, Stephanie and {McGee}, Sean L. and {Moretti}, Alessia and {Fritz}, Jacopo and {Gullieuszik}, Marco and {Jaff{\'e}}, Yara L. and {Franchetto}, Andrea and {Tomi{\v{c}}i{\'c}}, Neven and {Mingozzi}, Matilde and {Bettoni}, Daniela and {Wolter}, Anna},
        title = "{GASP XXX. The Spatially Resolved SFR-Mass Relation in Stripping Galaxies in the Local Universe}",
      journal = {\apj},
         year = 2020,
        month = aug,
       volume = {899},
       number = {2},
          eid = {98},
        pages = {98},
          doi = {10.3847/1538-4357/aba4ae},
archivePrefix = {arXiv},
       eprint = {2007.04996},
 primaryClass = {astro-ph.GA},
       adsurl = {https://ui.adsabs.harvard.edu/abs/2020ApJ...899...98V}
}

@article{abramson2011,
 adsurl = {https://ui.adsabs.harvard.edu/abs/2011AJ....141..164A},
 archiveprefix = {arXiv},
 author = {{Abramson}, Anne and {Kenney}, Jeffrey D.~P. and {Crowl}, Hugh H. and
{Chung}, Aeree and {van Gorkom}, J.~H. and {Vollmer}, Bernd and
{Schiminovich}, David},
 doi = {10.1088/0004-6256/141/5/164},
 eid = {164},
 eprint = {1101.4066},
 journal = {\aj},
 month = may,
 number = {5},
 pages = {164},
 primaryclass = {astro-ph.CO},
 title = {{Caught in the Act: Strong, Active Ram Pressure Stripping in Virgo Cluster Spiral NGC 4330}},
 volume = {141},
 year = {2011}
}

@ARTICLE{vulcani2020_sfr_resolved,
       author = {{Vulcani}, Benedetta and {Fritz}, Jacopo and {Poggianti}, Bianca M. and {Bettoni}, Daniela and {Franchetto}, Andrea and {Moretti}, Alessia and {Gullieuszik}, Marco and {Jaff{\'e}}, Yara and {Biviano}, Andrea and {Radovich}, Mario and {Mingozzi}, Matilde},
        title = "{GASP XXIV. The History of Abruptly Quenched Galaxies in Clusters}",
      journal = {\apj},
         year = 2020,
        month = apr,
       volume = {892},
       number = {2},
          eid = {146},
        pages = {146},
          doi = {10.3847/1538-4357/ab7bdd},
archivePrefix = {arXiv},
       eprint = {2003.02274},
 primaryClass = {astro-ph.GA},
       adsurl = {https://ui.adsabs.harvard.edu/abs/2020ApJ...892..146V}
}

@ARTICLE{chyzy2007_bfield,
       author = {{Chy{\.z}y}, K.~T. and {Ehle}, M. and {Beck}, R.},
        title = "{Magnetic fields and gas in the cluster-influenced spiral galaxy NGC 4254. I. Radio and X-rays observations}",
      journal = {\aap},
         year = 2007,
        month = nov,
       volume = {474},
       number = {2},
        pages = {415-429},
          doi = {10.1051/0004-6361:20077497},
archivePrefix = {arXiv},
       eprint = {0708.1533},
 primaryClass = {astro-ph},
       adsurl = {https://ui.adsabs.harvard.edu/abs/2007A&A...474..415C}
}

@ARTICLE{perley2017,
       author = {{Perley}, R.~A. and {Butler}, B.~J.},
        title = "{An Accurate Flux Density Scale from 50 MHz to 50 GHz}",
      journal = {\apjs},
         year = 2017,
        month = may,
       volume = {230},
       number = {1},
          eid = {7},
        pages = {7},
          doi = {10.3847/1538-4365/aa6df9},
archivePrefix = {arXiv},
       eprint = {1609.05940},
 primaryClass = {astro-ph.IM},
       adsurl = {https://ui.adsabs.harvard.edu/abs/2017ApJS..230....7P}
}

@ARTICLE{dressler1980,
       author = {{Dressler}, A.},
        title = "{Galaxy morphology in rich clusters: implications for the formation and evolution of galaxies.}",
      journal = {\apj},
         year = 1980,
        month = mar,
       volume = {236},
        pages = {351-365},
          doi = {10.1086/157753},
       adsurl = {https://ui.adsabs.harvard.edu/abs/1980ApJ...236..351D}
}

@ARTICLE{Makarove2014_hyperleda,
       author = {{Makarov}, Dmitry and {Prugniel}, Philippe and {Terekhova}, Nataliya and {Courtois}, H{\'e}l{\`e}ne and {Vauglin}, Isabelle},
        title = "{HyperLEDA. III. The catalogue of extragalactic distances}",
      journal = {\aap},
         year = 2014,
        month = oct,
       volume = {570},
          eid = {A13},
        pages = {A13},
          doi = {10.1051/0004-6361/201423496},
archivePrefix = {arXiv},
       eprint = {1408.3476},
 primaryClass = {astro-ph.GA},
       adsurl = {https://ui.adsabs.harvard.edu/abs/2014A&A...570A..13M}
}

@ARTICLE{crowl2006_ngc4522_sfr,
       author = {{Crowl}, Hugh H. and {Kenney}, Jeffrey D.~P.},
        title = "{The Stellar Population of Stripped Cluster Spiral NGC 4522: A Local Analog to K+A Galaxies?}",
      journal = {\apjl},
         year = 2006,
        month = oct,
       volume = {649},
       number = {2},
        pages = {L75-L78},
          doi = {10.1086/508344},
archivePrefix = {arXiv},
       eprint = {astro-ph/0608229},
 primaryClass = {astro-ph},
       adsurl = {https://ui.adsabs.harvard.edu/abs/2006ApJ...649L..75C}
}

@INCOLLECTION{beck2005_bfield,
       author = {{Beck}, Rainer},
        title = "{Magnetic Fields in Galaxies}",
    booktitle = {Cosmic Magnetic Fields},
         year = 2005,
       editor = {{Wielebinski}, Richard and {Beck}, Rainer},
       volume = {664},
        pages = {41},
          doi = {10.1007/3540313966_3},
       adsurl = {https://ui.adsabs.harvard.edu/abs/2005LNP...664...41B}
}

@ARTICLE{krause2020_bfield,
       author = {{Krause}, Marita and {Irwin}, Judith and {Schmidt}, Philip and {Stein}, Yelena and {Miskolczi}, Arpad and {Carolina Mora-Partiarroyo}, Silvia and {Wiegert}, Theresa and {Beck}, Rainer and {Stil}, Jeroen M. and {Heald}, George and {Li}, Jiang-Tao and {Damas-Segovia}, Ancor and {Vargas}, Carlos J. and {Rand}, Richard J. and {West}, Jennifer and {Walterbos}, Rene A.~M. and {Dettmar}, Ralf-J{\"u}rgen and {English}, Jayanne and {Woodfinden}, Alex},
        title = "{CHANG-ES. XXII. Coherent magnetic fields in the halos of spiral galaxies}",
      journal = {\aap},
         year = 2020,
        month = jul,
       volume = {639},
          eid = {A112},
        pages = {A112},
          doi = {10.1051/0004-6361/202037780},
archivePrefix = {arXiv},
       eprint = {2004.14383},
 primaryClass = {astro-ph.GA},
       adsurl = {https://ui.adsabs.harvard.edu/abs/2020A&A...639A.112K}
}

@ARTICLE{choi2022,
       author = {{Choi}, Woorak and {Kim}, Chang-Goo and {Chung}, Aeree},
        title = "{Ram Pressure Stripping of the Multiphase ISM: A Detailed View from TIGRESS Simulations}",
      journal = {\apj},
         year = 2022,
        month = sep,
       volume = {936},
       number = {2},
          eid = {133},
        pages = {133},
          doi = {10.3847/1538-4357/ac82ba},
archivePrefix = {arXiv},
       eprint = {2207.05263},
 primaryClass = {astro-ph.GA},
       adsurl = {https://ui.adsabs.harvard.edu/abs/2022ApJ...936..133C}
}

@ARTICLE{vanWeeren2017_cr_reaccel,
       author = {{van Weeren}, Reinout J. and {Andrade-Santos}, Felipe and {Dawson}, William A. and {Golovich}, Nathan and {Lal}, Dharam V. and {Kang}, Hyesung and {Ryu}, Dongsu and {Br{\`\i}ggen}, Marcus and {Ogrean}, Georgiana A. and {Forman}, William R. and {Jones}, Christine and {Placco}, Vinicius M. and {Santucci}, Rafael M. and {Wittman}, David and {Jee}, M. James and {Kraft}, Ralph P. and {Sobral}, David and {Stroe}, Andra and {Fogarty}, Kevin},
        title = "{The case for electron re-acceleration at galaxy cluster shocks}",
      journal = {Nature Astronomy},
         year = 2017,
        month = jan,
       volume = {1},
          eid = {0005},
        pages = {0005},
          doi = {10.1038/s41550-016-0005},
archivePrefix = {arXiv},
       eprint = {1701.01439},
 primaryClass = {astro-ph.HE},
       adsurl = {https://ui.adsabs.harvard.edu/abs/2017NatAs...1E...5V}
}

@ARTICLE{tabatabaei2017_spcidx_flat_sf,
       author = {{Tabatabaei}, F.~S. and {Schinnerer}, E. and {Krause}, M. and {Dumas}, G. and {Meidt}, S. and {Damas-Segovia}, A. and {Beck}, R. and {Murphy}, E.~J. and {Mulcahy}, D.~D. and {Groves}, B. and {Bolatto}, A. and {Dale}, D. and {Galametz}, M. and {Sandstrom}, K. and {Boquien}, M. and {Calzetti}, D. and {Kennicutt}, R.~C. and {Hunt}, L.~K. and {De Looze}, I. and {Pellegrini}, E.~W.},
        title = "{The Radio Spectral Energy Distribution and Star-formation Rate Calibration in Galaxies}",
      journal = {\apj},
         year = 2017,
        month = feb,
       volume = {836},
       number = {2},
          eid = {185},
        pages = {185},
          doi = {10.3847/1538-4357/836/2/185},
archivePrefix = {arXiv},
       eprint = {1611.01705},
 primaryClass = {astro-ph.GA},
       adsurl = {https://ui.adsabs.harvard.edu/abs/2017ApJ...836..185T}
}

@ARTICLE{condon1992_rc_review,
       author = {{Condon}, J.~J.},
        title = "{Radio emission from normal galaxies.}",
      journal = {\araa},
         year = 1992,
        month = jan,
       volume = {30},
        pages = {575-611},
          doi = {10.1146/annurev.aa.30.090192.003043},
       adsurl = {https://ui.adsabs.harvard.edu/abs/1992ARA&A..30..575C}
}

@ARTICLE{gioia1982_rc_spec,
       author = {{Gioia}, I.~M. and {Gregorini}, L. and {Klein}, U.},
        title = "{High frequency radio continuum observations of bright spiral galaxies}",
      journal = {\aap},
         year = 1982,
        month = dec,
       volume = {116},
       number = {1},
        pages = {164-174},
       adsurl = {https://ui.adsabs.harvard.edu/abs/1982A&A...116..164G}
}

@ARTICLE{rau2011_mult,
       author = {{Rau}, U. and {Cornwell}, T.~J.},
        title = "{A multi-scale multi-frequency deconvolution algorithm for synthesis imaging in radio interferometry}",
      journal = {\aap},
         year = 2011,
        month = aug,
       volume = {532},
          eid = {A71},
        pages = {A71},
          doi = {10.1051/0004-6361/201117104},
archivePrefix = {arXiv},
       eprint = {1106.2745},
 primaryClass = {astro-ph.IM},
       adsurl = {https://ui.adsabs.harvard.edu/abs/2011A&A...532A..71R}
}

@ARTICLE{vollmer2008_ngc4501,
       author = {{Vollmer}, B. and {Soida}, M. and {Chung}, A. and {van Gorkom}, J.~H. and {Otmianowska-Mazur}, K. and {Beck}, R. and {Urbanik}, M. and {Kenney}, J.~D.~P.},
        title = "{Pre-peak ram pressure stripping in the Virgo cluster spiral galaxy NGC 4501}",
      journal = {\aap},
         year = 2008,
        month = may,
       volume = {483},
       number = {1},
        pages = {89-106},
          doi = {10.1051/0004-6361:20078139},
archivePrefix = {arXiv},
       eprint = {0801.4874},
 primaryClass = {astro-ph},
       adsurl = {https://ui.adsabs.harvard.edu/abs/2008A&A...483...89V}
}

@BOOK{klein_Fletcher_2015_bfield,
       author = {{Klein}, U. and {Fletcher}, A.},
        title = "{Galactic and Intergalactic Magnetic Fields}",
         year = 2015,
       adsurl = {https://ui.adsabs.harvard.edu/abs/2015gimf.book.....K}
}

@ARTICLE{taylor2009_rm_mw,
       author = {{Taylor}, A.~R. and {Stil}, J.~M. and {Sunstrum}, C.},
        title = "{A Rotation Measure Image of the Sky}",
      journal = {\apj},
         year = 2009,
        month = sep,
       volume = {702},
       number = {2},
        pages = {1230-1236},
          doi = {10.1088/0004-637X/702/2/1230},
       adsurl = {https://ui.adsabs.harvard.edu/abs/2009ApJ...702.1230T}
}

@ARTICLE{beck2015,
       author = {{Beck}, Rainer},
        title = "{Magnetic fields in spiral galaxies}",
      journal = {\aapr},
         year = 2015,
        month = dec,
       volume = {24},
          eid = {4},
        pages = {4},
          doi = {10.1007/s00159-015-0084-4},
archivePrefix = {arXiv},
       eprint = {1509.04522},
 primaryClass = {astro-ph.GA},
       adsurl = {https://ui.adsabs.harvard.edu/abs/2015A&ARv..24....4B}
}

@ARTICLE{cortese2012_ngc4522_mass,
       author = {{Cortese}, L. and {Ciesla}, L. and {Boselli}, A. and {Bianchi}, S. and {Gomez}, H. and {Smith}, M.~W.~L. and {Bendo}, G.~J. and {Eales}, S. and {Pohlen}, M. and {Baes}, M. and {Corbelli}, E. and {Davies}, J.~I. and {Hughes}, T.~M. and {Hunt}, L.~K. and {Madden}, S.~C. and {Pierini}, D. and {di Serego Alighieri}, S. and {Zibetti}, S. and {Boquien}, M. and {Clements}, D.~L. and {Cooray}, A. and {Galametz}, M. and {Magrini}, L. and {Pappalardo}, C. and {Spinoglio}, L. and {Vlahakis}, C.},
        title = "{The dust scaling relations of the Herschel Reference Survey}",
      journal = {\aap},
         year = 2012,
        month = apr,
       volume = {540},
          eid = {A52},
        pages = {A52},
          doi = {10.1051/0004-6361/201118499},
archivePrefix = {arXiv},
       eprint = {1201.2762},
 primaryClass = {astro-ph.CO},
       adsurl = {https://ui.adsabs.harvard.edu/abs/2012A&A...540A..52C}
}

@ARTICLE{padovani_2018_cre,
       author = {{Padovani}, Marco and {Ivlev}, Alexei V. and {Galli}, Daniele and {Caselli}, Paola},
        title = "{Cosmic-ray ionisation in circumstellar discs}",
      journal = {\aap},
         year = 2018,
        month = jun,
       volume = {614},
          eid = {A111},
        pages = {A111},
          doi = {10.1051/0004-6361/201732202},
archivePrefix = {arXiv},
       eprint = {1803.09348},
 primaryClass = {astro-ph.HE},
       adsurl = {https://ui.adsabs.harvard.edu/abs/2018A&A...614A.111P}
}

@ARTICLE{orlando_2018_cre,
       author = {{Orlando}, E.},
        title = "{Imprints of cosmic rays in multifrequency observations of the interstellar emission}",
      journal = {\mnras},
         year = 2018,
        month = apr,
       volume = {475},
       number = {2},
        pages = {2724-2742},
          doi = {10.1093/mnras/stx3280},
archivePrefix = {arXiv},
       eprint = {1712.07127},
 primaryClass = {astro-ph.HE},
       adsurl = {https://ui.adsabs.harvard.edu/abs/2018MNRAS.475.2724O}
}

@ARTICLE{arshakian_2011_pol,
       author = {{Arshakian}, Tigran G. and {Beck}, Rainer},
        title = "{Optimum frequency band for radio polarization observations}",
      journal = {\mnras},
         year = 2011,
        month = dec,
       volume = {418},
       number = {4},
        pages = {2336-2342},
          doi = {10.1111/j.1365-2966.2011.19623.x},
archivePrefix = {arXiv},
       eprint = {1101.2631},
 primaryClass = {astro-ph.CO},
       adsurl = {https://ui.adsabs.harvard.edu/abs/2011MNRAS.418.2336A}
}

@ARTICLE{pfrommer_2010_mag_drap,
       author = {{Pfrommer}, Christoph and {Jonathan Dursi}, L.},
        title = "{Detecting the orientation of magnetic fields in galaxy clusters}",
      journal = {Nature Physics},
         year = 2010,
        month = jul,
       volume = {6},
       number = {7},
        pages = {520-526},
          doi = {10.1038/nphys1657},
archivePrefix = {arXiv},
       eprint = {0911.2476},
 primaryClass = {astro-ph.CO},
       adsurl = {https://ui.adsabs.harvard.edu/abs/2010NatPh...6..520P}
}

@ARTICLE{linzer_2024,
       author = {{Linzer}, Nora B. and {Kim}, Jeong-Gyu and {Kim}, Chang-Goo and {Ostriker}, Eve C.},
        title = "{Ultraviolet Radiation Fields in Star-forming Disk Galaxies: Numerical Simulations with TIGRESS-NCR}",
      journal = {\apj},
         year = 2024,
        month = nov,
       volume = {975},
       number = {2},
          eid = {173},
        pages = {173},
          doi = {10.3847/1538-4357/ad7733},
archivePrefix = {arXiv},
       eprint = {2409.05958},
 primaryClass = {astro-ph.GA},
       adsurl = {https://ui.adsabs.harvard.edu/abs/2024ApJ...975..173L}
}

@ARTICLE{vargas_2018_specindx,
       author = {{Vargas}, Carlos J. and {Mora-Partiarroyo}, Silvia Carolina and {Schmidt}, Philip and {Rand}, Richard J. and {Stein}, Yelena and {Walterbos}, Ren{\'e} A.~M. and {Wang}, Q. Daniel and {Basu}, Aritra and {Patterson}, Maria and {Kepley}, Amanda and {Beck}, Rainer and {Irwin}, Judith and {Heald}, George and {Li}, Jiangtao and {Wiegert}, Theresa},
        title = "{CHANG-ES X: Spatially Resolved Separation of Thermal Contribution from Radio Continuum Emission in Edge-on Galaxies}",
      journal = {\apj},
         year = 2018,
        month = feb,
       volume = {853},
       number = {2},
          eid = {128},
        pages = {128},
          doi = {10.3847/1538-4357/aaa47f},
archivePrefix = {arXiv},
       eprint = {1801.01892},
 primaryClass = {astro-ph.GA},
       adsurl = {https://ui.adsabs.harvard.edu/abs/2018ApJ...853..128V}
}

@ARTICLE{kim2017_tigress,
       author = {{Kim}, Chang-Goo and {Ostriker}, Eve C.},
        title = "{Three-phase Interstellar Medium in Galaxies Resolving Evolution with Star Formation and Supernova Feedback (TIGRESS): Algorithms, Fiducial Model, and Convergence}",
      journal = {\apj},
         year = 2017,
        month = sep,
       volume = {846},
       number = {2},
          eid = {133},
        pages = {133},
          doi = {10.3847/1538-4357/aa8599},
archivePrefix = {arXiv},
       eprint = {1612.03918},
 primaryClass = {astro-ph.GA},
       adsurl = {https://ui.adsabs.harvard.edu/abs/2017ApJ...846..133K}
}

@ARTICLE{stein_2023_changes,
       author = {{Stein}, M. and {Heesen}, V. and {Dettmar}, R. -J. and {Stein}, Y. and {Br{\"u}ggen}, M. and {Beck}, R. and {Adebahr}, B. and {Wiegert}, T. and {Vargas}, C.~J. and {Bomans}, D.~J. and {Li}, J. and {English}, J. and {Chy{\.z}y}, K.~T. and {Paladino}, R. and {Tabatabaei}, F.~S. and {Strong}, A.},
        title = "{CHANG-ES. XXVI. Insights into cosmic-ray transport from radio halos in edge-on galaxies}",
      journal = {\aap},
         year = 2023,
        month = feb,
       volume = {670},
          eid = {A158},
        pages = {A158},
          doi = {10.1051/0004-6361/202243906},
archivePrefix = {arXiv},
       eprint = {2210.07709},
 primaryClass = {astro-ph.GA},
       adsurl = {https://ui.adsabs.harvard.edu/abs/2023A&A...670A.158S}
}

@ARTICLE{Heesen_2019_spix,
       author = {{Heesen}, V. and {Whitler}, L. and {Schmidt}, P. and {Miskolczi}, A. and {Sridhar}, S.~S. and {Horellou}, C. and {Beck}, R. and {G{\"u}rkan}, G. and {Scannapieco}, E. and {Br{\"u}ggen}, M. and {Heald}, G.~H. and {Krause}, M. and {Paladino}, R. and {Nikiel-Wroczy{\'n}ski}, B. and {Wilber}, A. and {Dettmar}, R. -J.},
        title = "{Warped diffusive radio halo around the quiescent spiral edge-on galaxy NGC 4565}",
      journal = {\aap},
         year = 2019,
        month = aug,
       volume = {628},
          eid = {L3},
        pages = {L3},
          doi = {10.1051/0004-6361/201936046},
archivePrefix = {arXiv},
       eprint = {1907.07076},
 primaryClass = {astro-ph.GA},
       adsurl = {https://ui.adsabs.harvard.edu/abs/2019A&A...628L...3H}
}

@ARTICLE{Boselli_2018_vestige,
       author = {{Boselli}, A. and {Fossati}, M. and {Ferrarese}, L. and {Boissier}, S. and {Consolandi}, G. and {Longobardi}, A. and {Amram}, P. and {Balogh}, M. and {Barmby}, P. and {Boquien}, M. and {Boulanger}, F. and {Braine}, J. and {Buat}, V. and {Burgarella}, D. and {Combes}, F. and {Contini}, T. and {Cortese}, L. and {C{\^o}t{\'e}}, P. and {C{\^o}t{\'e}}, S. and {Cuillandre}, J.~C. and {Drissen}, L. and {Epinat}, B. and {Fumagalli}, M. and {Gallagher}, S. and {Gavazzi}, G. and {Gomez-Lopez}, J. and {Gwyn}, S. and {Harris}, W. and {Hensler}, G. and {Koribalski}, B. and {Marcelin}, M. and {McConnachie}, A. and {Miville-Deschenes}, M.~A. and {Navarro}, J. and {Patton}, D. and {Peng}, E.~W. and {Plana}, H. and {Prantzos}, N. and {Robert}, C. and {Roediger}, J. and {Roehlly}, Y. and {Russeil}, D. and {Salome}, P. and {Sanchez-Janssen}, R. and {Serra}, P. and {Spekkens}, K. and {Sun}, M. and {Taylor}, J. and {Tonnesen}, S. and {Vollmer}, B. and {Willis}, J. and {Wozniak}, H. and {Burdullis}, T. and {Devost}, D. and {Mahoney}, B. and {Manset}, N. and {Petric}, A. and {Prunet}, S. and {Withington}, K.},
        title = "{A Virgo Environmental Survey Tracing Ionised Gas Emission (VESTIGE). I. Introduction to the survey}",
      journal = {\aap},
         year = 2018,
        month = jun,
       volume = {614},
          eid = {A56},
        pages = {A56},
          doi = {10.1051/0004-6361/201732407},
archivePrefix = {arXiv},
       eprint = {1802.02829},
 primaryClass = {astro-ph.GA},
       adsurl = {https://ui.adsabs.harvard.edu/abs/2018A&A...614A..56B}
}

@ARTICLE{watts_2024_mauve,
       author = {{Watts}, Adam B. and {Cortese}, Luca and {Catinella}, Barbara and {Fraser-McKelvie}, Amelia and {Emsellem}, Eric and {Coccato}, Lodovico and {van de Sande}, Jesse and {Brown}, Toby H. and {Ascasibar}, Yago and {Battisti}, Andrew and {Boselli}, Alessandro and {Davis}, Timothy A. and {Groves}, Brent and {Thater}, Sabine},
        title = "{MAUVE: a 6 kpc bipolar outflow launched from NGC 4383, one of the most H I-rich galaxies in the Virgo cluster}",
      journal = {\mnras},
         year = 2024,
        month = may,
       volume = {530},
       number = {2},
        pages = {1968-1983},
          doi = {10.1093/mnras/stae898},
archivePrefix = {arXiv},
       eprint = {2404.12616},
 primaryClass = {astro-ph.GA},
       adsurl = {https://ui.adsabs.harvard.edu/abs/2024MNRAS.530.1968W}
}

@ARTICLE{cainella_2025_mauve,
       author = {{Catinella}, B. and {Cortese}, L. and {Sun}, J. and {Brown}, T. and {Emsellem}, E. and {Fraser-McKelvie}, A. and {Watts}, A.~B. and {Attwater}, A. and {Battisti}, A. and {Boselli}, A. and {Choi}, W. and {Chung}, A. and {da Cunha}, E. and {Davis}, T.~A. and {Ellison}, S. and {J{\'a}chym}, P. and {Jimenez-Donaire}, M.~J. and {Kolcu}, T. and {Lee}, B. and {McGregor}, J. and {Roberts}, I. and {Schinnerer}, E. and {Spekkens}, K. and {Thater}, S. and {Thilker}, D. and {van de Sande}, J. and {Villanueva}, V. and {Williams}, T.~G. and {Zabel}, N.},
        title = "{Multiphase Astrophysics to Unveil the Virgo Environment (MAUVE)}",
      journal = {The Messenger},
         year = 2025,
        month = sep,
       volume = {195},
        pages = {15-18},
          doi = {10.18727/0722-6691/5393},
archivePrefix = {arXiv},
       eprint = {2511.17028},
 primaryClass = {astro-ph.GA},
       adsurl = {https://ui.adsabs.harvard.edu/abs/2025Msngr.195...15C}
}

@ARTICLE{brown_2025,
       author = {{Brown}, Toby and {Brown}, Toby and {Brown}, Toby and {Brown}, Toby and {Brown}, Toby and {Brown}, Toby and {Brown}, Toby and {Brown}, Toby and {Brown}, Toby},
      journal = {\apj},
         year = 2025,
       volume = {submitted},
        pages = {},
}

@ARTICLE{Wang_2021_wallaby,
       author = {{Wang}, Jing and {Staveley-Smith}, Lister and {Westmeier}, Tobias and {Catinella}, Barbara and {Shao}, Li and {Reynolds}, T.~N. and {For}, Bi-Qing and {Lee}, Bumhyun and {Liang}, Ze-zhong and {Wang}, Shun and {Elagali}, A. and {D{\'e}nes}, H. and {Kleiner}, D. and {Koribalski}, B{\"a}rbel S. and {Lee-Waddell}, K. and {Oh}, S. -H. and {Rhee}, J. and {Serra}, P. and {Spekkens}, K. and {Wong}, O.~I. and {Bekki}, K. and {Bigiel}, F. and {Courtois}, H.~M. and {Hess}, Kelley M. and {Holwerda}, B.~W. and {McQuinn}, Kristen B.~W. and {Pandey-Pommier}, M. and {van der Hulst}, J.~M. and {Verdes-Montenegro}, L.},
        title = "{WALLABY Pilot Survey: The Diversity of Ram Pressure Stripping of the Galactic H I Gas in the Hydra Cluster}",
      journal = {\apj},
         year = 2021,
        month = jul,
       volume = {915},
       number = {1},
          eid = {70},
        pages = {70},
          doi = {10.3847/1538-4357/abfc52},
archivePrefix = {arXiv},
       eprint = {2104.13052},
 primaryClass = {astro-ph.GA},
       adsurl = {https://ui.adsabs.harvard.edu/abs/2021ApJ...915...70W}
}

@BOOK{Longair_2011_cre,
       author = {{Longair}, Malcolm S.},
        title = "{High Energy Astrophysics}",
         year = 2011,
       adsurl = {https://ui.adsabs.harvard.edu/abs/2011hea..book.....L}
}

@ARTICLE{hummel_1991_spectral_index,
       author = {{Hummel}, E. and {Dahlem}, M. and {van der Hulst}, J.~M. and {Sukumar}, S.},
        title = "{The large-scale radio continuum structure of the edge-on spiral galaxy NGC 891.}",
      journal = {\aap},
         year = 1991,
        month = jun,
       volume = {246},
        pages = {10},
       adsurl = {https://ui.adsabs.harvard.edu/abs/1991A&A...246...10H}
}

@ARTICLE{klein_1984_spectral_index,
       author = {{Klein}, U. and {Wielebinski}, R. and {Beck}, R.},
        title = "{A survey of the distribution of lambda 2.8 CM radio continuum in nearby galaxies. V. A small sample of edge-on galaxies.}",
      journal = {\aap},
         year = 1984,
        month = apr,
       volume = {133},
        pages = {19-26},
       adsurl = {https://ui.adsabs.harvard.edu/abs/1984A&A...133...19K}
}

@ARTICLE{Matplotlib,
       author = {{Hunter}, John D.},
        title = "{Matplotlib: A 2D Graphics Environment}",
      journal = {Computing in Science and Engineering},
         year = 2007,
        month = jan,
       volume = {9},
       number = {3},
        pages = {90-95},
          doi = {10.1109/MCSE.2007.55},
       adsurl = {https://ui.adsabs.harvard.edu/abs/2007CSE.....9...90H}
}

@ARTICLE{astropy2013,
       author = {{Astropy Collaboration} and {Robitaille}, Thomas P. and {Tollerud}, Erik J. and {Greenfield}, Perry and {Droettboom}, Michael and {Bray}, Erik and {Aldcroft}, Tom and {Davis}, Matt and {Ginsburg}, Adam and {Price-Whelan}, Adrian M. and {Kerzendorf}, Wolfgang E. and {Conley}, Alexander and {Crighton}, Neil and {Barbary}, Kyle and {Muna}, Demitri and {Ferguson}, Henry and {Grollier}, Fr{\'e}d{\'e}ric and {Parikh}, Madhura M. and {Nair}, Prasanth H. and {Unther}, Hans M. and {Deil}, Christoph and {Woillez}, Julien and {Conseil}, Simon and {Kramer}, Roban and {Turner}, James E.~H. and {Singer}, Leo and {Fox}, Ryan and {Weaver}, Benjamin A. and {Zabalza}, Victor and {Edwards}, Zachary I. and {Azalee Bostroem}, K. and {Burke}, D.~J. and {Casey}, Andrew R. and {Crawford}, Steven M. and {Dencheva}, Nadia and {Ely}, Justin and {Jenness}, Tim and {Labrie}, Kathleen and {Lim}, Pey Lian and {Pierfederici}, Francesco and {Pontzen}, Andrew and {Ptak}, Andy and {Refsdal}, Brian and {Servillat}, Mathieu and {Streicher}, Ole},
        title = "{Astropy: A community Python package for astronomy}",
      journal = {\aap},
         year = 2013,
        month = oct,
       volume = {558},
          eid = {A33},
        pages = {A33},
          doi = {10.1051/0004-6361/201322068},
archivePrefix = {arXiv},
       eprint = {1307.6212},
 primaryClass = {astro-ph.IM},
       adsurl = {https://ui.adsabs.harvard.edu/abs/2013A&A...558A..33A}
}

@ARTICLE{astropy2018,
       author = {{Astropy Collaboration} and {Price-Whelan}, A.~M. and {Sip{\H{o}}cz}, B.~M. and {G{\"u}nther}, H.~M. and {Lim}, P.~L. and {Crawford}, S.~M. and {Conseil}, S. and {Shupe}, D.~L. and {Craig}, M.~W. and {Dencheva}, N. and {Ginsburg}, A. and {VanderPlas}, J.~T. and {Bradley}, L.~D. and {P{\'e}rez-Su{\'a}rez}, D. and {de Val-Borro}, M. and {Aldcroft}, T.~L. and {Cruz}, K.~L. and {Robitaille}, T.~P. and {Tollerud}, E.~J. and {Ardelean}, C. and {Babej}, T. and {Bach}, Y.~P. and {Bachetti}, M. and {Bakanov}, A.~V. and {Bamford}, S.~P. and {Barentsen}, G. and {Barmby}, P. and {Baumbach}, A. and {Berry}, K.~L. and {Biscani}, F. and {Boquien}, M. and {Bostroem}, K.~A. and {Bouma}, L.~G. and {Brammer}, G.~B. and {Bray}, E.~M. and {Breytenbach}, H. and {Buddelmeijer}, H. and {Burke}, D.~J. and {Calderone}, G. and {Cano Rodr{\'\i}guez}, J.~L. and {Cara}, M. and {Cardoso}, J.~V.~M. and {Cheedella}, S. and {Copin}, Y. and {Corrales}, L. and {Crichton}, D. and {D'Avella}, D. and {Deil}, C. and {Depagne}, {\'E}. and {Dietrich}, J.~P. and {Donath}, A. and {Droettboom}, M. and {Earl}, N. and {Erben}, T. and {Fabbro}, S. and {Ferreira}, L.~A. and {Finethy}, T. and {Fox}, R.~T. and {Garrison}, L.~H. and {Gibbons}, S.~L.~J. and {Goldstein}, D.~A. and {Gommers}, R. and {Greco}, J.~P. and {Greenfield}, P. and {Groener}, A.~M. and {Grollier}, F. and {Hagen}, A. and {Hirst}, P. and {Homeier}, D. and {Horton}, A.~J. and {Hosseinzadeh}, G. and {Hu}, L. and {Hunkeler}, J.~S. and {Ivezi{\'c}}, {\v{Z}}. and {Jain}, A. and {Jenness}, T. and {Kanarek}, G. and {Kendrew}, S. and {Kern}, N.~S. and {Kerzendorf}, W.~E. and {Khvalko}, A. and {King}, J. and {Kirkby}, D. and {Kulkarni}, A.~M. and {Kumar}, A. and {Lee}, A. and {Lenz}, D. and {Littlefair}, S.~P. and {Ma}, Z. and {Macleod}, D.~M. and {Mastropietro}, M. and {McCully}, C. and {Montagnac}, S. and {Morris}, B.~M. and {Mueller}, M. and {Mumford}, S.~J. and {Muna}, D. and {Murphy}, N.~A. and {Nelson}, S. and {Nguyen}, G.~H. and {Ninan}, J.~P. and {N{\"o}the}, M. and {Ogaz}, S. and {Oh}, S. and {Parejko}, J.~K. and {Parley}, N. and {Pascual}, S. and {Patil}, R. and {Patil}, A.~A. and {Plunkett}, A.~L. and {Prochaska}, J.~X. and {Rastogi}, T. and {Reddy Janga}, V. and {Sabater}, J. and {Sakurikar}, P. and {Seifert}, M. and {Sherbert}, L.~E. and {Sherwood-Taylor}, H. and {Shih}, A.~Y. and {Sick}, J. and {Silbiger}, M.~T. and {Singanamalla}, S. and {Singer}, L.~P. and {Sladen}, P.~H. and {Sooley}, K.~A. and {Sornarajah}, S. and {Streicher}, O. and {Teuben}, P. and {Thomas}, S.~W. and {Tremblay}, G.~R. and {Turner}, J.~E.~H. and {Terr{\'o}n}, V. and {van Kerkwijk}, M.~H. and {de la Vega}, A. and {Watkins}, L.~L. and {Weaver}, B.~A. and {Whitmore}, J.~B. and {Woillez}, J. and {Zabalza}, V. and {Astropy Contributors}},
        title = "{The Astropy Project: Building an Open-science Project and Status of the v2.0 Core Package}",
      journal = {\aj},
         year = 2018,
        month = sep,
       volume = {156},
       number = {3},
          eid = {123},
        pages = {123},
          doi = {10.3847/1538-3881/aabc4f},
archivePrefix = {arXiv},
       eprint = {1801.02634},
 primaryClass = {astro-ph.IM},
       adsurl = {https://ui.adsabs.harvard.edu/abs/2018AJ....156..123A}
}

@ARTICLE{numpy,
       author = {{Harris}, Charles R. and {Millman}, K. Jarrod and {van der Walt}, St{\'e}fan J. and {Gommers}, Ralf and {Virtanen}, Pauli and {Cournapeau}, David and {Wieser}, Eric and {Taylor}, Julian and {Berg}, Sebastian and {Smith}, Nathaniel J. and {Kern}, Robert and {Picus}, Matti and {Hoyer}, Stephan and {van Kerkwijk}, Marten H. and {Brett}, Matthew and {Haldane}, Allan and {del R{\'\i}o}, Jaime Fern{\'a}ndez and {Wiebe}, Mark and {Peterson}, Pearu and {G{\'e}rard-Marchant}, Pierre and {Sheppard}, Kevin and {Reddy}, Tyler and {Weckesser}, Warren and {Abbasi}, Hameer and {Gohlke}, Christoph and {Oliphant}, Travis E.},
        title = "{Array programming with NumPy}",
      journal = {\nat},
         year = 2020,
        month = sep,
       volume = {585},
       number = {7825},
        pages = {357-362},
          doi = {10.1038/s41586-020-2649-2},
archivePrefix = {arXiv},
       eprint = {2006.10256},
 primaryClass = {cs.MS},
       adsurl = {https://ui.adsabs.harvard.edu/abs/2020Natur.585..357H}
}

@software{APLpy,
       author = {{Robitaille}, Thomas and {Bressert}, Eli},
        title = "{APLpy: Astronomical Plotting Library in Python}",
 howpublished = {Astrophysics Source Code Library, record ascl:1208.017},
         year = 2012,
        month = aug,
          eid = {ascl:1208.017},
archivePrefix = {ascl},
       eprint = {1208.017},
       adsurl = {https://ui.adsabs.harvard.edu/abs/2012ascl.soft08017R}
}

@ARTICLE{scipy,
       author = {{Virtanen}, Pauli and {Gommers}, Ralf and {Oliphant}, Travis E. and {Haberland}, Matt and {Reddy}, Tyler and {Cournapeau}, David and {Burovski}, Evgeni and {Peterson}, Pearu and {Weckesser}, Warren and {Bright}, Jonathan and {van der Walt}, St{\'e}fan J. and {Brett}, Matthew and {Wilson}, Joshua and {Millman}, K. Jarrod and {Mayorov}, Nikolay and {Nelson}, Andrew R.~J. and {Jones}, Eric and {Kern}, Robert and {Larson}, Eric and {Carey}, C.~J. and {Polat}, {\.I}lhan and {Feng}, Yu and {Moore}, Eric W. and {VanderPlas}, Jake and {Laxalde}, Denis and {Perktold}, Josef and {Cimrman}, Robert and {Henriksen}, Ian and {Quintero}, E.~A. and {Harris}, Charles R. and {Archibald}, Anne M. and {Ribeiro}, Ant{\^o}nio H. and {Pedregosa}, Fabian and {van Mulbregt}, Paul and {SciPy 1. 0 Contributors}},
        title = "{SciPy 1.0: fundamental algorithms for scientific computing in Python}",
      journal = {Nature Medicine},
         year = 2020,
        month = feb,
       volume = {17},
        pages = {261-272},
          doi = {10.1038/s41592-019-0686-2},
archivePrefix = {arXiv},
       eprint = {1907.10121},
 primaryClass = {cs.MS},
       adsurl = {https://ui.adsabs.harvard.edu/abs/2020NatMe..17..261V}
}

@BOOK{Osterbrock_2006_eline,
       author = {{Osterbrock}, Donald E. and {Ferland}, Gary J.},
        title = "{Astrophysics of gaseous nebulae and active galactic nuclei}",
         year = 2006,
       adsurl = {https://ui.adsabs.harvard.edu/abs/2006agna.book.....O}
}

@ARTICLE{vulcani_2018_sf_burst,
       author = {{Vulcani}, Benedetta and {Poggianti}, Bianca M. and {Gullieuszik}, Marco and {Moretti}, Alessia and {Tonnesen}, Stephanie and {Jaff{\'e}}, Yara L. and {Fritz}, Jacopo and {Fasano}, Giovanni and {Bettoni}, Daniela},
        title = "{Enhanced Star Formation in Both Disks and Ram-pressure-stripped Tails of GASP Jellyfish Galaxies}",
      journal = {\apjl},
         year = 2018,
        month = oct,
       volume = {866},
       number = {2},
          eid = {L25},
        pages = {L25},
          doi = {10.3847/2041-8213/aae68b},
archivePrefix = {arXiv},
       eprint = {1810.05164},
 primaryClass = {astro-ph.GA},
       adsurl = {https://ui.adsabs.harvard.edu/abs/2018ApJ...866L..25V}
}

@ARTICLE{Roberts_2024,
       author = {{Roberts}, I.~D. and {van Weeren}, R.~J. and {Lal}, D.~V. and {Sun}, M. and {Chen}, H. and {Ignesti}, A. and {Br{\"u}ggen}, M. and {Lyskova}, N. and {Venturi}, T. and {Yagi}, M.},
        title = "{Radio-continuum spectra of ram-pressure-stripped galaxies in the Coma Cluster}",
      journal = {\aap},
         year = 2024,
        month = mar,
       volume = {683},
          eid = {A11},
        pages = {A11},
          doi = {10.1051/0004-6361/202347977},
       adsurl = {https://ui.adsabs.harvard.edu/abs/2024A&A...683A..11R}
}

@ARTICLE{Roberts_2021,
       author = {{Roberts}, I.~D. and {van Weeren}, R.~J. and {McGee}, S.~L. and {Botteon}, A. and {Drabent}, A. and {Ignesti}, A. and {Rottgering}, H.~J.~A. and {Shimwell}, T.~W. and {Tasse}, C.},
        title = "{LoTSS jellyfish galaxies. I. Radio tails in low redshift clusters}",
      journal = {\aap},
         year = 2021,
        month = jun,
       volume = {650},
          eid = {A111},
        pages = {A111},
          doi = {10.1051/0004-6361/202140784},
archivePrefix = {arXiv},
       eprint = {2104.05383},
 primaryClass = {astro-ph.GA},
       adsurl = {https://ui.adsabs.harvard.edu/abs/2021A&A...650A.111R}
}

@ARTICLE{Roberts_2022,
       author = {{Roberts}, I.~D. and {van Weeren}, R.~J. and {Timmerman}, R. and {Botteon}, A. and {Gendron-Marsolais}, M. and {Ignesti}, A. and {Rottgering}, H.~J.~A.},
        title = "{LoTSS jellyfish galaxies. III. The first identification of jellyfish galaxies in the Perseus cluster}",
      journal = {\aap},
         year = 2022,
        month = feb,
       volume = {658},
          eid = {A44},
        pages = {A44},
          doi = {10.1051/0004-6361/202142294},
archivePrefix = {arXiv},
       eprint = {2112.08728},
 primaryClass = {astro-ph.GA},
       adsurl = {https://ui.adsabs.harvard.edu/abs/2022A&A...658A..44R}
}

@ARTICLE{Ignesti_2023,
       author = {{Ignesti}, A. and {Vulcani}, B. and {Botteon}, A. and {Poggianti}, B. and {Giunchi}, E. and {Smith}, R. and {Brunetti}, G. and {Roberts}, I.~D. and {van Weeren}, R.~J. and {Rajpurohit}, K.},
        title = "{Radio continuum tails in ram pressure-stripped spiral galaxies: Experimenting with a semi-empirical model in Abell 2255}",
      journal = {\aap},
         year = 2023,
        month = jul,
       volume = {675},
          eid = {A118},
        pages = {A118},
          doi = {10.1051/0004-6361/202346517},
archivePrefix = {arXiv},
       eprint = {2305.19941},
 primaryClass = {astro-ph.GA},
       adsurl = {https://ui.adsabs.harvard.edu/abs/2023A&A...675A.118I}
}

@ARTICLE{vallee_1990_virgo_rm,
       author = {{Vallee}, J.~P.},
        title = "{A Possible Excess Rotation Measure and Large-Scale Magnetic Field in the Virgo Supercluster of Galaxies}",
      journal = {\aj},
         year = 1990,
        month = feb,
       volume = {99},
        pages = {459},
          doi = {10.1086/115343},
       adsurl = {https://ui.adsabs.harvard.edu/abs/1990AJ.....99..459V}
}

@ARTICLE{brentjens_2005_rm_synthesis,
       author = {{Brentjens}, M.~A. and {de Bruyn}, A.~G.},
        title = "{Faraday rotation measure synthesis}",
      journal = {\aap},
         year = 2005,
        month = oct,
       volume = {441},
       number = {3},
        pages = {1217-1228},
          doi = {10.1051/0004-6361:20052990},
archivePrefix = {arXiv},
       eprint = {astro-ph/0507349},
 primaryClass = {astro-ph},
       adsurl = {https://ui.adsabs.harvard.edu/abs/2005A&A...441.1217B}
}

@ARTICLE{condon_1998_nvss,
       author = {{Condon}, J.~J. and {Cotton}, W.~D. and {Greisen}, E.~W. and {Yin}, Q.~F. and {Perley}, R.~A. and {Taylor}, G.~B. and {Broderick}, J.~J.},
        title = "{The NRAO VLA Sky Survey}",
      journal = {\aj},
         year = 1998,
        month = may,
       volume = {115},
       number = {5},
        pages = {1693-1716},
          doi = {10.1086/300337},
       adsurl = {https://ui.adsabs.harvard.edu/abs/1998AJ....115.1693C}
}

@ARTICLE{padovani_2021_cre,
       author = {{Padovani}, Marco and {Bracco}, Andrea and {Jeli{\'c}}, Vibor and {Galli}, Daniele and {Bellomi}, Elena},
        title = "{Spectral index of synchrotron emission: insights from the diffuse and magnetised interstellar medium}",
      journal = {\aap},
         year = 2021,
        month = jun,
       volume = {651},
          eid = {A116},
        pages = {A116},
          doi = {10.1051/0004-6361/202140799},
archivePrefix = {arXiv},
       eprint = {2106.10929},
 primaryClass = {astro-ph.HE},
       adsurl = {https://ui.adsabs.harvard.edu/abs/2021A&A...651A.116P}
}
    \bibliographystyle{aasjournalv7}
    
    
    
    \end{document}